%% file: main.tex
\RequirePackage{fix-cm}
\documentclass[twocolumn,epjc3]{svjour3}  
\smartqed  
\RequirePackage{graphicx}
\journalname{Eur. Phys. J. A}

\usepackage{graphicx}
\usepackage{amsmath}
\usepackage{amssymb}
\usepackage[dvipsnames]{xcolor}
\usepackage{slashed}
\usepackage[normalem]{ulem}
\usepackage{xspace}

\usepackage{hyperref}
\usepackage{enumerate}
\usepackage{mathtools}
\usepackage{physics}
\usepackage{upgreek}
\usepackage{comment}
\DeclarePairedDelimiter{\ceil}{\lceil}{\rceil}
\DeclarePairedDelimiter{\nint}{\lfloor}{\rceil}
\usepackage{subfig}
\usepackage[above,below]{placeins}
\usepackage{float}
\usepackage{booktabs}
\usepackage{multirow}
\usepackage[export]{adjustbox}

\newcommand{\SpB}{\texttt{SPHINCS\_BSSN}\xspace}
\newcommand{\spB}{\texttt{SPHINCS\_BSSN}}

\newcommand{\Lo}{\texttt{LORENE}\xspace}
\newcommand{\comp}{\texttt{CompOSE}\xspace}
\newcommand{\binns}{\texttt{BinNS}\xspace}
\newcommand{\McL}{\texttt{McLachlan}\xspace}
\newcommand{\Extract}{\texttt{Extract}\xspace}
\newcommand{\WeylScal}{\texttt{WeylScal}\xspace}
\newcommand{\Multipole}{\texttt{Multipole}\xspace}

\include{own_latex_commands_i}

\newcommand{\ftinsert}[1]{\color{red}[FT: #1]\color{black}\xspace}

\newcommand{\ftcomment}[1]{\color{red}[FT: #1]\color{black}\xspace}

\newcommand{\spl}{\texttt{SPHINCS\_ID}\xspace}
\newcommand{\massp}{m_\mathrm{p}}

\newcommand{\SKR}[1]{{\color{blue}[SKR: #1]}}

\begin{document}

\title{Simulating neutron star mergers with the Lagrangian Numerical Relativity code SPHINCS\_BSSN
}


\author{ Peter Diener\thanksref{e1,addr1,addr2}
        \and
        Stephan Rosswog\thanksref{addr3}
        \and
        Francesco Torsello\thanksref{addr3} 
}

\thankstext{e1}{e-mail: diener@cct.lsu.edu}


\institute{Center for Computation \& Technology, Louisiana State University, Baton Rouge, LA 70803, USA \label{addr1}
           \and
            Department of Physics \& Astronomy, Louisiana State University, Baton Rouge, LA 70803, USA\label{addr2}
           \and
           Department of Astronomy and Oskar Klein Centre, Stockholm University, AlbaNova, SE-10691 Stockholm, Sweden\label{addr3}
}

\date{Received: date / Accepted: date}

\maketitle

\begin{abstract}
We present the first neutron star merger simulations performed 
with the newly developed Numerical Relativity code \spB. This code
evolves the spacetime  on a mesh using the BSSN formulation, but
matter is evolved via Lagrangian particles according to a high-accuracy 
version of general-relativistic Smooth Particle Hydrodynamics (SPH). Our code contains a 
number of new methodological elements compared to other Numerical
Relativity codes. The main focus here is on the new elements that
were introduced to model neutron star mergers. These include
a) a refinement (fixed in time) of the spacetime-mesh, 
b) corresponding changes in the particle--mesh mapping algorithm 
and c) a novel way to construct SPH initial data for binary
systems via the recently developed ``Artificial Pressure Method." This latter method makes use of
the spectral initial data produced by the library \Lo, and is implemented in a new code called \spl. 
While our main focus is on introducing these new methodological elements
and documenting the current status of \spB, we also show as a first application a set of neutron star merger simulations employing ``soft" ($\Gamma=2.00$)
and ``stiff" ($\Gamma=2.75$) polytropic equations of state.

\end{abstract}

\noindent
\section{Introduction}
\label{intro}

With the first observation of a merging binary black hole in 2015 \cite{abbott16a},
gravitational wave detections have become an active part of observational
astronomy. The first detection of a merging neutron star binary
\cite{abbott17a,abbott17b,abbott17c,abbott17d,smartt17,tanaka17,villar17}, in both gravitational waves {(GWs) and 
electromagnetic (EM) radiation, followed in 2017 and this watershed event (hereafter referred to as GW170817) settled many
long-standing open questions. It showed, among other things, that neutron star mergers
are able to launch relativistic jets and can produce
short gamma ray bursts \cite{kasliwal17,goldstein17,savchenko17,mooley18} and it confirmed the long-held suspicion \cite{symbalisty82,eichler89,rosswog99,freiburghaus99b} that neutron star mergers are
major production sites of r-process elements \cite{tanvir17,kasen17,tanaka17}. For an excellent recent review on r-process nucleosynthesis see \cite{cowan21}. The inspiral phase of GW170817 has also provided
interesting constraints on the tidal deformability of neutron stars \cite{abbott17b,raithel18,malik18} and therefore on the
properties of cold nuclear matter. The combined GW-/EM-detection further allowed
to tightly constrain the propagation speed of gravitational waves 
\cite{abbott17c} and to measure the Hubble parameter \cite{abbott17a}.

To connect multi-messenger observations to the physical processes that
govern the merger and post-merger phases, one needs 3D numerical simulations
that include the relevant physics ingredients. Despite the fact that the
first fully relativistic simulations of a binary neutron star merger were
performed more than two decades ago \cite{shibata00}, the 
simulation of binary neutron star mergers has remained, until today, a formidable
computational physics challenge \cite{baiotti17,shibata17b,shibata19a,radice20}. 
Part of this is related to
the multitude of involved physics
ingredients, such as strong-field gravity, relativistic
hydrodynamics, the hot nuclear matter equation of state (EOS) and neutrino transport. The other part is of purely 
numerical origin and includes, for example, dealing with 
the very sharp transition from high-density neutron star 
matter to vacuum, the accurate evolution of a spacetime,
the treatment of singularities and horizons,
or the huge range of length and 
time scales  in the long-term evolution of ejected 
matter.

Until very recently all Numerical Relativity codes that solve the full set of Einstein equations used an Eulerian 
hydrodynamics framework 
\cite{alcubierre08,baumgarte10,rezzolla13a,shibata17}. 
While these approaches have delivered a
wealth of precious insights into the dynamics and
physics of compact binary mergers, 
they are also not completely free of limitations. For
example, following the small amounts of merger ejecta,
$\sim1\%$ of the binary mass, is a serious challenge 
for such approaches. This is because the advection of matter
is not exact and its accuracy depends on the numerical
resolution (which usually deteriorates with the distance
from the collision site). Moreover, vacuum is usually treated
as a low density atmosphere (but still often of the density of
a  low mass white dwarf star) which can impact the  properties 
of the ejecta. Since the small amounts of
ejecta  are responsible for the entire EM signature, 
we have invested a fair amount of effort
into developing a new methodology that is particularly well-suited
for following such ejecta from a relativistic binary merger.

With a focus on ejecta properties and 
to increase the methodological diversity, we have recently developed
\cite{rosswog21a} the first Numerical Relativity code that solves the 
full set of Einstein equations on a computational mesh, but evolves 
the fluid via Lagrangian particles according to a high-accuracy version
of the Smooth Particle Hydrodynamics (SPH) method. This code has been developed from
scratch and it has 
delivered a number of results that are in excellent agreement with 
established Eulerian codes, see \cite{rosswog21a}. Here we describe the further development of
our new code with a particular focus on simulating 
the first fully relativistic binary neutron star (BNS) merger 
simulations with a Lagrangian hydrodynamics code.

Our paper is structured as follows. Sec. \ref{sec:metho} is dedicated to
the various new methodological elements of our code \SpB (``Smooth Particle Hydrodynamics In Curved Spacetime using BSSN"). 
In Sec. \ref{sec:hydro} we summarize the major hydrodynamics ingredients,
in Sec. \ref{sec:spacetime} we describe our new (fixed) mesh refinement
approach to evolve spacetime  and in 
Sec. \ref{sec:particle_mesh} we describe how the particles and the mesh
interact with each other. Sec. \ref{subsec:ID} is dedicated to 
the detailed description of how we set up relativistic binary systems with
(almost) equal
baryon-number SPH particles based on initial configurations obtained 
with the \Lo library \cite{Gourgoulhon:2000nn,lor,lorene}, using the new code \spl. In Sec. \ref{sec:results} we present our first 
BNS simulation results, in Sec.~\ref{sec:performance} we briefly describe the performance of the current code version and in Sec. \ref{sec:summary} we provide a concise summary with an
outlook on future work.

\section{Methodology}
\label{sec:metho}

Within  \SpB \cite{rosswog21a} we evolve the spacetime, like the more common Eulerian approaches, on a mesh using the BSSN formulation \cite{shibata95,baumgarte99}, but we 
evolve the  fluid via Lagrangian particles in a relativistic SPH  framework, see \cite{monaghan05,rosswog09b,price12a,rosswog15b} for general
reviews of the method. Note that all previous SPH approaches  used approximations
to relativistic gravity, either  Newtonian plus GW-emission backreaction forces
\cite{rosswog99,rosswog02a,lee10a},  Post-Newtonian 
hydrodynamics approaches 
\cite{ayal00,faber00,faber01} that implemented the formalism developed in \cite{blanchet90}, used 
a fixed background metric \cite{laguna93a,liptai19} or 
the conformal flatness approximation \cite{oechslin02,bauswein09}, originally 
suggested in \cite{isenberg80,wilson95a}. The latter approach obtains at each time slice
a static solution to the relativistic field equations and therefore the spacetime
is devoid of gravitational waves. 
For this reason GW-radiation reaction accelerations  have to be added ``by hand" 
in order to drive a binary system towards coalescence.

\SpB is to the best of our knowledge the first Lagrangian hydrodynamics code
that solves the full Einstein field equations.
The code has been documented in detail in the original paper \cite{rosswog21a} and it has been extensively tested 
with shock tubes, oscillating neutron stars in Cowling approximation, oscillating neutron stars in fully dynamical spacetimes
and with unstable neutron stars that either transit from an unstable branch to a stable one or collapse into a black hole. In all
of these tests very good agreement with established Eulerian approaches was found.

Here, we present simulations of the first neutron star mergers with \SpB and to this end we needed
some additional code  enhancements. These are, first, the use of a refined mesh
(currently fixed in time; our original version used a uniform Cartesian mesh), and
second, related modifications to the particle-to-mesh mapping algorithm. The third
new element concerns the construction of initial configurations using an
adaptation of the recently developed ``Artificial Pressure Method" (APM) 
\cite{rosswog20a,rosswog21a} to the case of binary systems. Based on binary solutions
obtained with the library \Lo \cite{lorene}, the APM represents the matter distribution
with optimally placed SPH particles of (nearly) equal baryonic mass.
We focus here mostly on the new
methodological elements of \spB, 
for some technical details
we refer  the reader to the original paper \cite{rosswog21a}.

As conventions, we use $G=c=1$, with $G$ gravitational constant and $c$ speed of light, metric signature ($-,+,+,+$), greek indices run over $0,\ldots,3$ and latin indices over $1,\ldots,3$.
Contravariant spatial indices of a vector quantity $w$ at particle $a$ are denoted as $w^i_a$  and covariant ones will be 
written as $(w_i)_a$. 

\subsection{Time evolution code}
\label{subsec:evolution}

We describe the Lagrangian particle hydrodynamics part of our evolution code  in
Sec.~\ref{sec:hydro} and the spacetime evolution in Sec.~\ref{sec:spacetime}.  A
new algorithm to couple the particles and the mesh
is explained in Sec.~\ref{sec:particle_mesh}. This new
algorithm follows  the ideas of ``Multidimensional Optimal Order Detection" (MOOD) method
\cite{diot13}.

\subsubsection{Hydrodynamics}
\label{sec:hydro}

The simulations are performed in an \textit{a priori} chosen ``computing frame," the line element 
and proper time are given by $ds^2= g_{\mu \nu} \, dx^\mu \, dx^\nu$ and $d\tau^2= - ds^2$ 
and the line element in a 3+1 split of spacetime reads
\be
ds^2= -\alpha^2 dt^2 + \gamma_{ij} (dx^i + \beta^i dt) (dx^j + \beta^j dt),
\ee
where $\alpha$ is the lapse function, $\beta^i$ the shift vector and $\gamma_{ij}$ the spatial 3-metric.
A particle's proper time $\tau$ is related to coordinate time $t$ by
$\Theta d\tau = dt$,
where a generalization of the Lorentz factor
\be
\Theta\equiv \frac{1}{\sqrt{-g_{\mu\nu} v^\mu v^\nu}} \quad {\rm with} \quad v^\alpha=\frac{dx^\alpha}{dt}
\label{eq:theta_def}
\ee
is used. 
This relates to the four-velocity $U^\nu$, normalized to $U^\mu U_\mu= -1$, via
\be
v^\mu= \frac{dx^\mu}{dt}= \frac{dx^\mu}{d\tau} \frac{d\tau}{dt}= \frac{U^\mu}{\Theta}= \frac{U^\mu}{U^0}.
\label{eq:v_mu}
\ee
The equations of motion can be derived from the Lagrangian $L= - \int T^{\mu \nu} U_\mu U_\nu \sqrt{-g}\, dV$,
where $g$ is the determinant of the spacetime metric
and we  use the stress--energy tensor of an ideal fluid
\be
T^{\mu \nu}= (\rho+P)U^\mu U^\nu + P g^{\mu \nu}.
\label{eq:Tmunu}
\ee
Here $P$ is the fluid pressure and
the local energy density (for clarity including the speed of light) is given by
\be
\rho= \rho_{\rm rest} + \frac{u \rho_{\rm rest}}{c^2}= n m_0 c^2 \left(1 + \frac{u}{c^2}\right).
\ee
The  specific internal energy (i.e., per rest mass) is abbreviated as $u$, and $n$ is the 
baryon number density as measured  in the rest frame of the fluid. 
The quantity $m_0$ is the average baryon mass, the exact value of which 
depends on the nuclear composition of the matter. If the matter is constituted 
by free protons of mass $m_p$ and mass fraction $X_p$, free neutrons ($m_n$, $X_n$) 
and a distribution of nuclei where species $i$ has a mass fraction $X_i$, 
a proton number $Z_i$, neutron number $N_i$, mass number $A_i= Z_i + N_i$ 
and binding energies $B_i$, then the average baryon mass is given by
\be
m_0= X_p m_p + X_n m_n + \sum_i X_i \frac{Z_i m_p + N_i m_n - B_i}{A_i}.
\ee
In practice, however, the deviations of the exact value of $m_0$ 
from the atomic mass unit ($m_u= 931.5$ MeV) are small. 
For example, for pure neutrons the deviation is below $0.9 \%$ 
and for a strongly bound nucleus such as iron  the deviation of the average mass   
from $m_u$ would only be $\approx 0.2 \%$.
Therefore, we use in the following 
$m_0 \approx m_u$. We further use from now on the convention 
that all energies are measured in units of $m_0 c^2$ (and 
then use again $c= 1$).
This means practically, that our pressure
is measured in the rest mass energy units of $m_0$, that is, it is the
physical pressure divided by $m_0$, and thus the $\Gamma$-law
equation of state reads $P= (\Gamma-1) n u$. Note that the specific
energy $u$ is, with our conventions, dimensionless and therefore it is
not scaled.

Important choices for every numerical hydrodynamics method are the fluid variables that  are evolved 
in time. We use a density variable that is very similar to what is used in Eulerian approaches \cite{alcubierre08,baumgarte10,rezzolla13a,shibata16}, which, 
with our conventions, reads
\be
N= \sqrt{-g}\, \Theta\, n.
\label{eq:N_def}
\ee
If we decide that each SPH particle $a$ carries a fixed
baryon number $\nu_a$, we can at each step of the
evolution calculate the density at the position of a
particle $a$ via a  summation (rather than by explicitly
solving a continuity equation)
\be
N_a= \sum_b \nu_b\, W(|\vec{r_a} - \vec{r}_b|,h_a),
\label{eq:N_sum}
\ee
where the smoothing length $h_a$ characterizes the support size 
of the SPH smoothing kernel $W$, see below. Here and in all other 
SPH-summations the sum runs in principle over all particles, but since 
the kernel has compact support, it contains only a moderate number 
of particles (in our case exactly 300). We refer to these contributing 
particles as ``neighbours."  As a side remark, we note
that non-zero baryon numbers $\nu_b$ and positive definite SPH kernels $W$ ensure
strictly positive density values at particle positions which makes 
it safe to divide by $N$ in the equations below.

As momentum variable, 
we choose the canonical momentum per baryon which reads (a detailed 
step-by-step derivation of the equations can be found in Sec.~4.2 of 
\cite{rosswog09b})
\be
(S_i)_a = (\Theta \mathcal{E} v_i)_a,
\label{eq:can_mom}
\ee
where $\mathcal{E}= 1 + u + P/n$ is the relativistic enthalpy per baryon.
This quantity evolves according to
\be
\frac{d(S_i)_a}{dt}  =  \left(\frac{d(S_i)_a}{dt}\right)_{\rm hyd} +  \left(\frac{d(S_i)_a}{dt}\right)_{\rm met}
\label{eq:dSdt_full}
\ee
with the hydrodynamic part being
\be
\left(\frac{d(S_i)_a}{dt}\right)_{\rm hyd}= -\sum_b \nu_b \left\{ \frac{P_a}{N_a^2}  D^a_i  +  
\frac{P_b}{N_b^2} D^b_i \right\}
\label{eq:dSdt_hydro}
\ee
and the gravitational part
\be 
\left(\frac{d(S_i)_a}{dt}\right)_{\rm met}= \left(\frac{\sqrt{-g}}{2N} T^{\mu \nu} \frac{\p g_{\mu \nu}}{\p x^i}\right)_a.
\label{eq:dSdt_metric}
\ee
In the hydrodynamic terms we have used the abbreviations
\be
D^a_i \equiv   \sqrt{-g_a} \;  \frac{\p W_{ab}(h_a)}{\p x_a^i} \quad {\rm and} \quad 
D^b_i \equiv    \sqrt{-g_b} \; \frac{\p W_{ab}(h_b)}{\p x_a^i}.
\label{eq:kernel_grad}
\ee
The canonical energy per baryon reads
\be
e_a= \left(S_i v^i + \frac{1 + u}{\Theta}\right)_a = \left(\Theta \mathcal{E} v_i v^i + \frac{1 + u}{\Theta}\right)_a,
\label{eq:can_en}
\ee
and is evolved according to
\be
\frac{d e_a}{dt}= \left(\frac{d e_a}{dt}\right)_{\rm hyd}  + \left(\frac{de_a}{dt}\right)_{\rm met},
\label{eq:energy_equation}
\ee
with
\be
\left(\frac{d e_a}{dt}\right)_{\rm hyd} = -\sum_b \nu_b \left\{ \frac{P_a}{N_a^2}  \;  v_b^i   \; D^a_i +  
\frac{P_b}{N_b^2} \;  v_a^i \; D^b_i \right\}
\label{eq:dedt_hydro}
\ee
and
\be
\left(\frac{de_a}{dt}\right)_{\rm met}= -\left(\frac{\sqrt{-g}}{2N} T^{\mu \nu} \frac{\p g_{\mu \nu}}{\p t}\right)_a.
\label{eq:dedt_metric}
\ee
Note that the physical gradients in the hydrodynamic evolution equations 
are numerically expressed in the above SPH equations (\ref{eq:dSdt_hydro}) and (\ref{eq:dedt_hydro})
by terms involving gradients of the SPH smoothing kernel $W$,  see Eq.~(\ref{eq:kernel_grad}).
This is the most frequently followed strategy in SPH. It is, however, possible
to use more accurate gradient approximations that involve the inversion
of a $3\times3$-matrix. This gradient version was first used in an astrophysical context
by \cite{garcia_senz12} and it possesses the same anti-symmetry with respect to exchanging
particle identities as the standard kernel gradient approach, which makes it possible to ensure numerical conservation exactly in a straightforward way; see, for example, Sec. 2.3 in \cite{rosswog09b} for a detailed discussion of conservation in SPH. This alternative gradient approximation
has been extensively tested in \cite{rosswog15b} and \cite{rosswog20a} and was found
to very substantially increase the overall accuracy of SPH.\footnote{A numerical gradient accuracy measurement showed an improvement of 10 orders of magnitude when
using the matrix inversion gradients, see 
Fig.~1 in \cite{rosswog15b}.}  Following \cite{rosswog15b}, one can simply replace the
quantity $D_i^a$ in Eq.~(\ref{eq:kernel_grad}) by
\be
\tilde{D}_i^a= \sqrt{-g}_a \sum_{d=1}^3 C_{id}(\vec{r}_a,h_a) \; (\vec{r}_b - \vec{r}_a)^d \; W_{ab}(h_a),
\label{eq:IA_gradient}
\ee
and correspondingly for $\tilde{D}_i^b$, where the ``correction matrix" (accounting for 
the local particle distribution) is given by 
\be
C_{ki} (\vec{r},h)= \left( \sum_b \frac{\nu_b}{N_b} (\vec{r}_b - \vec{r})^k (\vec{r}_b - \vec{r})^i W(|\vec{r}-\vec{r}_b|,h)\right)^{-1}.
\label{eq:corr_mat}
\ee
For the simulations in this paper we 
use $\tilde{D}_i^a$ and $\tilde{D}_i^b$ (instead of
$D_i^a$ and $D_i^b$) in 
Eqs.~(\ref{eq:dSdt_hydro}) and (\ref{eq:dedt_hydro}).

With our choice of variables, the  SPH equations retain the ``look-and-feel" of 
Newtonian SPH, although the  variables have a different interpretation. 
As a corollary of this choice, we have to  recover the physical 
variables $n$, $u$, $v^i$, $P$ from the numerically evolved variables 
$N$, $S_i$, $e$ at every step. This ``recovery step" is done in a 
similar way as in Eulerian approaches, the detailed 
procedure that we use is described in Sec.~2.2.4 of \cite{rosswog21a}.

We use a modern artificial viscosity approach to handle shocks, 
where---following the original suggestion of von Neumann and Richtmyer
\cite{vonneumann50}---the physical pressure $P$ is augmented by a viscous pressure 
contribution $Q$. Here we briefly summarize the main ideas, but 
we refer to \cite{rosswog21a}, Sec.~2.2.3, for the explicit expressions.
For the form of the viscous pressure $Q$ we follow \cite{liptai19},
but we make two important changes. First, instead of using ``jumps" in 
quantities between particles (i.e., differences of quantities at 
particle positions) we perform a  slope-limited reconstruction
of these terms to the midpoint between the particles and use the 
difference of reconstructed values from both sides in the artificial 
pressure. This reconstruction in artificial viscosity terms is a new
element in SPH, but it has been shown to be very beneficial in Newtonian 
hydrodynamics \cite{frontiere17,rosswog20a}.
The second change concerns the additional time-dependence
of the amount of dissipation applied.
The expression for the viscous pressure $Q$ contains a parameter which needs to be of order unity for dealing with shocks. 
This parameter can be made time-dependent \cite{morris97,rosswog00,cullen10,rosswog15b} so that it has a close to vanishing value where it is not needed. In steering
this dissipation parameter, we follow the idea of \cite{rosswog20b} to 
 monitor the entropy evolution of each particle.
Since we are simulating an ideal fluid, the evolution should conserve 
a particle's entropy perfectly unless it encounters a shock. Since, in 
our SPH version, entropy conservation is not actively enforced, we 
can use it to monitor the quality of the flow. If a particle either enters 
a shock or becomes ``noisy" for numerical reasons its entropy will 
not be conserved exactly and we use this non-conservation to
steer the exact amount of dissipation that needs to be applied. 
For details of the method we refer to the original paper
\cite{rosswog20b} and to \cite{rosswog21a} for the
\spB-specific implementation.

The SPH equations require a smoothing kernel
function. We have implemented a large variety of
different SPH kernel functions, but here we employ exclusively 
the Wendland $C^6$-smooth kernel \cite{wendland95} that we 
have also used in our original paper \cite{rosswog21a}. This 
kernel has provided excellent results in extensive test series 
\cite{rosswog15b,rosswog20a}. This kernel needs, however, a 
large particle number in its support for good estimates of 
densities and gradients and we therefore assign to each particle a 
smoothing length so that exactly 300 particles contribute 
in the density estimate, Eq.~(\ref{eq:N_sum}). 
This number has turned out to be a good compromise between 
accuracy and computational effort and it is enforced 
{\em exactly} as a further measure the keep the numerical 
noise very low. In practice, we achieve this via a  very
fast ``recursive coordinate bisection" tree-structure \cite{gafton11}
that we use to efficiently search for neighbour particles. For further
implementation details concerning the smoothing length adaptation, we refer to \cite{rosswog20a}.

\subsubsection{Spacetime evolution on a structured mesh}
\label{sec:spacetime}

We have implemented two frequently used variants of the 
BSSN equations in \spB, the ``$\Phi$-method''
\cite{shibata95,baumgarte99} and the ``$W$-method''
\cite{marronetti08,tichy07}. The corresponding code 
was extracted  from the \texttt{McLachlan} thorn
\cite{brown08} in the Einstein Toolkit
\cite{ETK:web,loeffler12}, and we built our own
wrappers to call all the needed functions. The
complete set of BSSN equations is rather lengthy
and will therefore not be reproduced here. It can be
found in Numerical Relativity text books~\cite{alcubierre08,baumgarte10,rezzolla13a,shibata16,baumgarte21} 
and also in the original \SpB paper \cite{rosswog21a}. For all the tests 
presented later in the paper we use the ``$\Phi$-method.'' We expect our simulations 
to be insensitive to this choice, as the two methods
differ mostly in their treatment of black hole
punctures. Hence  the simulations could just as well have been done using the ``$W$-method''.

Our original implementation \cite{rosswog21a} evolved 
the spacetime on a uniform Cartesian grid, but for 
the complex geometry of a neutron star merger a 
(vertex-centered) fixed mesh re\-fine\-ment scheme 
is much more efficient. In this scheme, the first 
refinement level consists of a coarse grid whose outer boundaries 
represent the physical boundaries of the spacetime simulation. 
Each next, finer level has the same number of
grid points, but the boundaries placed at only half of 
the previous level, and consequently this  refinement level has 
twice the resolution. Any number of refinement levels can be 
specified.

As we are using finite differences (FD) to calculate 
the spatial derivatives for the spacetime evolution, 
we need to surround the finer grids with a number of 
ghost grid points that are filled with values 
from the previous, coarser grid via an interpolation
(``prolongation") operator. 
This is illustrated for two levels of refinement
in Figure~\ref{fig:grid}. 
When prolongating in 3D, the {\em fine grid points} can be divided 
into four classes depending on where they are located  on a 
cube defined by eight coarse grid points: 
1) at the corners
(vertices), 2) at the center of edges, 
3) at the center of faces and 4) at the center of the cube.
Hence we do
not need a fully general interpolation scheme and it is
sufficient, and also more efficient, to implement only a small
subset specialized for those four specific cases. In the current
code version, we have linear and cubic polynomial interpolation 
implemented. For linear
interpolation, the interpolated value is simply the 
value of the coinciding coarse grid point for the 
corner case. When the fine grid point is located on
the edge of the cube, it is exactly between two coarse
grid points and linear interpolation reduces to the
average of those two coarse grid points. Similarly
when the fine grid point is located on the face or
center of a cube, it is always located at the
exact center of four (face) or eight (center) coarse
grid points and linear interpolation reduces to just
the average of those coarse grid points.
For the edge-centered fine grid points (for simplicity
we only write the $x$-direction here) the cubic interpolation stencil
consists of four coarse grid points located at $x_i-3\delta x$,
$x_i-\delta x$, $x_i+\delta x$ and $x_i+3\delta x$ with values
$\vec{f_{\mathrm{e}}}=(f_{i-3},f_{i-1},f_{i+1},f_{i+3})$, where 
$\delta x$ is the fine grid spacing. In this case, the
unique interpolating cubic polynomial corresponds to weights 
$\vec{w}_{\mathrm{e}}=(-1/16,9/16,9/16,-1/16)$ so
that the interpolated value at $x_i$ is 
$f_i=\vec{w}_{\mathrm{e}}\cdot\vec{f_{\mathrm{e}}}$. The 
interpolating operators in the $y$- and $z$-directions are 
similarly defined.
For interpolation of face-centered points, the stencil, 
$\vec{f}_{\mathrm{f}}$, will consist of $4\times 4$ coarse grid
points that we flatten into a row vector of length 16 and the
weights are simply the vectorization of the outer product of the
weights for the edge interpolation (producing a column vector of
length 16) $\vec{w}_{\mathrm{f}}=\mathrm{vec}(\vec{w}_{\mathrm{e}}\bigotimes
\vec{w}_{\mathrm{e}})$ and the interpolated value is
$f_i=\vec{w}_{\mathrm{f}}\cdot\vec{f_{\mathrm{f}}}$. In practice the
flattening is done in the $x$-direction first, but the order does not matter
since the outer product of weights for the edge interpolation is symmetric.

Finally, for interpolation of the cube centered points, the stencil,
$\vec{f}_{\mathrm{c}}$, will consist of $4\times 4\times 4$ coarse grid points
flattened into a row vector of length 64 and, again, the weights are
simply the vectorization of the outer product of the weights for the 
edge interpolation $\vec{w}_{\mathrm{c}}=\mathrm{vec}(\vec{w}_{\mathrm{e}}\bigotimes
\vec{w}_{\mathrm{e}}\bigotimes \vec{w}_{\mathrm{e}})$ (this time producing
a column vector of length 64) and the interpolated value is $f_i=\vec{w}_{\mathrm{c}}\cdot\vec{f_{\mathrm{c}}}$.
\begin{figure}[!t]
    \centering
    \includegraphics[width=8.5cm]{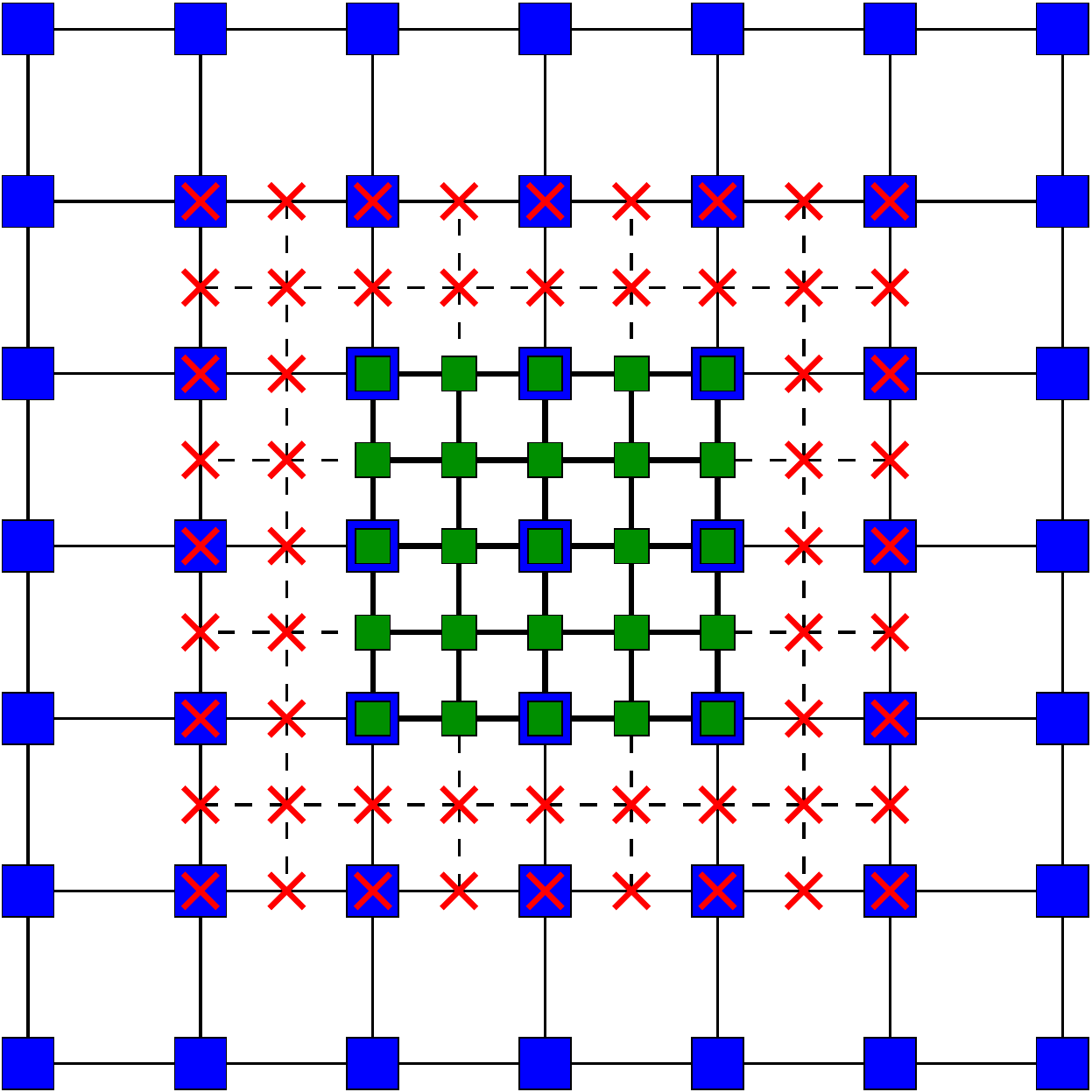}
    \caption{An illustration of 2 levels of grid refinement used in the code. The coarse 
    grid points are shown as blue squares, the fine grid points as green ones and the 
    ghost points (needed for the finite difference operators near the boundary of the 
    fine grid) are marked as red crosses.}
    \label{fig:grid}
\end{figure}

During the evolution, we update the values in the coarse grid
points wherever possible with values from the fine (more accurate) grid (``restriction"). As 
coarse grid points that need restriction always coincide with a
fine grid point, restriction simply consists of copying the value 
from the fine grid point to the coarse grid point.

We integrate our coupled system of hydrodynamics and BSSN equations via a 3rd order
Total Variation Diminishing (TVD) Runge--Kutta integrator
\cite{gottlieb98}. During each substep, the right-hand-sides (RHS) for
the BSSN variables are calculated using finite differences (FD) 
where possible, that is, everywhere except at the ghost points (taking into account the 
size of the FD stencil) on all refinement levels. The state 
vector (consisting of all evolved variables on all possible 
grid points) is then evolved forward in time on all refinement levels where the 
RHS has been computed. After that, the state vector is updated in
the ghost zones on each fine grid, via prolongation from the next-coarser
grid, in a loop over refinement levels starting from the 
coarsest level. Finally, the solution on the coarser grids are 
updated via restriction from the finer grids in a second loop over 
refinement levels starting from the finest refinement level. 

When interpolating metric information from the grid to the 
particles, we first
find out which refinement level to interpolate from. Obviously, we
want to use the finest possible refinement level in order
to get the most accurate metric information on the particles. We 
therefore start on the finest refinement level and check whether
the considered particle is located inside this grid. If it is, we perform 
the interpolation the same way as described in \cite{rosswog21a}.
If it is not, we move on to the next refinement level. We repeat 
until the interpolation can be performed or we reach the coarsest 
refinement level. Handling particles that leave the coarsest grid
is not necessary for the cases presented here, but it will be 
implemented for future studies of merger ejecta.

\subsubsection{Particle--mesh coupling: A MOOD approach}
\label{sec:particle_mesh}

A crucial step in our approach is the mapping of  $T_{\mu\nu}$, originally known at the particle positions, to the mesh (``P2M-step")
and the mapping of the metric acceleration terms
$(dS_i/dt)_{\rm met}$ and $(de/dt)_{\rm met}$ from the mesh to the particle positions (``M2P-step"), see Eqs.~\eqref{eq:dSdt_metric} and \eqref{eq:dedt_metric}. In this paper, we provide
a further refinement of the P2M-step, see below, whereas the M2P-step is the same as in our original paper \cite{rosswog21a}.

We map a quantity $A$ known at particle positions $\vec{r}_p=(x_p,y_p,z_p)$ to the grid point $\vec{r}_g$
via 
\begin{equation}
A_g= A(\vec{r}_g)=  \frac{\sum_p V_p A_p \,\Psi_g(\vec{r}_p)}{\sum_p V_p \, \Psi_g(\vec{r}_p)},
\label{eq:P2M_mapping}
\end{equation}
where $V_p= \nu_p/N_p$ is a measure of the particle volume. 
We construct the  functions $\Psi$ as  tensor products of
1D shape functions
\begin{equation}
\Psi_g(\vec{r}_p)= \Phi\left(\frac{|x_p-x_g|}{l_p}\right) \Phi\left(\frac{|y_p-y_g|}{l_p}\right) \Phi\left(\frac{|z_p-z_g|}{l_p}\right),
\label{eq:T_product}
\end{equation}
where $l_p=V_p^{1/3}$. After extensive experimenting, we had settled in the 
original paper on a hierarchy of shape functions that 
have been developed in the context of vortex methods \cite{cottet00}.

The interpolation quality of the shape functions is closely related 
to the order with which they fulfill the so-called moment conditions \cite{cottet00}
\be
\sum_q x_q^\alpha \Phi(x_q-x)= x^\alpha \quad {\rm for} \quad 0 \le |\alpha| \le m-1,
\ee
for points located at $x_q$. This  means that the interpolation is exact for polynomials of a degree less than or equal to $m-1$ and such an interpolation is said to be ``of order $m$" \cite{cottet00}.
Good interpolation quality, however, does not
automatically guarantee the smoothness of the shape
functions, understood as the number of continuous derivatives.
In fact, a number of shape functions that are commonly used,
e.g.  in particle-mesh methods for plasmas,
are of low smoothness only and this can
introduce a fair amount of noise in simulations. 
Being of higher order $m$, however, comes at the price of a larger stencil 
and therefore at some computational expense. Positive definite 
functions can only be maximally of order two \cite{monaghan92}, 
for higher orders the shape functions need to contain negative parts which 
can become problematic when the particle distribution within the 
kernel is far from being isotropic. A particularly disastrous 
case is encountering a sharp (e.g. stellar) surface. Here,
violent Gibbs-phenomenon like oscillations can occur
that can lead to unphysical results. For this reason
we have implemented a hierarchy of kernels, so that 
the highest order shape functions can be used when it
is safe, and less accurate, but more robust functions 
are used when it is not. How this is decided and
implemented is described below. 

Last, but not least, another characteristics of shape functions 
is their highest involved polynomial  degree 
(``degree").\footnote{We mention the degree here only for completeness.} 
In this work we use smooth shape functions of high 
order that were constructed by Cottet et al. \cite{cottet14}. 
We follow their notation of using $\Lambda_{p,r}$ for a 
function that is of order $p+1$ (i.e., reproduces polynomials up to order $p$) and of smoothness $C^r$.
Our chosen shape functions are the following}:
\begin{enumerate}[(i)]
    \item the $\Lambda_{4,4}$-kernel (order 5, regularity $C^4$ and degree 9) \cite{cottet14}  
    \vspace{-0mm}
     \begin{align}
       &\Lambda_{4,4} (|x|)   = \nonumber \\[2mm] 
       &\hspace{-8mm} \left\{\begin{array}{lll}
                1 - \dfrac{5}{4}|x|^2 + \dfrac{1}{4} |x|^4 - \dfrac{100}{3} |x|^5 + \dfrac{455}{4} |x|^6 &\\[3mm]
               -\dfrac{295}{2} |x|^7 + \dfrac{345}{4} |x|^8 - \dfrac{115}{6}|x|^9, & |x| <  1,\\[4mm]
         -199 + \dfrac{5485}{4} |x|  - \dfrac{32975}{8} |x|^2  &\\[3mm]
         + \dfrac{28425}{4} |x|^3 -\dfrac{61953}{8}|x|^4 + \dfrac{33175}{6} |x|^5 &\\[3mm]
         - \dfrac{20685}{8} |x|^6 + \dfrac{3055}{4} |x|^7 - \dfrac{1035}{8} |x|^8 &\\[3mm]
                + \dfrac{115}{12} |x|^9, & 1 \le |x| <  2,\\[4mm]     
          5913 - \dfrac{89235}{4} |x| + \dfrac{297585}{8} |x|^2 &\\[3mm]
          -\dfrac{143895}{4} |x|^3 + \dfrac{177871}{8} |x|^4- \dfrac{54641}{6} |x|^5 &\\[3mm]
          + \dfrac{19775}{8}|x|^6- \dfrac{1715}{4} |x|^7 + \dfrac{345}{8} |x|^8 &\\[3mm]
           - \dfrac{23}{12} |x|^9, & 2 \le |x| <  3, \\[4mm]  
      0, & {\rm else}. 
        \end{array}\right. 
    \end{align}
    \item   $\Lambda_{2,2}$ kernel (order 3, smoothness $C^2$, degree 5 \cite{cottet14} 
    \begin{align}
            \hspace{-5mm} \Lambda_{2,2} (|x|) = \left\{\begin{array}{llr}
        1 - |x|^2 - \dfrac{9}{2} |x|^3 + \dfrac{15}{2}|x|^4 &\\[3mm]
        - 3 |x|^5, & \; |x| <  1,\\[4mm]
        -4 + 18 |x| -29|x|^2 &\\ [2mm]
        + \dfrac{43}{2}|x|^3 - \dfrac{15}{2} |x|^4 + |x|^5 ,  & \; 1\leq |x| <2,\\[4mm]
        0, & \; {\rm else}.
       \end{array}\right. 
    \end{align}
    \item  and, finally, the $M_4$ kernel (order 2, smoothness $C^2$, degree 3) \cite{cottet00}
    \begin{align}
        \hspace{-6mm} M_4 (x) = \left\{\begin{array}{llr}
\dfrac{1}{6}\big(2-|x|\big)^3 - \dfrac{2}{3}\big(1-|x|\big)^3, & |x| <  1,\\[4mm]
\dfrac{1}{6}\big(2-|x|\big)^3,  & 1 \leq |x| < 2,\\[4mm]
0, & {\rm else.}
\end{array}\right. 
    \end{align}
\end{enumerate}
Note that $\Lambda_{4,4}$ and $\Lambda_{2,2}$ are, differently 
from usual SPH-kernels, not positive definite. The supports of these 
kernels are sketched (for a 2D example)
in Fig.~\ref{fig:support}.  We  
apply a hierarchy of these kernels starting with $\Lambda_{4,4}$ followed by $\Lambda_{2,2}$ and we use
the safest (positive definite, but least accurate) kernel $M_4$ as a ``parachute."

In the original paper we applied a heuristic  method based on the particle content of the neighbour cells to decide 
which kernel to use. Instead of this, we use here a Multidimensional Optimal Order Detection (MOOD) method.
The main idea is to use a ``repeat-until-valid" approach, that is, to use
the most accurate kernel that does not lead to any artifacts. To detect artifacts we check whether
the resulting grid values of $T^g_{\mu\nu}$ are outside of the 
range of the values of the contributing particles, see below. 
This strategy is actually similar to MOOD approaches that are used in hydrodynamics; see, for example, \cite{diot13}. Specifically,
we proceed according to the following steps:
\begin{enumerate}[(i)]
    \item Start with the highest order kernel $\Lambda_{4,4}$ for the mapping.
    \item Check whether the $\Lambda_{4,4}$-result is acceptable: if  all components of $T^g_{\mu\nu}$  at a given grid point are {\em inside} of 
    the bracket $[T^{\rm min}_{\mu\nu},T^{\rm max}_{\mu\nu}]$, 
    where the maximum and minimum values refer to the particles 
    inside the $\Lambda_{4,4}$ support, we accept the 
    $\Lambda_{4,4}$-mapping. Otherwise, we consider the mapping as questionable and proceed to the next kernel, $\Lambda_{2,2}$.\footnote{In practice, all three kernel options are calculated in the same loop, so that no re-mapping is needed, we
    only need to choose the highest-order, but still valid option.}
    \item Check whether the $\Lambda_{2,2}$-result is acceptable: as in the previous step, we check whether the grid-result is outside of the
    bracket given by the particles in the support of $\Lambda_{2,2}$. If it is not, the $\Lambda_{2,2}$-mapping is accepted. 
    \item If also the $\Lambda_{2,2}$-result is
    not acceptable, we resort to our ``parachute" solution, the positive definite $M_4$-mapping.
\end{enumerate}

\begin{figure}[!tp] 
\centering
   \hspace*{-2mm}\includegraphics[width=8.7cm]{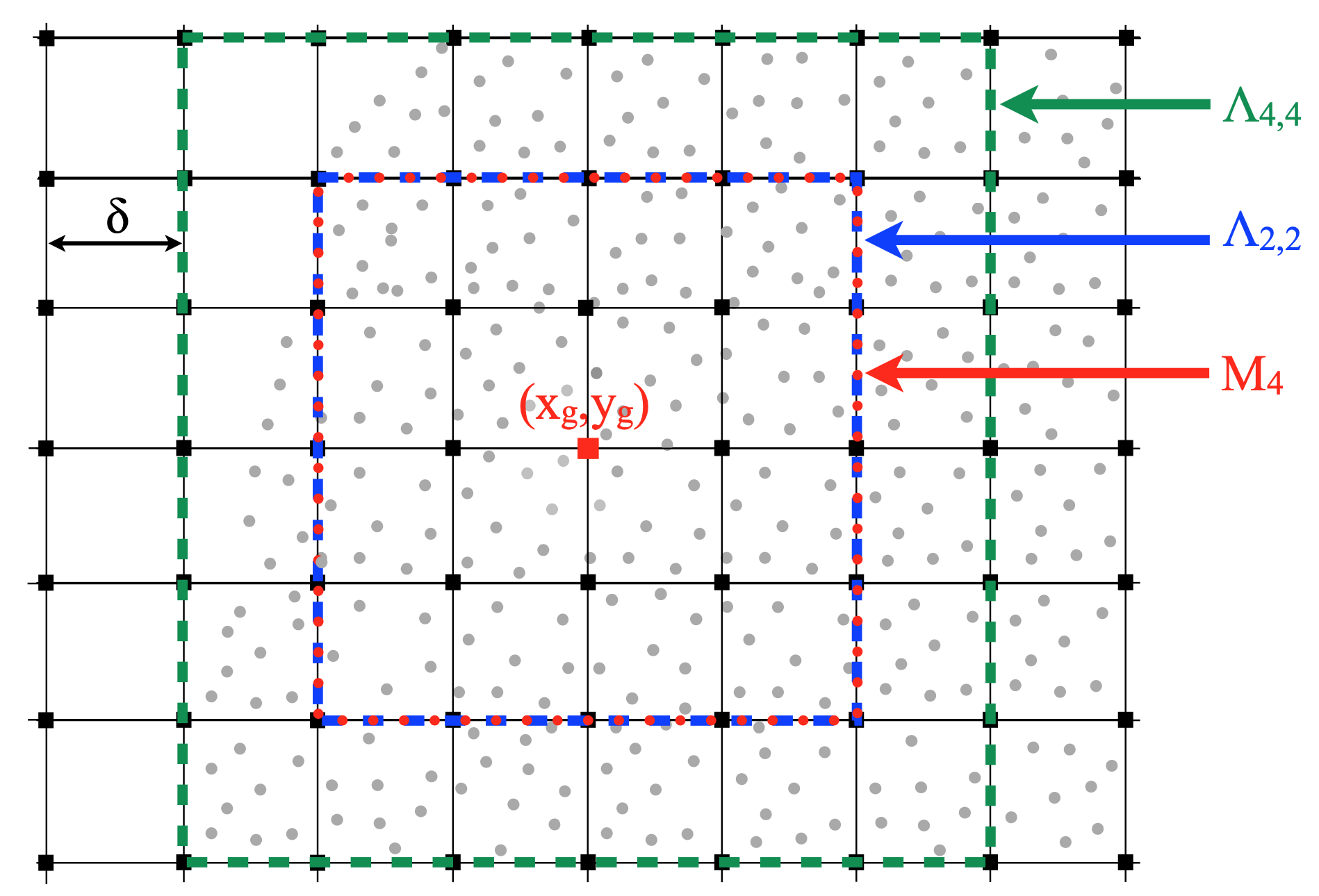}
   \caption{2D sketch of the support sizes for the different kernels that are used in the MOOD-mapping
                 for a specific grid point (red square).
                 The green square indicates the support of the $\Lambda_{4,4}$-kernel, blue refers to $\Lambda_{2,2}$ and red to the ``parachute" kernel $M_4$.}
   \label{fig:support}
\end{figure}

\begin{figure}[ht]
    \centering
    \includegraphics[width=0.5\textwidth]{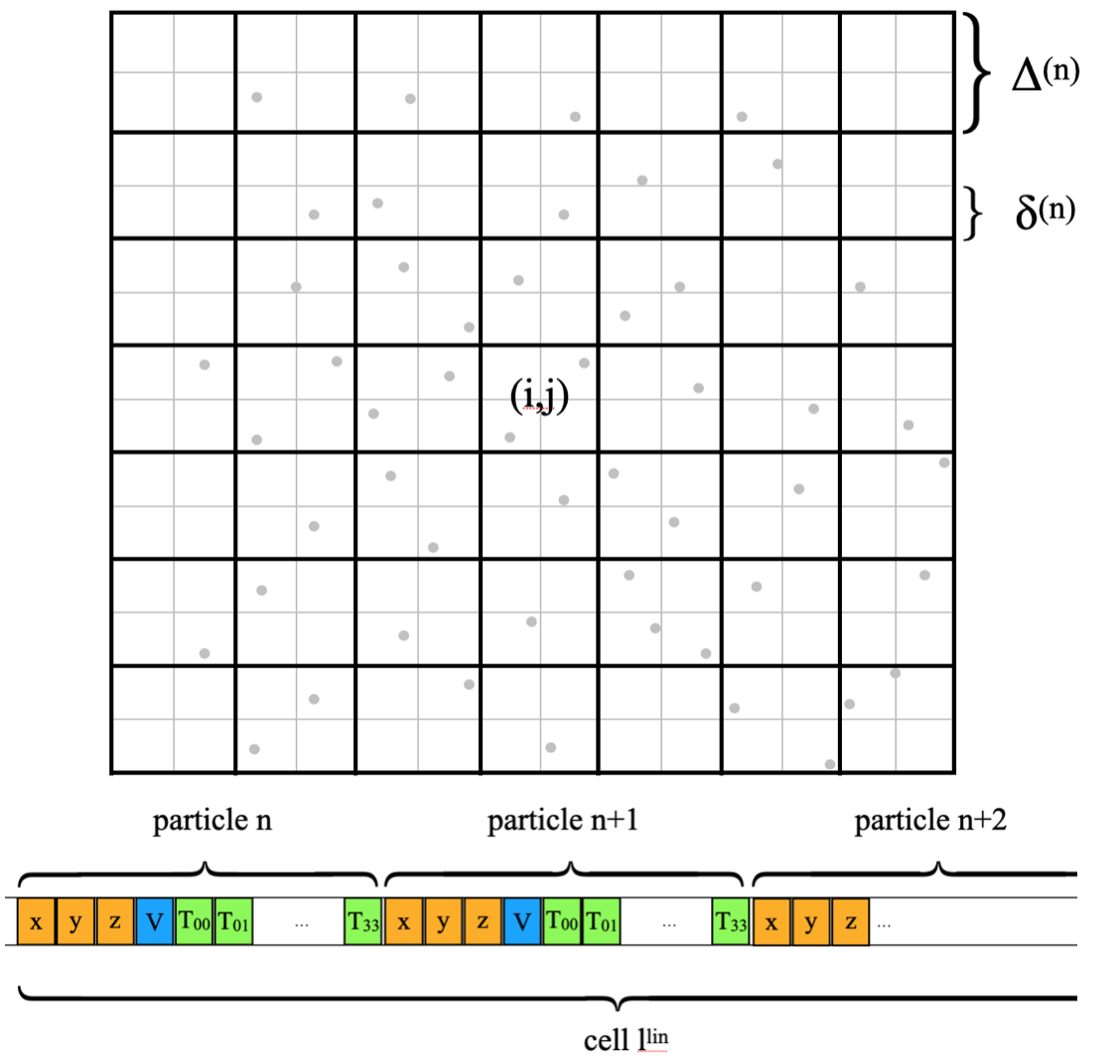}
    \caption{For an efficient mapping of the particle-$T_{\mu\nu}$ to the gravity grid on refinement level $n$ (on which the spacetime is evolved; grid spacing $\delta^{(n)}$; thin lines), we employ an additional hash grid with twice the grid-spacing ($\Delta^{(n)}$; thick lines). This is illustrated for the 2D-case in the upper part of the figure. The actual particle data needed for the mapping is stored in a cache-friendly linear array, see the lower part of the figure and the main text for more details.
    }
    \label{fig:hash_grid}
\end{figure}

\paragraph{Efficient implementation via a hash grid.} The tasks involved in the P2M-step are a) identify all particles that contribute
to any given grid point, see Eq.~(\ref{eq:P2M_mapping}), that is, those
particles that are within the (tensor-product) support of the grid point's kernel, b) loop over all grid points and add 
the particle $T_{\mu\nu}$-contributions.\\

We have implemented an efficient P2M-step that involves a hash grid at each 
refinement level $n$. The mesh size of our hash grid $\Delta^{(n)}$ is chosen to be
twice the mesh size of the gravity grid, $\delta^{(n)}$, see upper part of
Fig.~\ref{fig:hash_grid} for an illustrating sketch in 2D. 
The reason why we chose a factor of two is that when updating the stress--energy 
tensor at a given gravity grid point, we need to collect the contributing particles from 
the hash grid. We do not want to have to check too many hash grid cells, so the hash 
grid cells should be larger than the gravity grid cells. On the other hand, they should 
not be too large, since then we would loop over many particles that in the end do not 
contribute to the grid point. A factor of two fulfils these criteria and makes the involved 
details (e.g., getting the hash grid indices from the gravity grid indices) very simple.

For performance reasons we
store the data needed in the P2M-mapping process in a simple linear 
cache-array. The hash grid is only ``virtual" in the sense that its structure is only 
needed to identify particles belonging to the same hash grid cell, but the data 
is actually stored in the linear cache-array. We first identify the 3D indices
$(i,j,k)$ of the hash grid cell that contains the particle.
These indices are then translated 
into a single index in our linear cache-array via 
$l^{\rm lin}= i + (j-1)n_x + (k-1)n_xn_y$, where $n_x,n_y, n_z$ are 
the numbers of hash grid cells in the $x$-, $y$- 
and $z$-direction on the current mesh refinement level. The index $l^{\rm lin}$ 
labels a segment in the cache-array that contains all the data associated with the particle 
in hash grid cell $(i,j,k)$ that is needed for the mapping. Since all data is stored in exactly the order
in which it is needed, this approach guarantees virtually
perfect cache efficiency.\\
The particles are filled into this hash grid as follows:
\bi
\item In one first linear loop over the particles, each particle determines in which hash cell $(i,j,k)$ it is
located. Each of the hash cells keeps track of how many particles it contains. 
\item After this loop, we can quickly count the number of entries, so that each ($i,j,k$) hash cell knows how many entries
there are in the cache-array corresponding to the previous cells. In other words, after this (very fast) counting step,
each hash grid cell knows the starting and finishing index
that define the cache-array section storing all the properties of the particles contained in this ($i,j,k$) hash cell. Hence, all subsequent summations can be performed very efficiently.
\item In another linear loop, those particle properties  that are needed during
the actual mapping are filled into the 1D array in exactly the same order in which they  will be addressed during the mapping step (position, particle volume, stress--energy tensor components): 
$(x,y,z,V,T_{00},T_{0x}, ... ,T_{zz})$. The resulting cache-array is sketched in the bottom part of
Fig.~\ref{fig:hash_grid}. This cache-array approach has the advantage that the array has a fixed maximum length of 14 times the SPH particle number. We apply this mapping sequentially for each grid refinement level. If all particles are contained
in the grid of a given level, the cache-array is completely filled, otherwise it is shorter since it does not contain entries from the particles outside the grid. Most importantly, the cache-array  has to be
allocated only once during the simulation, its size is known at compile time, and it can be used for every refinement
level as a memory-efficient, ``read-only" data structure in the P2M-step.
\item The actual contribution-loop for each grid cell is then performed by checking the particle content
of each potentially contributing hash cell and, if applicable,  the particle contribution is added according
to Eq.~(\ref{eq:P2M_mapping}).
\ei
Compared to our initial, straightforward implementation, the above described cache-efficient P2M-step is  more than 20 times faster for the simulations shown in this paper. 

\subsubsection{Code performance}
\label{sec:performance}
\begin{table*}
\caption{Breakdown of the time spent in different parts of the code.
The different parts of the code are, in order: computation of the
right-hand-sides (RHSs) in the hydro (SPH) and BSSN equations, mapping of
the stress--energy tensor from the particles to the grid (P2M), mapping of 
the metric from the grid to the particles (M2P), restriction operators, 
prolongation operators, and finally the state vector update (see 
Sec.~\ref{subsec:evolution} for more details). Note that not all parts of
the code are listed, so numbers do not add up to 100\%. The symbols LR, MR and HR 
refer to low, medium and high resolution, see Sec.~\ref{sec:initial_setup}.}
\centering
\begin{tabular}{cccccccc}
\toprule
    resolution & hydro RHS & BSSN RHS & P2M & M2P & restriction & prolongation & update state vector \\
\midrule
    LR & 29.5\% & 33.6\% & 12.1\% & 1.8\% & 1.4\% & 5.8\% & 8.7\% \\
    MR & 31.3\% & 31.9\% & 17.2\% & 1.7\% & 0.6\% & 3.5\% & 6.5\% \\
    HR & 35.9\% & 23.1\% & 14.1\% & 1.7\% & 0.6\% & 3.6\% & 7.1\% \\
\bottomrule
\end{tabular}
\label{tab:performance}
\end{table*}

The code is written in modern Fortran, except for the C++ routines 
which we 
extracted from \McL from the Einstein Toolkit in order
to be able to evolve the spacetime. Currently the code is
parallelized for shared memory architectures using OpenMP directives
and pragmas. As the time spent in different parts of the code
depends on the particle distribution, it is impossible to uniquely
quantify the performance, but based on the simulations presented
here, we can give some representative numbers.

In \autoref{tab:performance} we give a breakdown of how the time is 
spent in various important parts of the code for representative low,
medium and high resolution runs. As can be seen, the major part of
the time is spent in the hydrodynamics and spacetime evolution
routines with additional significant time spent in the mapping of
the stress--energy tensor to the grid.
Running the code on compute nodes equipped with 128 AMD EPYC 7763
processors it initially takes 1.3, 3.2, 7.8 minutes to evolve one
code unit of time at low, medium and high resolution respectively. 
This is maintained for the duration of the inspiral but as smaller
time steps are required for hydrodynamic stability during and 
after the merger, the times required to evolve 1 code unit of time
at the end are 2.2, 5.1 and 11.3 minutes at low, medium and high
resolution, respectively. The memory usage for one of the medium 
resolution runs presented here is about 30 GB.\\
The OpenMP scaling is decent, but can certainly be improved. Scaling
experiments show a speedup of about 43 on 128 cores.

\subsection{Initial data}
\label{subsec:ID}

A crucial ingredient for every relativistic simulation are initial data (ID) that 
both satisfy the
constraint equations and accurately describe the physical system of interest. In the following, we describe how we compute such
ID for BNS, that we subsequently
evolve with \spB.

\subsubsection{Quasi-equilibrium BNS with \Lo}
\label{subsec:lorene}

The library \Lo computes ID for relativistic BNS under the assumption of ``quasi-equilibrium," see  \cite{Gourgoulhon:2000nn} and \cite[Sec.~9.4]{gourgoulhon20123+1} for more details. This assumption states that the radial velocity component of the stars is negligibly small compared
to the azimuthal one and the orbital evolution is essentially realized via
a sequence of circular orbits. Equivalently, one assumes that a helical Killing vector field exists.
This assumption is reasonably well justified since at a separation of 
$\sim 50$ km, the time derivative of the orbital period (at the second post-Newtonian level) is about $2\%$ of the period itself \cite{Gourgoulhon:2000nn}.

\Lo allows to compute ID for corotational and irrotational binaries. Since 
any neutron star viscosity is too small to spin up the neutron stars
during the rapid final inspiral stages to corotation \cite{bildsten92,kochanek92,Gourgoulhon:2000nn}, the irrotational case
is generally considered more realistic, but mergers with rapidly
spinning neutron stars can lead to interesting effects and the corresponding 
simulations are now feasible \cite{bernuzzi14,dietrich15,dudi21}. In this paper, we restrict ourselves to irrotational binaries.
Another assumption that is made in \Lo is the conformal flatness of the 
spatial metric $\gamma_{ij}$. This is a commonly made approximation
which substantially simplifies the solving of the constraint equations.

\Lo provides spectral solutions that can be evaluated at any point. By default,
the code \binns in \newline\Lo allows to export the \Lo ID to a Cartesian grid. However,
this is not sufficient for our purposes, since we need to evaluate the solution
not only on our refined mesh, but also at the positions of the SPH particles. Hence, 
we have extended \binns to evaluate the spectral data at any given
spacetime point. We have further linked the relevant functions in 
\binns to our own code that sets up the ID for \SpB and is called \spl.
Since in the original version of \Lo, \binns was able to handle only BNS with single polytropic EOS, we extended it to read and export also configurations with piecewise polytropic and tabulated EOS such as those available in the \comp database \cite{compose}. In other words, all the needed information that \Lo provides about the BNS is accessible within \spl. Despite having these capabilities, we restrict ourselves to purely polytropic equations of state in this first study.

\subsubsection{The ``Artificial Pressure Method" for binary systems}
\label{subsec:apm}

SPH initial conditions are a delicate subject. The particle distribution has to,
of course, accurately represent the physical system under consideration (e.g., densities and velocities),
but there are additional requirements concerning how the particles are actually distributed, see, for example,
\cite{rosswog15b,diehl_rockefeller_fryer_riethmiller_statler_2015,arth_2019}.
The particle arrangement should be stable, so that in an equilibrium situation,
the particles ideally should not move at all. Many straightforward particle arrangements (such as
cubic lattices), however,  are {\em not} stable and particles quickly start to move 
if set up in this way. The reason is a ``self-regularization" mechanism built into SPH that tries to locally optimize the particle distribution (see, e.g., Sec. 3.2.3 in \cite{rosswog15c} for more details). To make things even harder, the particles should be distributed
very regularly, but {\em not} in configurations with pronounced symmetries,
since the latter can lead to artifacts, for example if a shock propagates along
such a symmetry direction.\footnote{See, e.g., Fig. 17 in \cite{rosswog15b} for an illustration.} Last, but not least, the SPH particles should ideally
have practically the same masses/baryon numbers, since for large differences numerical
noise (i.e., small scale fluctuations in the particle velocities) occurs that can have
a negative impact on the interpolation accuracy.

All of these issues are addressed by the ``Artificial Pressure Method" (APM) 
to optimally place SPH particles. The method has been suggested in a Newtonian 
context \cite{rosswog20a} and was recently generalized to the case of general 
relativistic neutron stars \cite{rosswog21a}. In this work, we go one step 
further and apply the APM in the construction of relativistic binary systems.

The main idea of the APM is to start from a
distribution of equal-mass particles and let the
particles find the positions in which they minimize
their density error with respect to a desired 
density profile. In other words, each particle
has to measure in which direction the density error
is decreasing and move accordingly.

In practice, this is achieved by
assigning to each particle $a$ an ``artificial pressure," $\pi_a$, 
that is based on the discrepancy between actual,
measured SPH-density, $\rho_a$,  and the desired profile 
value at the position of the particle, $\rho^P(\vec{r}_a)$:
\be
\pi_a= {\rm max}\left(1 + \frac{\rho_a-\rho^P(\vec{r}_a)}{\rho^P(\vec{r}_a)}, 0.1 \right)
\ee
where the max-function has been used to avoid negative pressures. For more details 
of the method we refer to the original paper \cite{rosswog20a}.
This pressure is then used in a position update formula
that is modelled after a fluid momentum equation (see \cite{rosswog20a} for the derivation):
\be
{\delta \vec{r}_a}^{\rm APM}= -\frac{1}{2}h_a^2\nu_0 \sum_b \frac{\pi_a+\pi_b}{N_b} \nabla_a W_{ab}(h_a),
\ee
where the quantities have the same meaning as in the fluid section,
Sec.~\ref{sec:hydro}, and $\nu_0$ is the baryon number desired for each particle (equal to the total baryon mass divided by the chosen particle number). Iterations with such position updates drive the particles
towards positions where they minimize their density error. It does, however, not 
necessarily guarantee that the particle distribution is locally regular
and provides a good interpolation accuracy (for corresponding criteria see, 
e.g., Sec. 2.1 in \cite{rosswog15c}).
To achieve this, we add a small additional ``regularization term" 
\cite{gaburov11}\footnote{Note the typo in our original \SpB paper \cite{rosswog21a}:
in Eq.~(100) the summation symbol is missing.}
\be
{\delta \vec{r}_a}^{\rm reg}= h_a^4 \sum_b W_{ab}(h_a) \hat{e}_{ab},
\ee
where $\hat{e}_{ab}= (\vec{r}_a-\vec{r}_b)/|\vec{r}_a-\vec{r}_b|$,
so that the final position update reads
\be
\vec{r}_a \rightarrow  \vec{r}_a + (1-\zeta){\delta \vec{r}_a}^{\rm APM} + \zeta {\delta \vec{r}_a}^{\rm reg}.
\ee
The parameter $\zeta$ determines how important the local regularity requirement on the particle distribution is compared to the accuracy with which the desired density profile is reproduced. Since our main emphasis is on reproducing the desired density, the value of $1-\zeta$ should not be too much below unity, but the exact parameter value is only of minor importance. We find good results for $\zeta=0.05$, which we use throughout this paper. We start the APM iteration with an initial particle distribution obtained by first placing particles on spherical surfaces, as described in \ref{subsec:app-spherical}, and then slightly moving them away from the surfaces by a small random displacement. We noticed that the use of such random displacements led to better results of the APM procedure.

\begin{figure*}[!ht]
    \centering
    \subfloat[Initial particle distribution used for the APM iteration. The particles are placed on spherical surfaces first (see \ref{subsec:app-spherical} for more details), and then  randomized by a small amount.]{\includegraphics[width=\textwidth,valign=b]{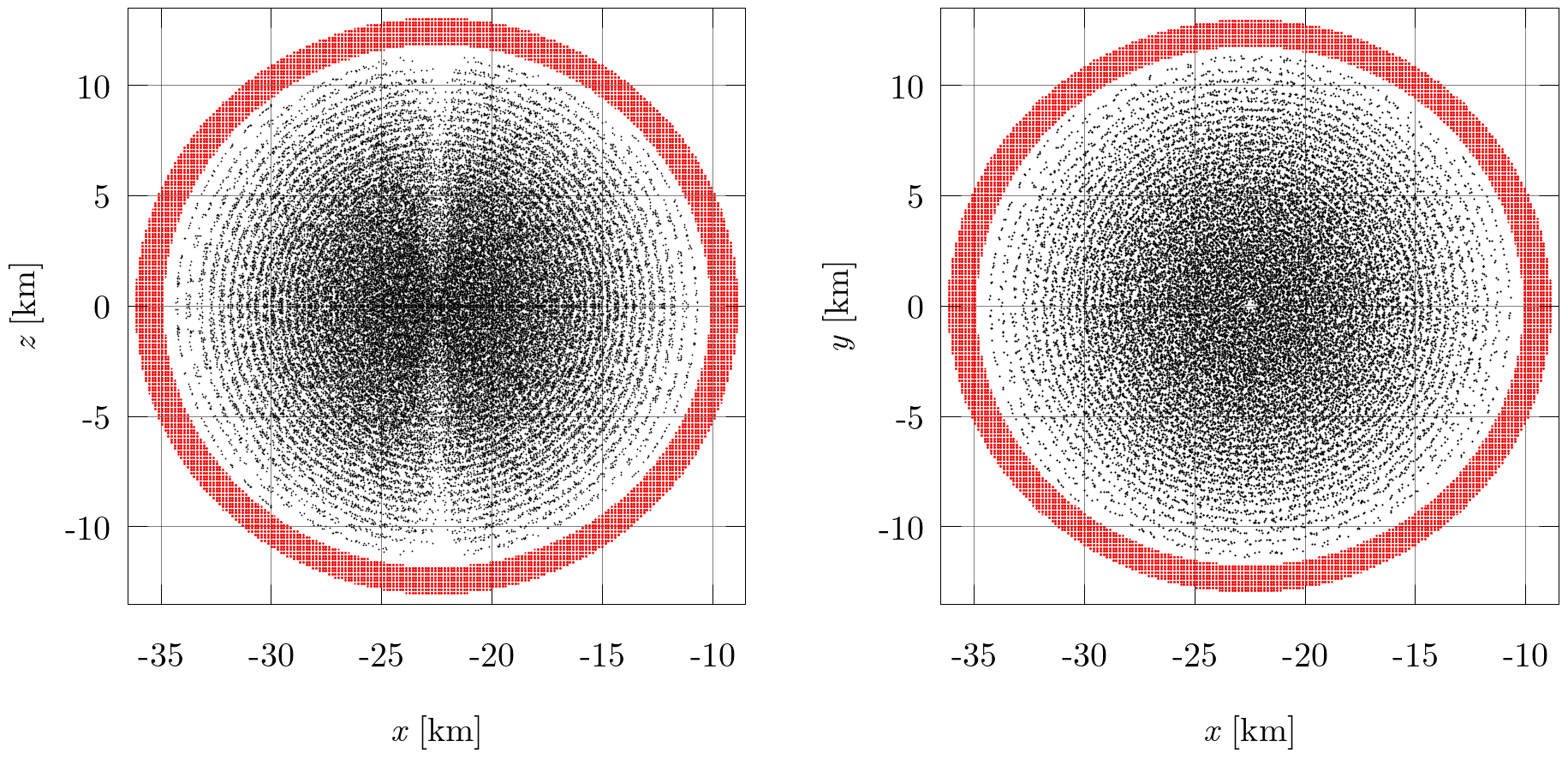}} \\
    \subfloat[At the end of the APM iteration, the particles
    form a very regular, ``glass-like" structure. ]{\includegraphics[width=\textwidth,valign=t]{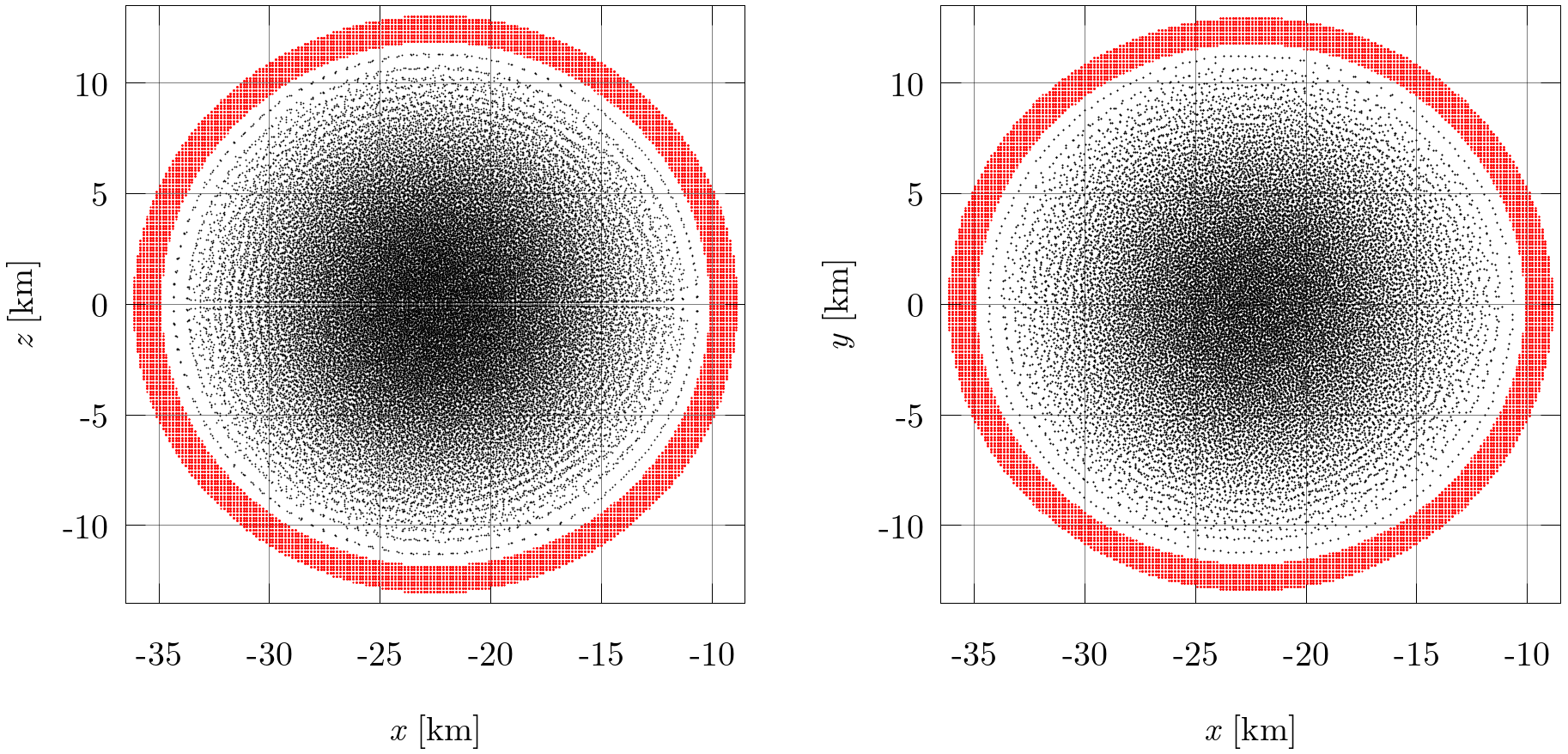}}
    \caption{Shown are the particle distributions before (a) and after (b) the APM iteration for one neutron star of our run \texttt{LR\_2x1.3\_G2.00}, see \autoref{tab:run_params}.
    The physical SPH particles are shown in black and the ghost/boundary particles (discarded after the APM iteration) are shown in red. The plots are cuts through the $xz$ and $xy$ plane, within $y \in [-1.18,1.18]$km and $z \in [-1.18,1.18]$km.    }
    \label{fig:apm}
\end{figure*}

A crucial ingredient of the method is the placement of the ``boundary" 
or ``ghost" particles. These serve as a shield around the star to prevent the real particles from exiting 
 the surface of the star during the APM iteration. 
The artificial pressure that we assign to these 
ghost particles is a few times higher (we use three times)
than the maximum value inside the star. This is to ensure
that particles that approach the stellar surface from the 
inside, see an increasing pressure gradient which keeps
them inside the stellar surface.
Since the stars in a close binary system are tidally deformed, 
the ghost particles have to be placed in a way that allows the 
real particles to model the surface geometry, see \ref{subsec:app-ghost} 
for more details on how this is
achieved. Once the APM iteration has
converged, the ghost particles are discarded. \autoref{fig:apm} 
shows the ghost particles (red) placed around the real particles (black) on 
spherical surfaces before and after applying the APM, for one star of 
one of our simulations (run \texttt{LR\_2x1.3\_G2.00}, see \autoref{tab:run_params}).

\begin{figure*}[!ht]
    \centering
    \includegraphics[width=\textwidth, valign=t]{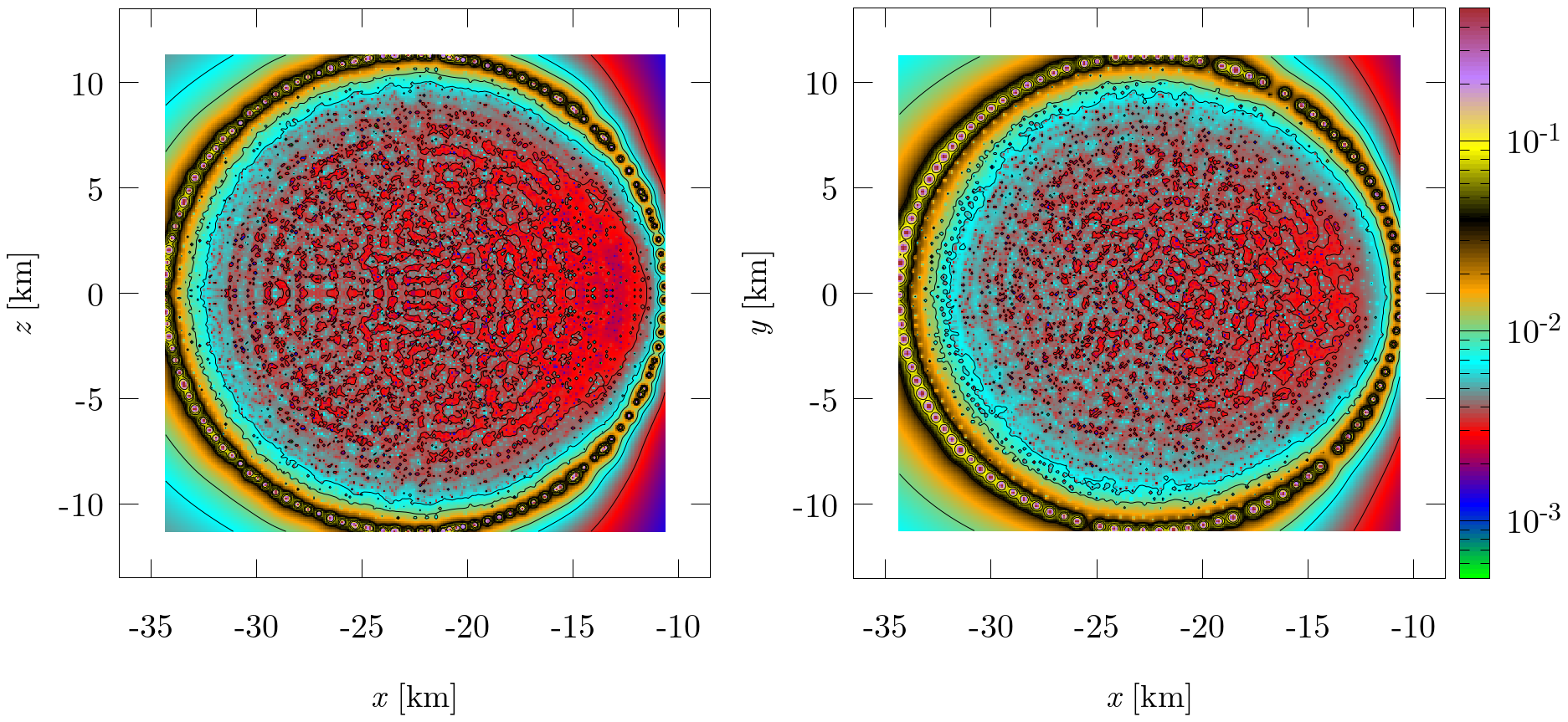}
    \caption{Typical relative density errors (i.e., relative difference between the SPH kernel density estimate at particle positions and the
    \Lo density) after the APM iteration on the $xz$ and $xy$ planes (the contours are computed using particles in $y\in [-0.44,0.44]$km and $z\in [-0.44,0.44]$km, 
    respectively). The same star as in \autoref{fig:apm} (\texttt{LR\_2x1.3\_G2.00}, see \autoref{tab:run_params}) is shown. Typical density errors in the bulk of the star are below 1\%, whereas in the hard-to-resolve surface layers, errors can reach 10\%.}
    \label{fig:apm-error}
\end{figure*}

After each step of the APM iteration, the particle positions are reset so that the center of mass of the particle distribution coincides (within machine precision) with the stellar center of mass given by \Lo. In addition, at each step of the APM iteration, the particle positions are reflected about the $xy$ plane, to impose exact equatorial-plane symmetry.

Once the APM iteration with particles of equal ba\-ry\-on 
number, $\nu_0$, has converged, we perform one single, 
final correction of the {\em individual} particle baryon
numbers, $\nu_a$. To this end we first calculate the SPH particle
number density,\footnote{This is just the SPH-density formula, \eqref{eq:N_sum}, but weighing each particle with unity rather than with its
baryon number.}
\be
\tilde{N}_a= \sum_b W_{ab}(h_a),
\ee
and then assign to each particle the ``desired" baryon number
\be
\label{eq:nu}
\nu_a^{\rm des}= \frac{N_a}{\tilde{N}_a},
\ee
where $N_a$ is the density according to \Lo. The baryon number assigned to each particle is a ``cap\-ped" version of $\nu_a^{\rm des}$ in \eqref{eq:nu}, 
so that $\nu_a$ remains in the interval $[\sqrt{2}\nu_o,\nu_o/\sqrt{2}]$
and the baryon number ratio is $\le 2$.
This correction step changes the baryon masses only
moderately, but improves the density estimate in the outer layers  by 
roughly an order of magnitude. 

This method allows to obtain a low baryon number ratio on each star, separately. In order to have a low baryon number ratio \emph{overall}, that is, across both stars, we set the particle numbers within each star to have a similar ratio as the binary mass ratio.
For more details, see \ref{subsec:app-spherical}.

 \autoref{fig:apm-error} shows contour plots of the relative difference between the \Lo mass density and the SPH kernel estimate of the mass density performed on the final particle distribution (including the one update on the baryon number). 
In the bulk of the stars the errors are lower than $\sim 1$\%.
Only in the surface layers are the errors larger. Here, the very steep physical density gradients are difficult to capture at finite resolution with nearly equal SPH particle masses. These layers will adjust slightly at the beginning of a simulation, trying to find a true numerical equilibrium.

As the last comment, we note that, for equal-mass BNS, we use the APM to place particles within one star only. The particles in the second star are obtained by simply reflecting those on the first star with respect to the $yz$ plane. In this way, the symmetry of the system is preserved also at the level of the particle distribution.

\subsubsection{Initial values for the SPH particles}
\label{subsec:sphid}

Once the final particle locations have been found,
we need to assign  particle properties according to
the\newline \Lo solution. The computing frame fluid velocity in Eq.~\eqref{eq:v_mu} is related to the fluid velocity with respect to the Eulerian observer $v^i_\mathrm{Eul}$ (provided by \Lo) by
\begin{align}
    v^0 = 1, \quad v^i = \alpha\, v^i_\mathrm{Eul} - \beta^i.
\end{align}
 The generalized Lorentz factor $\Theta$ can then be computed from $v^i$ using Eq.~\eqref{eq:theta_def}. The baryon number per particle $\nu $ is determined  
 as described in \ref{subsec:app-spherical} and Sec.~\ref{subsec:apm}. The smoothing length $h$ of each particle is computed so that each particle has exactly 300 contributing neighbours in the density estimate, as in \spB. Then, knowing $\nu$ and $h$, the density variable $N$ can be computed using Eq.~\eqref{eq:N_sum}, and the local rest frame baryon number density $n$ is computed inverting Eq.~\eqref{eq:N_def},
\begin{align}
    n= \dfrac{N}{\sqrt{-g}\; \Theta}.
\end{align}
The local rest frame baryon mass density is then
\begin{align}
    \rho_{\rm rest}=n\, m_{0},
\end{align}
where $m_{0}$ is the average baryon mass.
The specific internal energy $u$ and the pressure $P$ are then computed using the EOS, starting from $\rho_{\rm rest}$.

\subsubsection{BSSN initial data on the refined mesh}
\label{subsec:bssnid}

The ID for the BSSN variables is computed straightforwardly by first 
importing the \Lo ID for the standard 3+1, or ADM, variables to each 
level of the mesh refinement hierarchy, and then computing the BSSN variables 
from them using a routine extracted from the \McL thorn from the Einstein 
Toolkit \cite{loeffler12}.\footnote{As is common practice,
we  refer to the Arnowitt--Deser--Misner formalism \cite{misner73} 
compactly as ``ADM."} For the sake of clarity, we note that, for the runs 
shown in this paper, we use the initial values for the lapse function 
and the shift vector that LORENE provides.

\section{Simulations}
\label{sec:results}

\begin{table*}
\caption{Parameters of the performed simulations. The masses are {\em gravitational} masses in solar units. $\Gamma$ is the polytropic exponent and $K$ the corresponding constant in code units. The initial orbital angular velocity $\Omega_0$ is in units of rad/s. The quantity $h_{\rm min}$ is the minimal smoothing length during the evolution, $\Delta_g^{\rm min}$ is the grid resolution at the finest refinement level, and they are both given in m. $n_{\rm grid}$ is the number of grid points on each refinement level. All simulations start from a coordinate separation of $a_0= 30.4748$ (corresponding to 45 km) and use 7 levels of grid refinement, with the outermost refinement level's boundary at $\approx 2268$ km.}
\label{tab:run_params}       
\centering
\begin{tabular}{ccccccccc}
\toprule
Masses [$M_\odot$] & $\Gamma$ & $K$ & $\Omega_0$ [rad/s] & 
\# particles & $h_{\rm min}$ [m] & $n_{\rm grid}$ & $\Delta_g^{\rm min}$ [m] & Name  \\
\midrule
2 $\times$ 1.3 & 2.00 & 100 &  1774 & 
$5 \times 10^5$ & 40 & $121^3$ & 590 & \texttt{LR\_2x1.3\_G2.00}\\
2 $\times$ 1.3 & 2.00 & 100 &  1774 & 
$1 \times 10^{6} $ &  183 & $143^3$ & 500 & \texttt{MR\_2x1.3\_G2.00}\\
2 $\times$ 1.3 & 2.00 & 100 &  1774 & 
$2 \times 10^{6} $ & 165 & $193^3$ & 370 & \texttt{HR\_2x1.3\_G2.00}\\
\midrule
2 $\times$ 1.3 & 2.75 & $3\times 10^4$ &  1772 & 
$5 \times 10^{5}$ &  338  & $121^3$ & 590 & \texttt{LR\_2x1.3\_G2.75}\\
2 $\times$ 1.3 & 2.75 & $3\times 10^4$ &  1772 & 
$1 \times 10^{6}$ & 274  & $143^3$ & 500 & \texttt{MR\_2x1.3\_G2.75}\\
2 $\times$ 1.3 & 2.75 & $3\times 10^4$ &  1772 & 
$2 \times 10^{6}$ & 224  & $193^3$ & 370 & \texttt{HR\_2x1.3\_G2.75}\\
\midrule
2 $\times$ 1.4 & 2.00 & 100 & 1827 &  
$5 \times 10^{5}$ & 53  & $121^3$ & 590 & \texttt{LR\_2x1.4\_G2.00}\\
2 $\times$ 1.4 & 2.00 & 100 & 1827 &  
$1 \times 10^{6}$ & 70  & $143^3$ & 500 & \texttt{MR\_2x1.4\_G2.00}\\
2 $\times$ 1.4 & 2.00 & 100 & 1827 & 
$2 \times 10^{6}$ & 25  & $193^3$ & 370 & \texttt{HR\_2x1.4\_G2.00}\\
\midrule
2 $\times$ 1.4 & 2.75  & $3\times 10^4$ &  1823 & 
$5 \times 10^{5} $ & 307  & $121^3$ & 590 & \texttt{LR\_2x1.4\_G2.75}\\
2 $\times$ 1.4 & 2.75 & $3\times 10^4$ &  1823 & 
$1 \times 10^{6}$ & 245  & $143^3$ & 500 & \texttt{MR\_2x1.4\_G2.75}\\
2 $\times$ 1.4 & 2.75 & $3\times 10^4$ &  1823 & 
$2 \times 10^{6} $ & 192  & $193^3$ & 370 & \texttt{HR\_2x1.4\_G2.75}\\

\bottomrule
\end{tabular}
\end{table*}

In our original paper, we had focused on test
cases where the outcomes are accurately known.
These included shock tubes (exact result known), 
oscillating neutron stars in Cowling approximation and in
dynamically evolved space times (in both cases oscillation 
frequencies are accurately known)
and, finally, the evolution of an unstable neutron star that,
depending on small initial perturbations, either transitions 
into a stable configuration or collapses and forms a black 
hole (results known from independent numerical approaches, e.g., \cite{font02,cordero09,bernuzzi10}). 
In all of these benchmarks our results were in 
excellent agreement with established results.

Here, we want to take the next step towards more astrophysically 
motivated questions. In particular, we want to address for
the first time the merger of two neutron stars with \spB.

\subsection{Initial Setup}
\label{sec:initial_setup}
In this first study, we simulate two binary systems with $2\times 1.3$ 
\Msun and $2\times 1.4$ \msun, each time with a soft ($\Gamma= 2.00$, $K= 100$ in code units;
$M_{\rm max}^{\rm TOV}\approx 1.64$ \msun) 
and a stiff ($\Gamma= 2.75$, $K= 30~000$ in code units; $M_{\rm max}^{\rm TOV}\approx 2.43$ \msun) EOS. Both equations
of state are admittedly highly idealized, and in one of our next 
steps we will include the physical EOSs that are provided
by the \comp database \cite{compose}. Each of the simulations 
are run at three different resolutions: a) low-resolution (LR) 
with $5 \times 10^5$ SPH particles and $121^3$ grid 
points on every refinement level, b) medium-resolution 
(MR) with $10^6$ SPH  particles and $143^3$ grid 
points on each refinement level
and c) high-resolution (HR) with $2\times 10^6$ SPH 
particles and $193^3$ grid points on each 
refinement level. All simulations start from a coordinate
separation of 45 km, and employ seven refinement levels out 
to 1536 code units ($\approx$ 2268 km) in each direction.
The parameters of the performed simulations are summarized in \autoref{tab:run_params}, the quoted numbers are  accurate to $\sim$ 1\%.\footnote{Our ``soft" 1.3 \Msun star has a gravitational mass of 1.312 \Msun (1.458 \Msun baryonic), the ``soft" 1.4 \Msun star has 1.409 \Msun (1.590 \Msun baryonic); our ``stiff" 1.3 \Msun star has a gravitational mass of 1.309 \Msun (1.475 \Msun baryonic) and the ``stiff" 1.4 \Msun star has 1.403 \msun (1.600 \Msun baryonic).} 
Note that we also show the 
minimum smoothing length, $h_{\rm min}$, during a simulation. This is, of course, a quantity that adapts automatically according to the dynamics and that is not determined beforehand. For example,
the $h_{\rm min}= 40$ m in run \texttt{LR\_2x1.3\_G2.00} is so small because the central object at very late times ($\sim$ 18 ms)
collapses to a black hole. This is a very low resolution result,
it should be interpreted with caution.

\subsection{Results}
\begin{figure*}
  \includegraphics[width=0.99\textwidth]{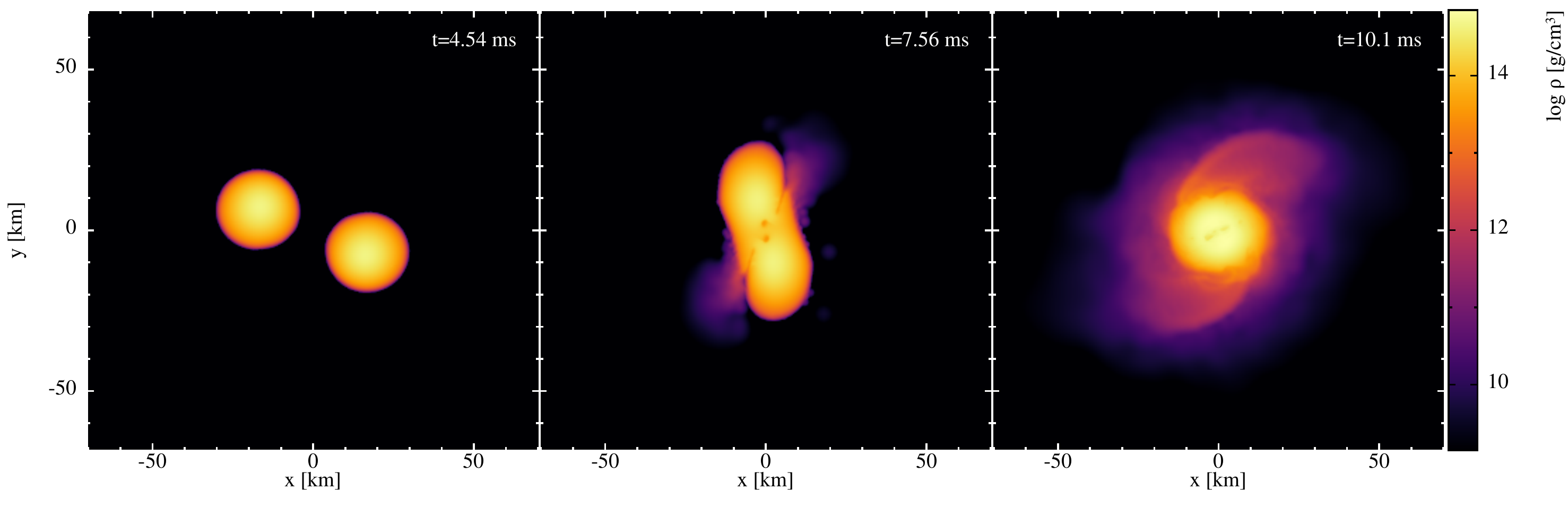}
  \includegraphics[width=0.994\textwidth]{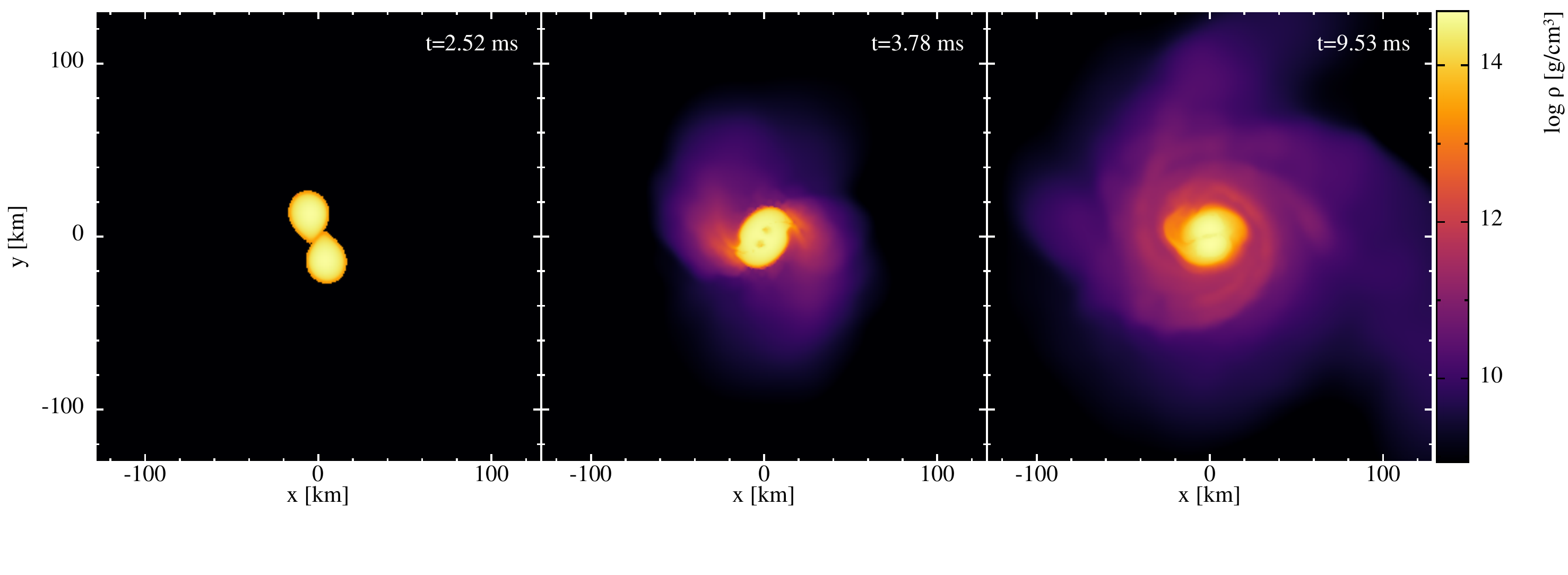}
  \includegraphics[width=0.99\textwidth]{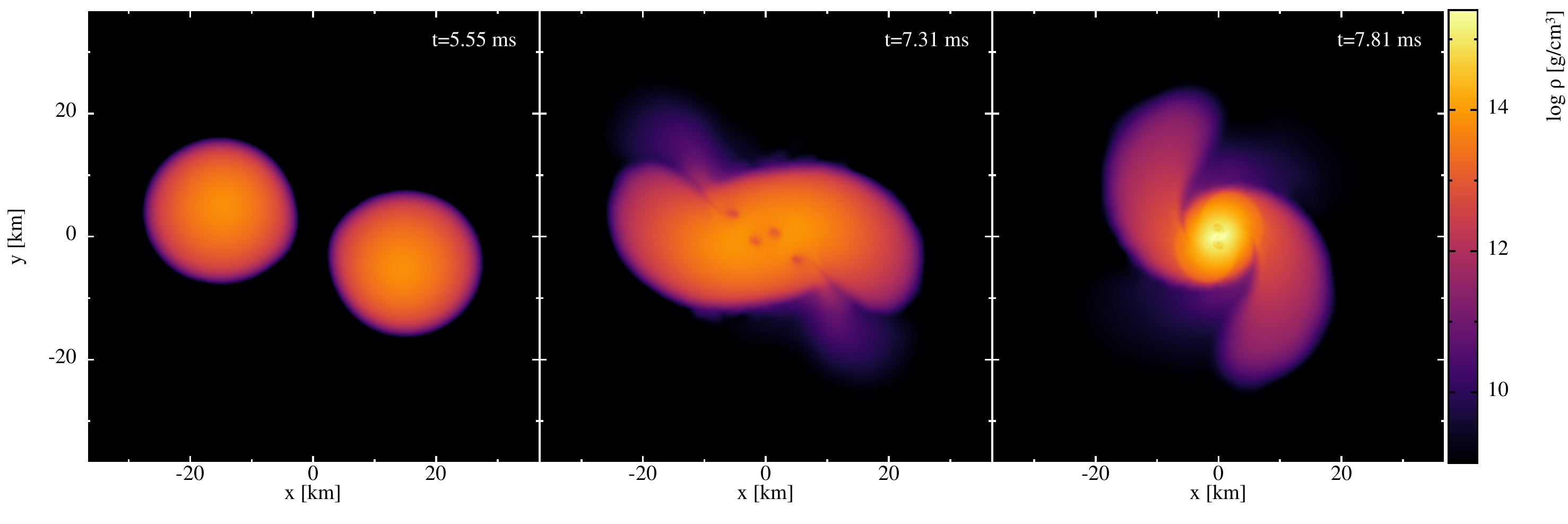}
  \includegraphics[width=0.99\textwidth]{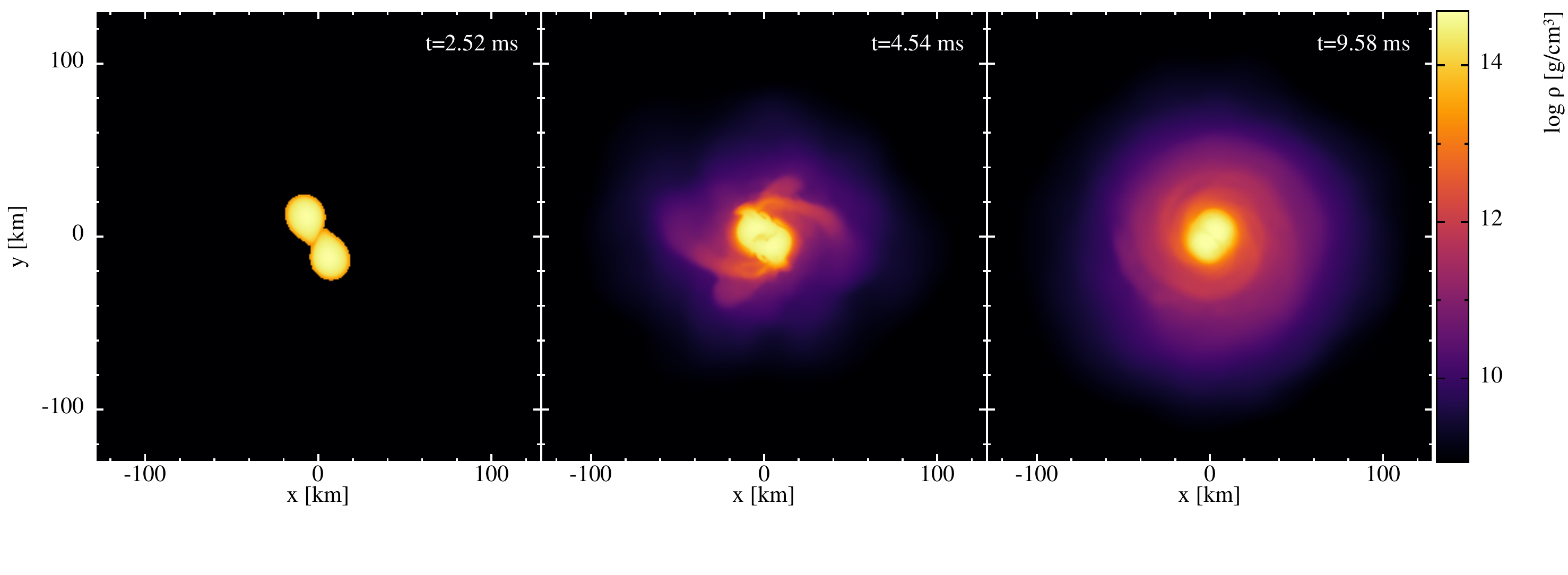}   
\caption{Density evolution of the simulations (top to bottom) \texttt{HR\_2x1.3\_G2.00},  \texttt{HR\_2x1.3\_G2.75},  \texttt{HR\_2x1.4\_G2.00} and  \texttt{HR\_2x1.4\_G2.75}, see \autoref{tab:run_params} for the corresponding parameters.}
\label{fig:dens_evol}       
\end{figure*}
\begin{figure*}
  \centerline{
  \includegraphics[width=0.55\textwidth]{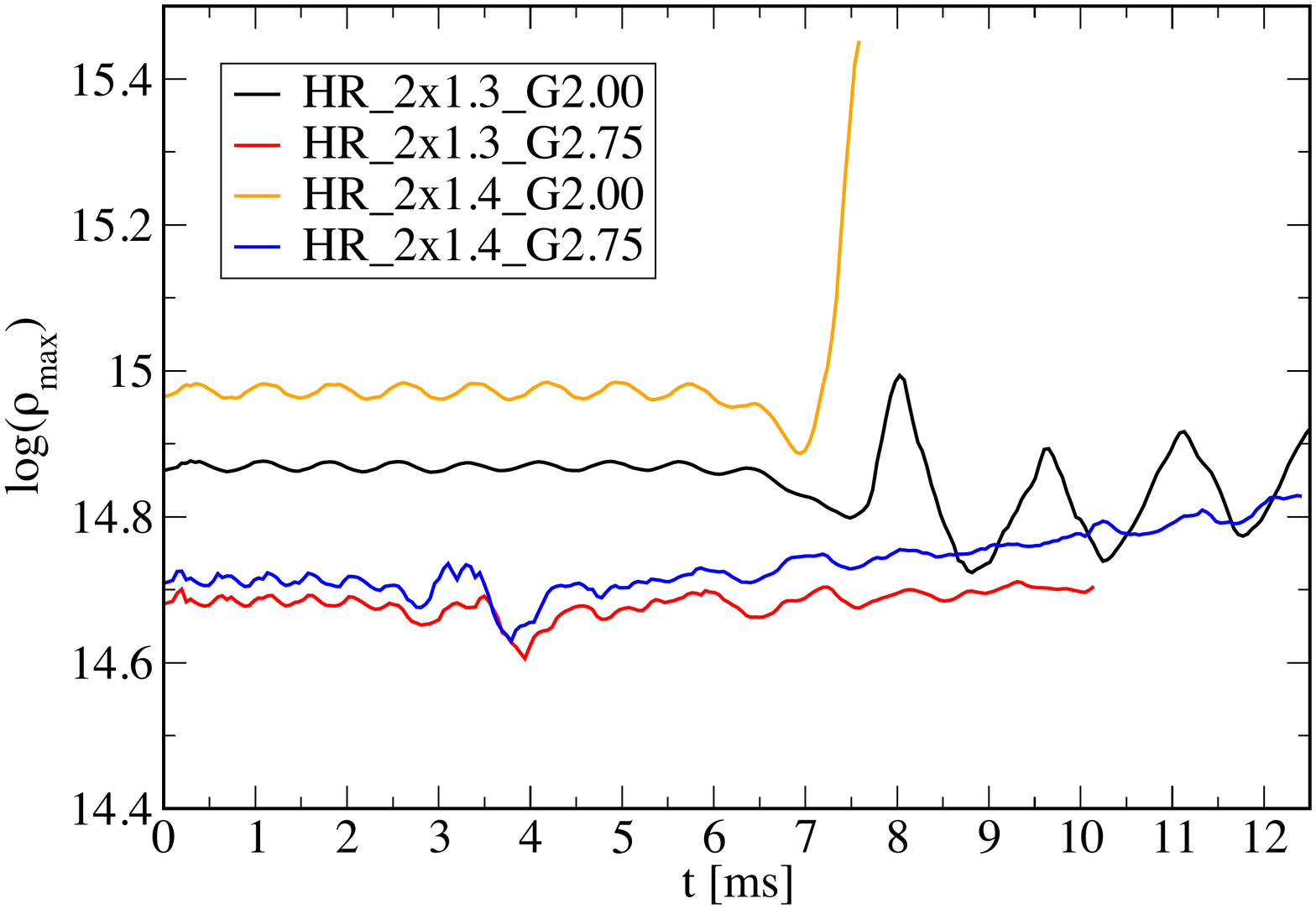} \hspace*{-1cm}
  \includegraphics[width=0.55\textwidth]{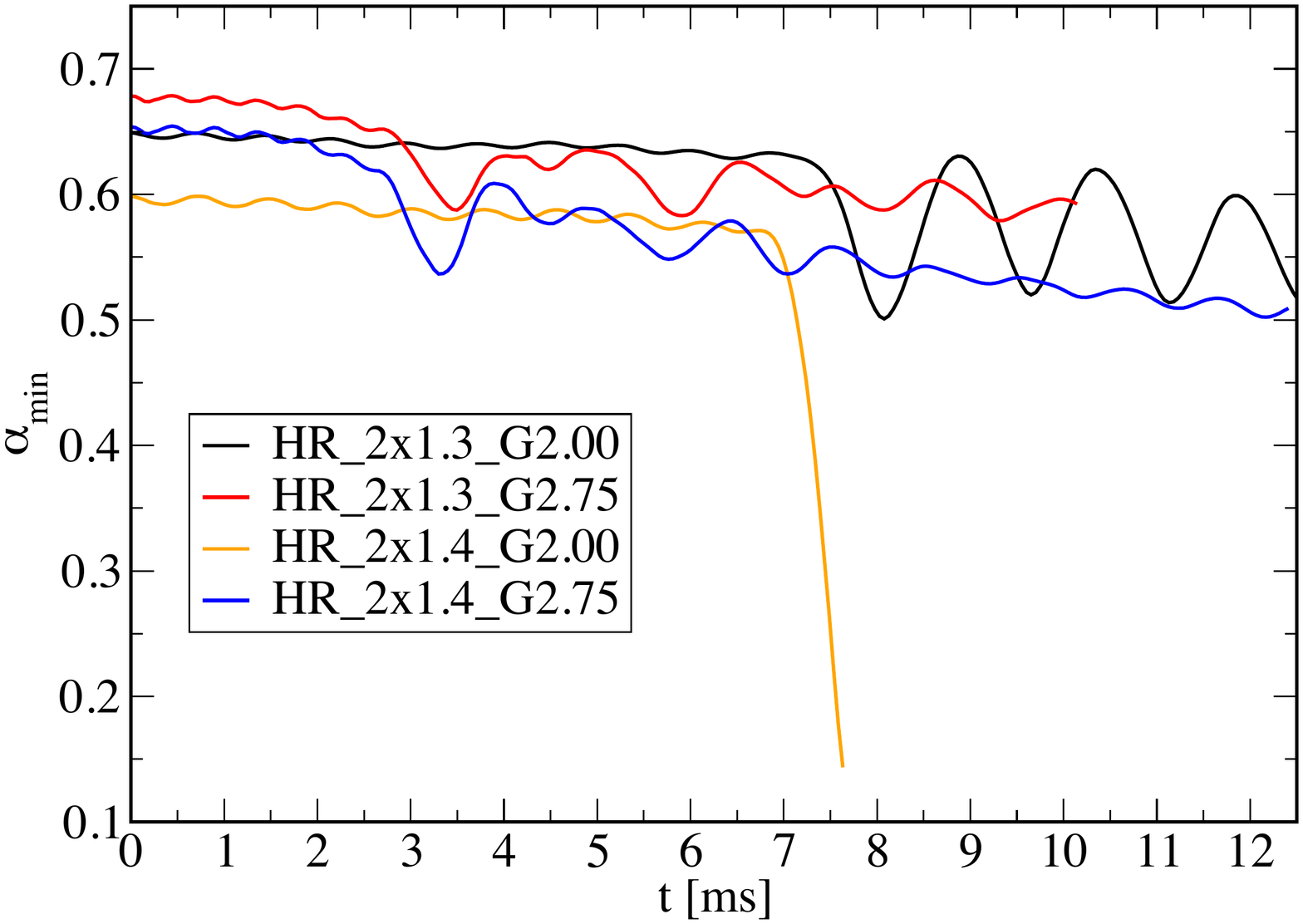}}
  \vspace*{-0.5cm}
\caption{Evolution of the maximum density and the minimum lapse value
for the HR-runs.} 
\label{fig:Rho_Lapse}       
\end{figure*}
\begin{figure*}
  \includegraphics[width=0.9\textwidth]{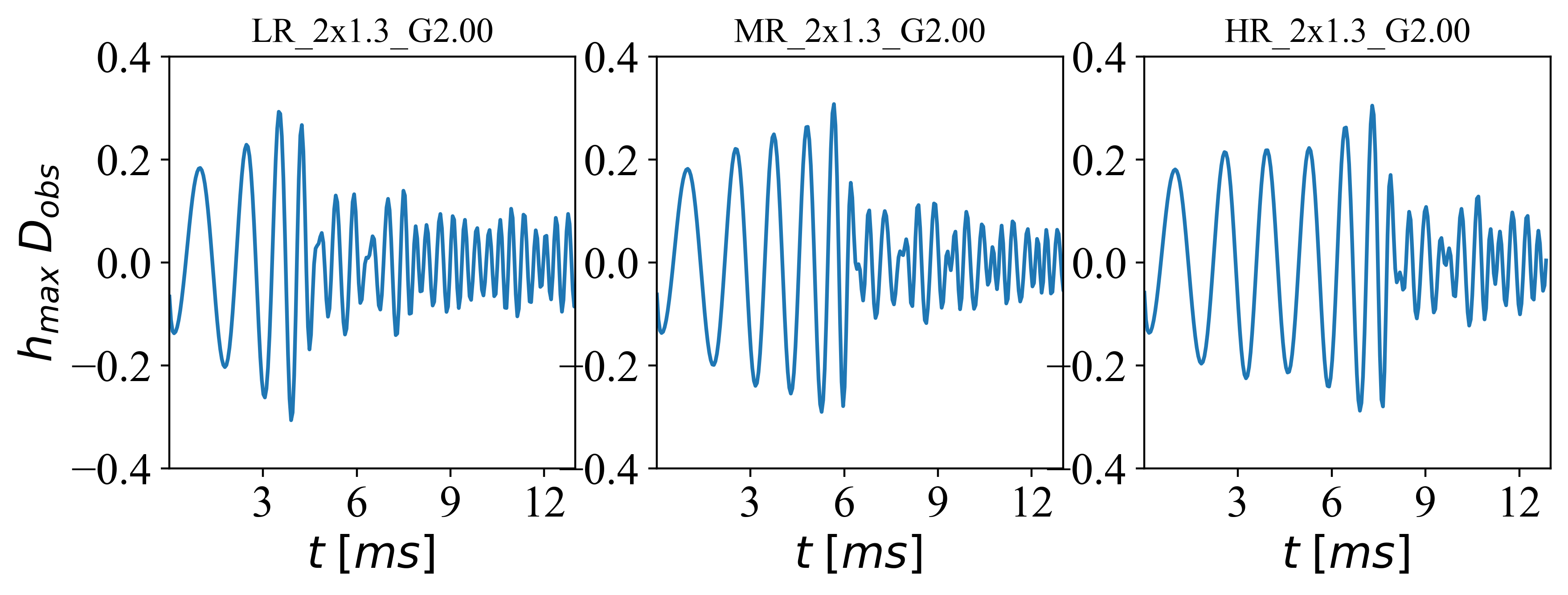}
  \includegraphics[width=0.9\textwidth]{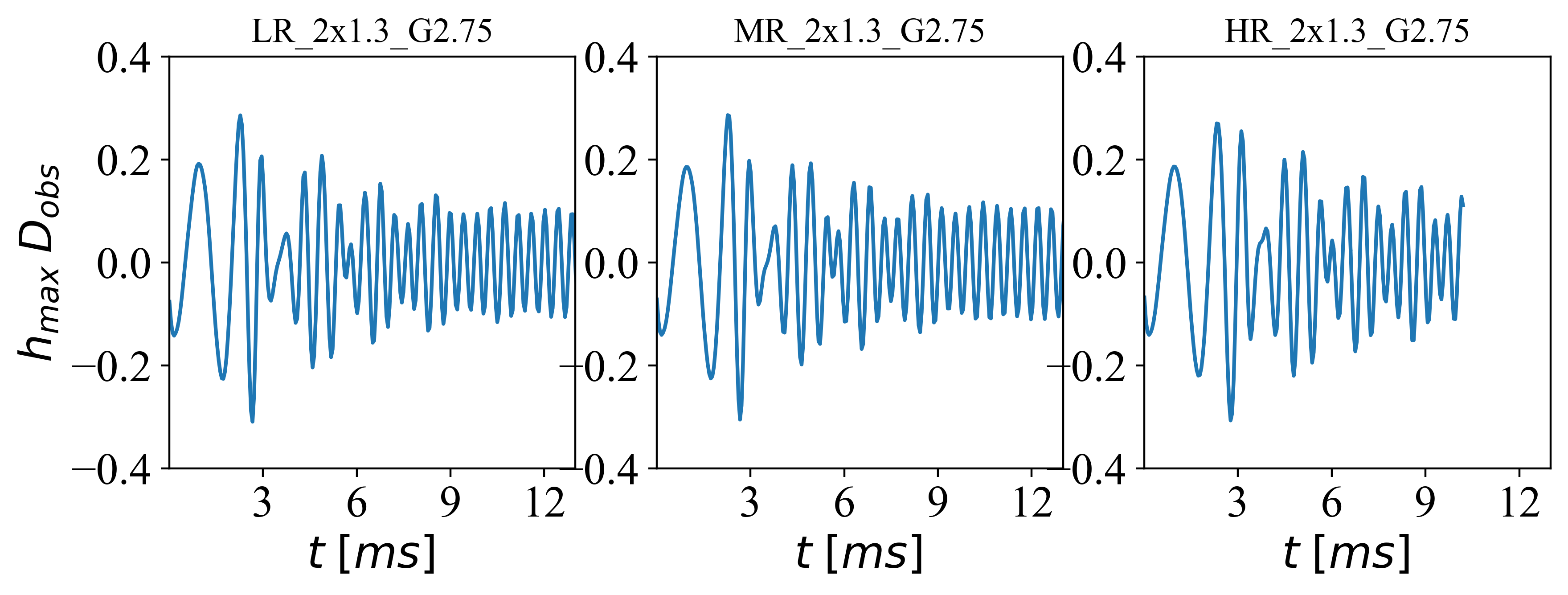}
  \includegraphics[width=0.9\textwidth]{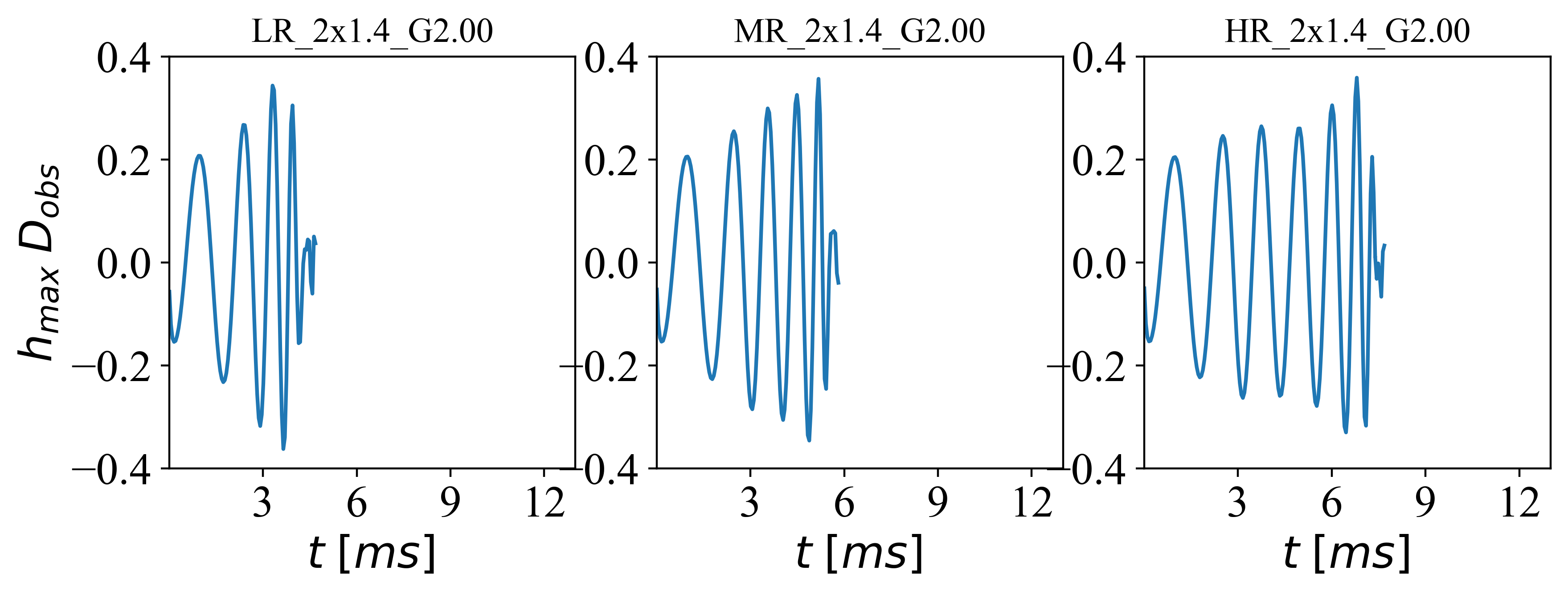}
  \includegraphics[width=0.9\textwidth]{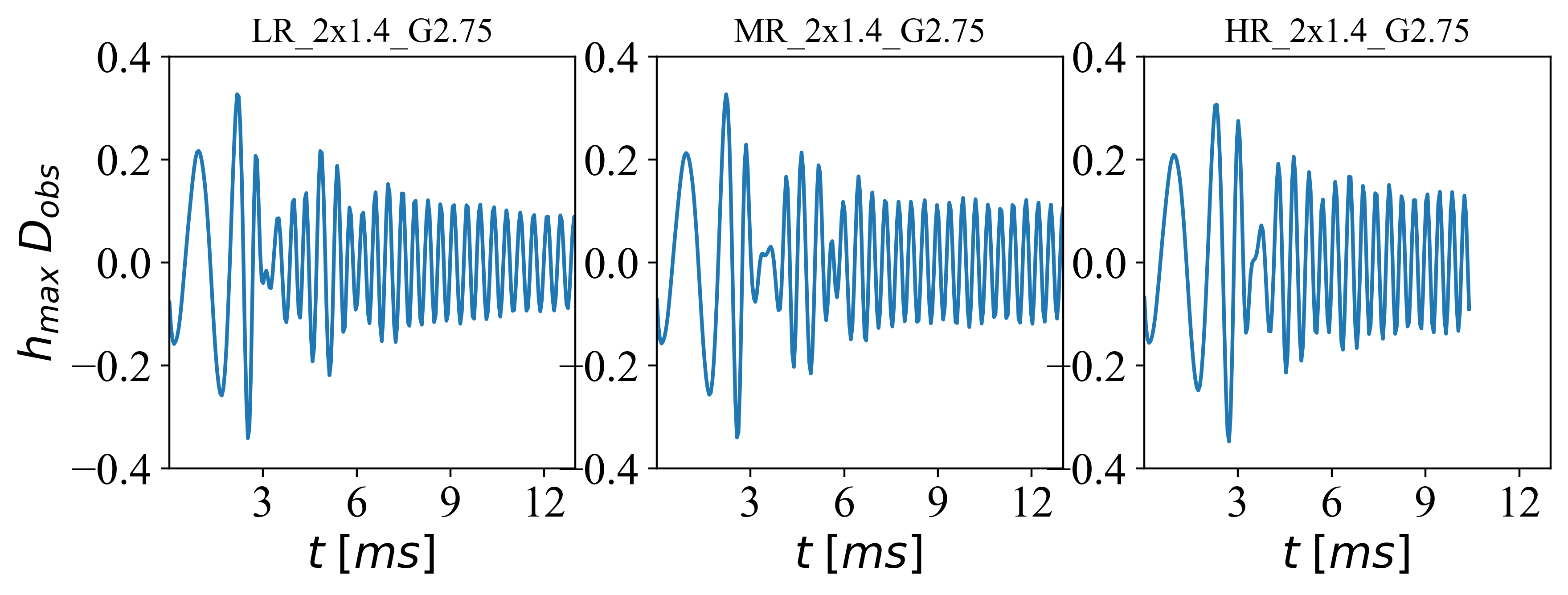}   
\caption{Gravitational wave amplitudes computed with the quadrupole formula, see \ref{sec:App1}, for the simulations (top to bottom) \texttt{HR\_2x1.3\_G2.00},  \texttt{HR\_2x1.3\_G2.75},  \texttt{HR\_2x1.4\_G2.00} and  \texttt{HR\_2x1.4\_G2.75}, see \autoref{tab:run_params}. The quantity $D_{\rm obs}$ is the distance to the observer in code units, and $h_{\rm max}$ is the maximal amplitude for an observer located on the $z$ axis.}
\label{fig:GWs}       
\end{figure*}

In Fig.~\ref{fig:dens_evol} we show snapshots of the
rest-mass density evolution of the HR runs of our different 
binaries, with $t= 0$ corresponding to the simulation start.
Fig.~\ref{fig:Rho_Lapse} shows the evolution of the maximum
density (left) and minimum lapse value (right) of the
corresponding runs and Fig.~\ref{fig:GWs} shows 
the quadrupole GW amplitudes $h_{\rm max}$ (for an observer
located along the rotation axis; see \ref{sec:App1}) 
times the distance to the observer $D_{\rm obs}$. These
quadrupole approximation results are written out ``on the
fly" and they can be compared to the more accurate results
that can be extracted in a post-processing step from the
spacetime evolution, see below.

%
%
As expected, the cases with the soft EOS, where the stars are more
compact and therefore closer to the point mass limit, show a
substantially longer inspiral and chirp signal (lines 1 and 3 vs 
lines 2 and 4 in Fig.~\ref{fig:GWs}). 
Being more compressible, their peak density evolution
is also more impacted by the merger, see the left panel of 
Fig.~\ref{fig:Rho_Lapse}, and this is also reflected in the
evolution of the minimum lapse value (right panel).
In the cases with the stiff 
EOS, in contrast, where the stars are larger and  more mass is
at larger radii from the stellar centre, tidal effects are much 
more pronounced and effectively
represent a short-range, attractive interaction \cite{damour09b,radice20}.
Therefore
our binaries with stiff EOS merge within less than two orbital
revolutions from the chosen initial separation.  Being rather
incompressible, the post-merger central densities are actually 
only moderately above the central densities of the initial individual stars, see Fig.~\ref{fig:Rho_Lapse}, left panel, and
also the lapse oscillations (right panel) are small.
All of the remnants seem to evolve to more compact configurations,
but of the shown cases only \texttt{HR\_2x1.3\_G2.00} collapses
to a black hole during the simulated times.

\begin{figure*}
  \includegraphics[width=0.48\textwidth]{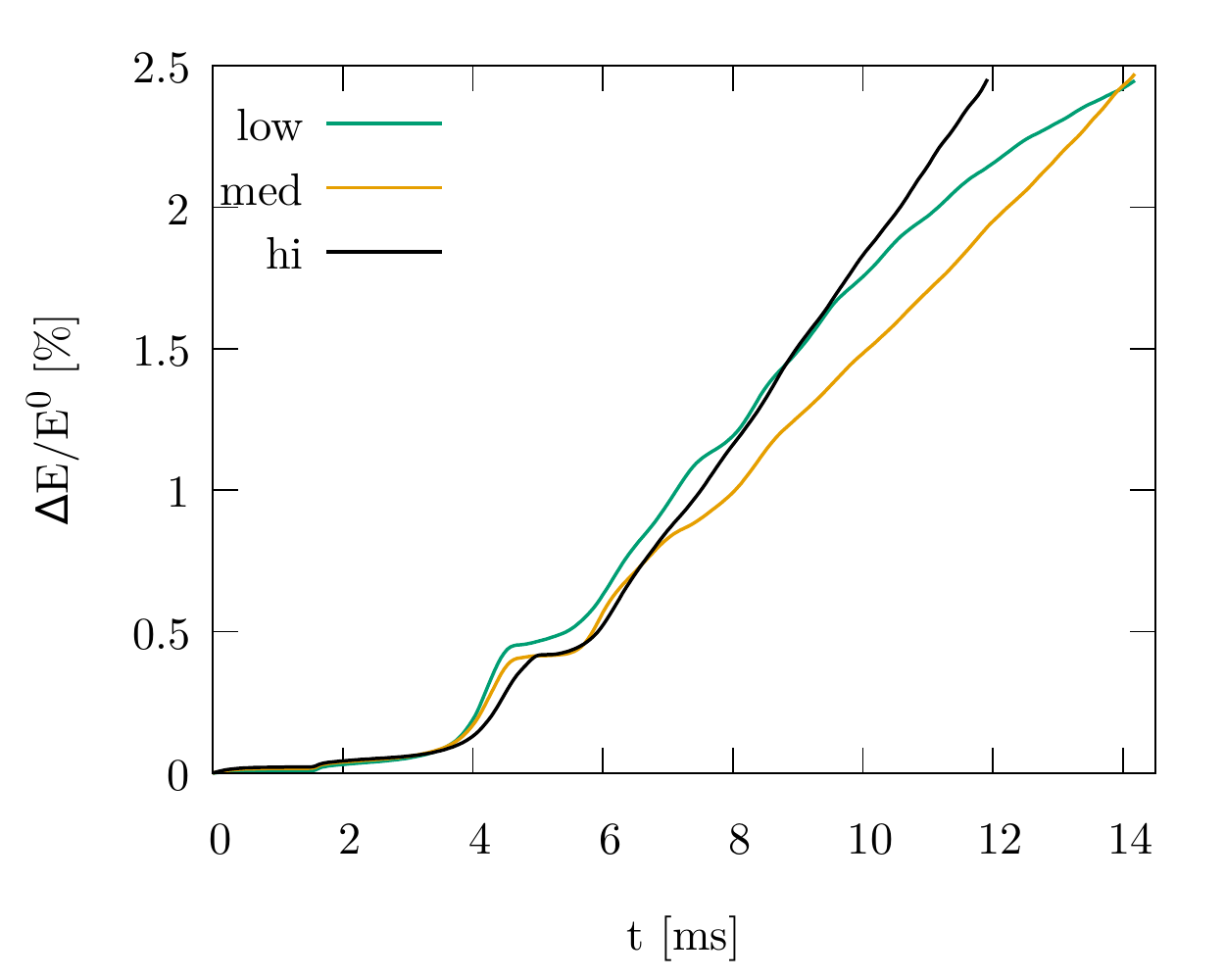}
  \includegraphics[width=0.48\textwidth]{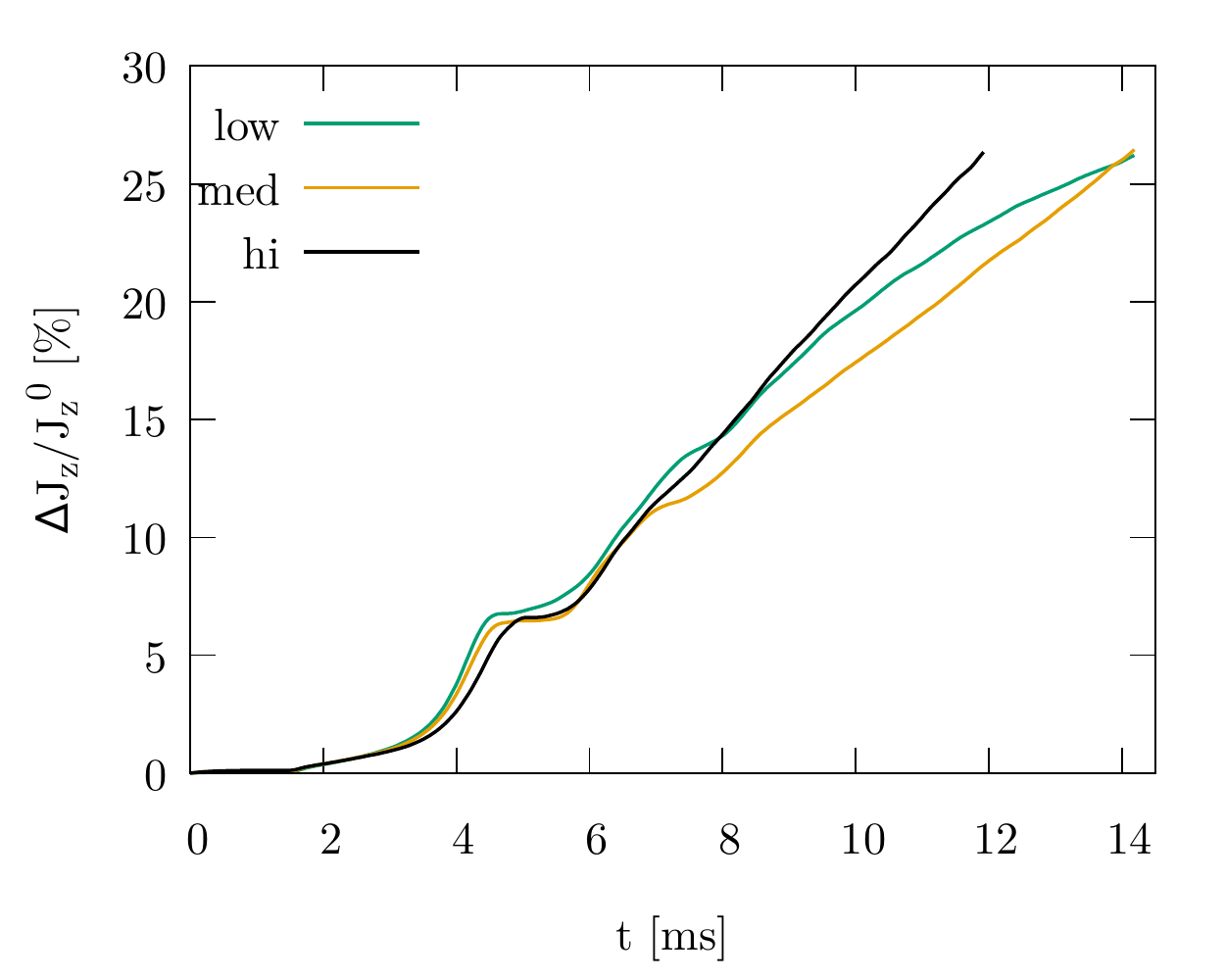}
  \caption{The accumulated radiated energy, $\Delta E$ (left plot), and
  $z$-component of the angular momentum, $\Delta J_z$ (right plot) in percent of 
  the initial values, $E^0$ and ${J_z}^0$ for the system with two 1.4 \Msun
  neutron stars and $\Gamma=2.75$ extracted at the detector at coordinate
  radius R=300. In both plots the green, orange, black curves are for the low, 
  medium and high resolution runs, respectively.}
   \label{fig:radEJ}
\end{figure*}

For the simulations afforded in this study, numerical resolution 
still has a noticeable effect on the inspiral as, for example, 
demonstrated by simulation \texttt{HR\_2x1.3\_G2.00}, which takes about one 
 orbit more until merger than \texttt{LR\_2x1.3\_G2.00}
(upper right vs upper left in Fig.~\ref{fig:GWs}).
The HR runs with the soft equation of state do show some
signs of amplitude variations that could be attributed to 
eccentricity, whereas the low and medium resolutions do not seem
to show that. We certainly do expect some eccentricity due to the
assumptions going into the ID construction. As to why it
is more pronounced in the HR case, it is possible that the low and
medium resolution runs are short enough so that signs of
eccentricity are washed out. Finally, these extracted waveforms 
are calculated from gauge dependent quantities and the 
amplitude variations could be entirely spurious.
All investigated systems, apart from the $2\times 1.4$ \Msun cases with $\Gamma=2.00$, leave a stable remnant (at least on the simulation time 
scale) and therefore keep emitting gravitational waves. The exceptional
case, see row three in Fig.~\ref{fig:dens_evol}, undergoes a prompt
collapse to a black hole within $\sim$ 1 ms  after merger (Fig.~\ref{fig:Rho_Lapse}) which efficiently 
shuts off the gravitational wave emission (row three in Fig.~\ref{fig:GWs}). 
A collapse to a black hole is expected when the binary mass exceeds 
$\sim 1.5 \times M_{\rm max}^{\rm TOV}$ 
\cite{hotokezaka11,bauswein13b,koeppel19,kashyap21},
so the collapse in \texttt{HR\_2x1.4\_G2.00} is actually expected given that
$M_{\rm max}^{\rm TOV}$ is only 1.64 \Msun and the initial ADM mass of the system is 2.9\msun.\\
We have, in addition to using the quadrupole formula, also extracted
gravitational waves based on the spacetime data using both thorn
\Extract and (in combination) thorns \WeylScal4 and \Multipole from the
Einstein Toolkit~\cite{loeffler12}. The procedure was to read in metric 
data from our third coarsest refinement level at the times we had checkpoint
data, run the wave extraction tools and finally use the formulae in 
\ref{app:etk} to calculate the strain as well as the radiated 
energy and angular momentum. We used detectors at spheres of coordinate
radii 50, 100, 150, 200, 250 and 300. We see very good agreement between
the results from \Extract and \WeylScal4, but as the \WeylScal4 waveforms
are less noisy we only report on those results in the following.\\
In Fig.~\ref{fig:radEJ} we plot the radiated energy, $\Delta E$, and
$z$-component of the angular momentum, $\Delta J_z$, as function of time for
the system with two 1.4 \Msun neutron stars and $\Gamma=2.75$. The quantities
are plotted as percentages of the initial ADM values of the spacetime, $E^0$
and ${J_z}^0$. We can clearly not yet claim convergence of these results as the
rate of energy and angular momentum emission after merger increases 
significantly from the medium to high resolution runs and the behavior of the
low resolution run is substantially different showing a decreasing rate 
of emission at late times.

In Table~\ref{tab:radEJ}, we list the final radiated energy at the end of our
simulations, $\Delta E$, and$z$-component of the angular momentum, $\Delta J_z$,
again as percentages of the initial ADM energy $E^0$ and angular momentum 
$\Delta J_z^0$ of the spacetime for a sample of our simulations. Note that in
the simulations listed, the merger remnant has not yet collapsed to a black 
hole, hence the systems are still emitting strong gravitational waves. In 
addition, as we cannot claim that the quantities are converged, we can 
are unable to perform a conclusive comparison with, for example, the results
of~\cite{zappa18}, but for now we can only state that our simulations appear
to be consistent. 

\begin{table}
\caption{Percentage of radiated energy, $\Delta E$, and $z$-component of 
angular momentum, $\Delta J_z$, with respect to the initial ADM energy, $E^0$,
and ADM angular momentum, $J_z^0$ for a sample of our simulations. As all 
simulations were not run for the same amount of time, we also list the 
simulation duration, $t_{\mathrm{end}}$.}
\centering
\begin{tabular}{cccc}
\toprule
    Name & $t_{\mathrm{end}}$ [ms] & $\Delta E/E^0$ [\%] & $\Delta J_z/\Delta J_z^0$ [\%] \\
\midrule
    \texttt{LR\_2x1.3\_G2.00} & 14.78 & 1.7 & 19.6 \\
    \texttt{MR\_2x1.3\_G2.00} & 14.78 & 1.3 & 17.4 \\
    \texttt{LR\_2x1.3\_G2.75} & 14.78 & 1.7 & 20.8 \\
    \texttt{MR\_2x1.3\_G2.75} & 14.78 & 1.7 & 20.9 \\
    \texttt{LR\_2x1.4\_G2.75} & 14.78 & 2.5 & 26.9 \\
    \texttt{MR\_2x1.4\_G2.75} & 14.78 & 2.6 & 28.0 \\
    \texttt{HR\_2x1.4\_G2.75} & 12.46 & 2.6 & 29.5 \\
\bottomrule
\end{tabular}
\label{tab:radEJ}
\end{table}

\begin{figure*}
   \includegraphics[width=0.34\textwidth]{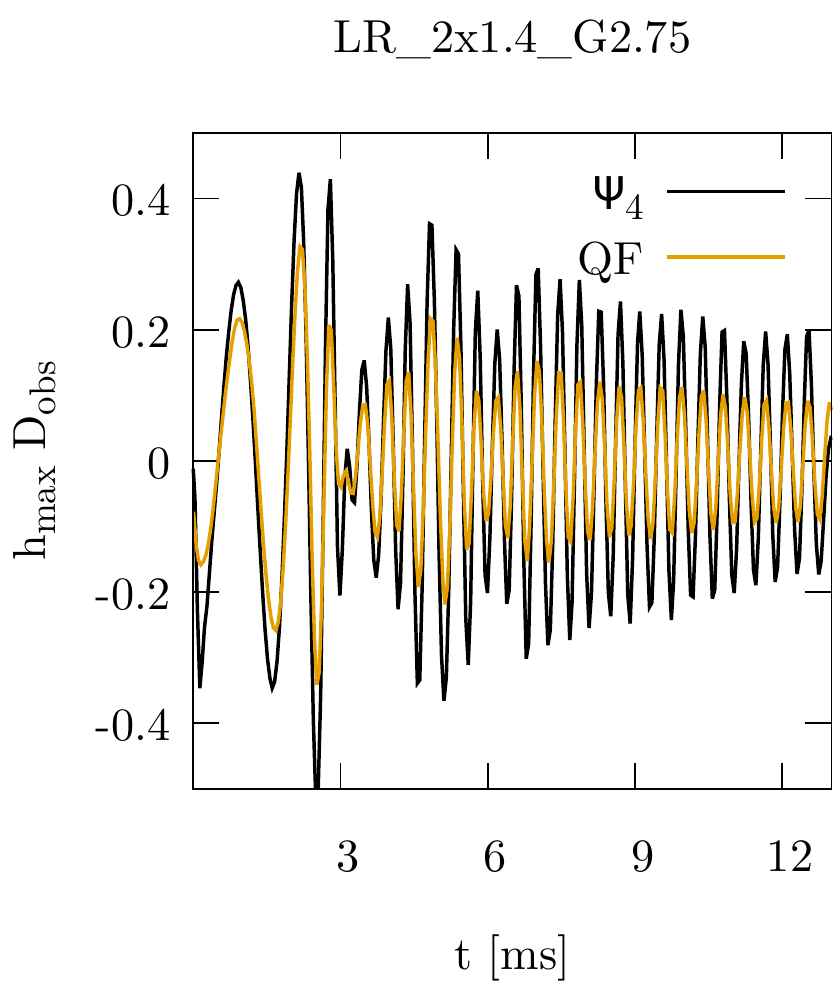}
   \hspace{-2.0em}
   \includegraphics[width=0.34\textwidth]{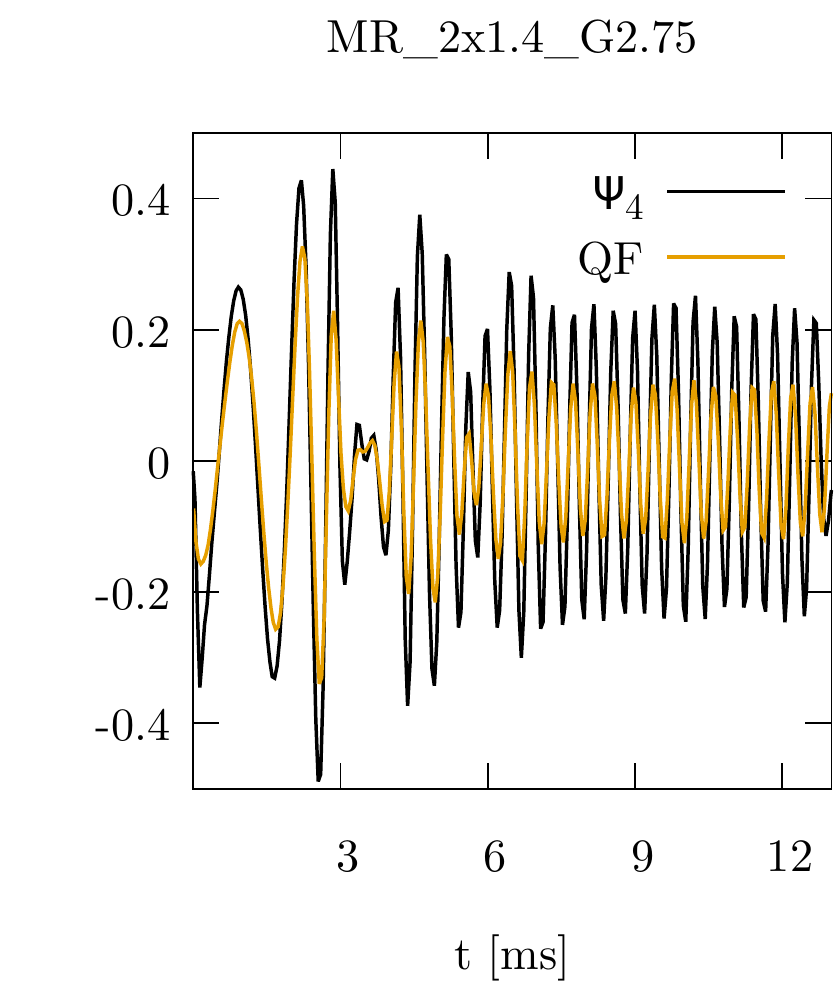}
   \hspace{-2.0em}
   \includegraphics[width=0.34\textwidth]{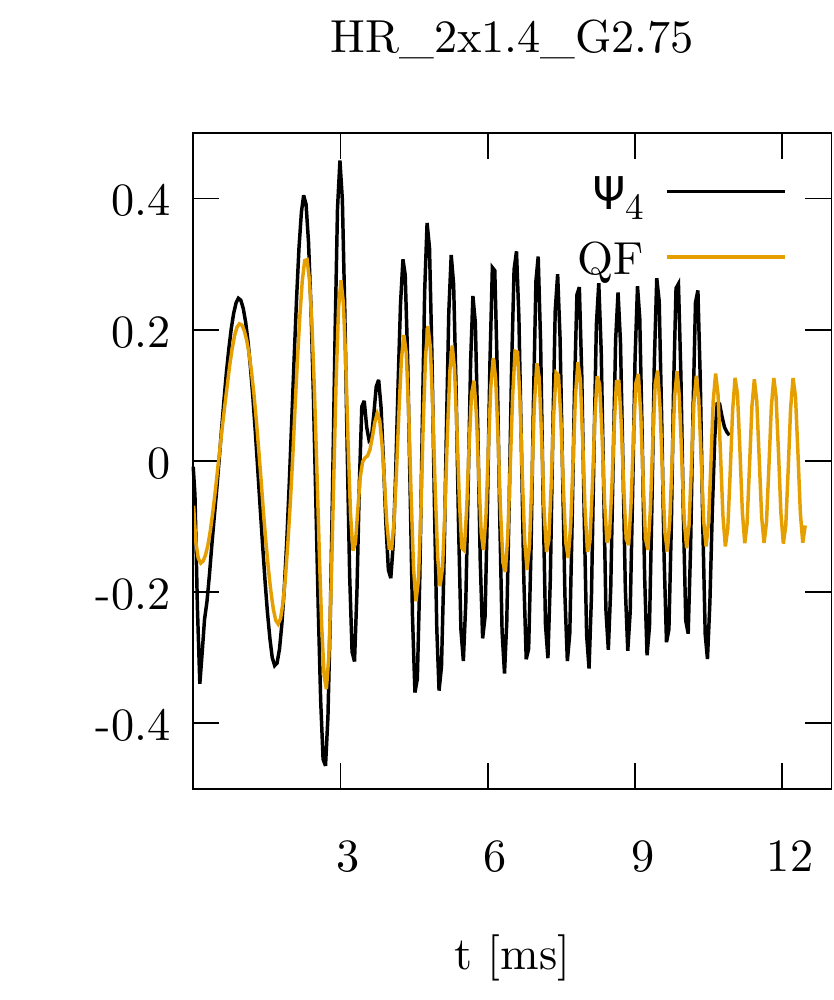}
   \caption{Comparison of the maximal strain amplitude as calculated by
   the extraction of $\Psi_4$ (black) and the quadrupole formula (QF, orange)
   for the simulations with two 1.4\Msun neutron stars and $\Gamma=2.75$. The 
   strain from $\Psi_4$ is the sum of spin-weight -2 modes from $\ell=2$ to 4 
   evaluated on the z-axis ($\theta=0$) and shifted in time by 1.54 ms in
   order to align it with the quadrupole waveform. The quantity $D_{\rm obs}$ 
   is the distance to the observer (code units), and $h_{\rm max}$ is the 
   maximum amplitude for an observer located on the $z$ axis.}
   \label{fig:strain_compare}
\end{figure*}

In Fig.~\ref{fig:strain_compare} we show a comparison of the maximal
strain amplitude extracted using either the quadrupole formula or the Weyl
scalar $\Psi_4$. Here the strain from $\Psi_4$ is the sum of spin weight $-2$
spherical harmonic modes from $\ell=2$ to 4 evaluated on the $z$-axis 
($\theta=0$ and $\phi=\pi/2$ to match the observer orientation) and then
shifted in time (by about 1.54 ms) to account for the signal travel time
to the detector. As can be seen, the quadrupole formula consistently 
underestimates the amplitude (especially after the merger; less so during
the inspiral). This is in agreement with~\cite{baiotti09} where it was 
found that the amplitudes could be over- or underestimated by more than 
50\% depending on the definition of density used. On the other hand,
also in agreement with~\cite{baiotti09}, the frequencies are well captured
by the quadrupole formula.

\begin{figure}
  \hspace*{-0.4cm}\includegraphics[width=0.51\textwidth]{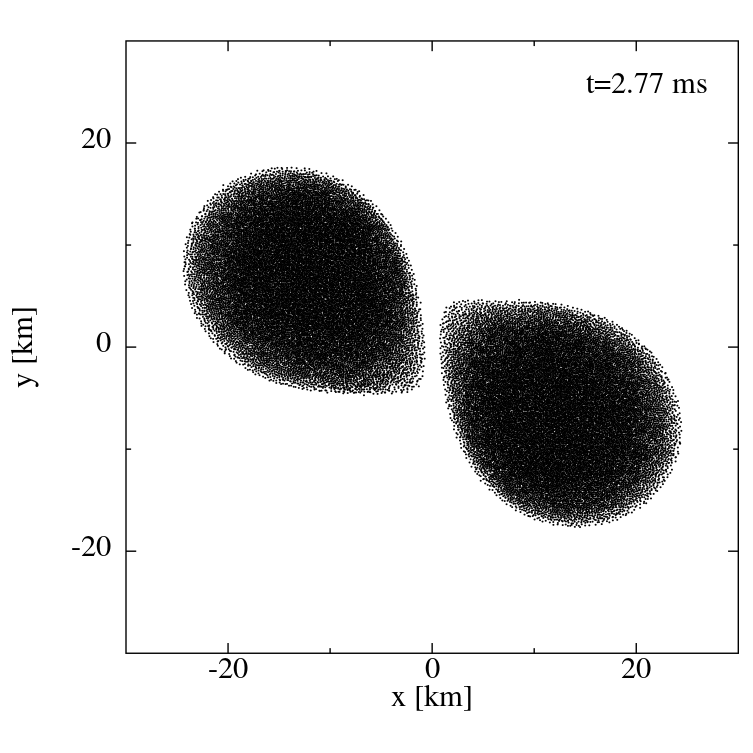}
\caption{Particle positions (within $|z| < $ 1 km) in  run \texttt{HR\_2x1.4\_G2.75}, see \autoref{tab:run_params}, prior to merger. Modelling the transition between the
neutron star matter and vacuum represents a serious numerical challenge for Eulerian hydrodynamic approaches. In our approach, however, the stellar surface does not require any special treatment and it remains smooth and well-behaved during the inspiral without any ``outlier particles."}
\label{fig:particle_pos}       
\end{figure}

We also want to briefly illustrate a major advantage of our
methodology: the treatment of the neutron star surface.
Contrary to traditional Eulerian approaches, the sharp 
transition between the high-density neutron star and 
the surrounding vacuum does not pose any challenge for 
our method and no special treatment such as an ``artificial
atmosphere" representing vacuum is needed. The particles
merely adjust their positions according to the balance of
gravity and pressure gradients to find their true numerical
equilibrium and vacuum simply corresponds to the absence of particles. Throughout the inspiral, the neutron star surface remains smooth and well-behaved without any ``outliers," see Fig.~\ref{fig:particle_pos}. This illustrates 
the quality of both the evolution code and the initial
particle setup, see Sec.~\ref{subsec:apm}.

\begin{figure*}
  \centering

  \includegraphics[width=1\textwidth]{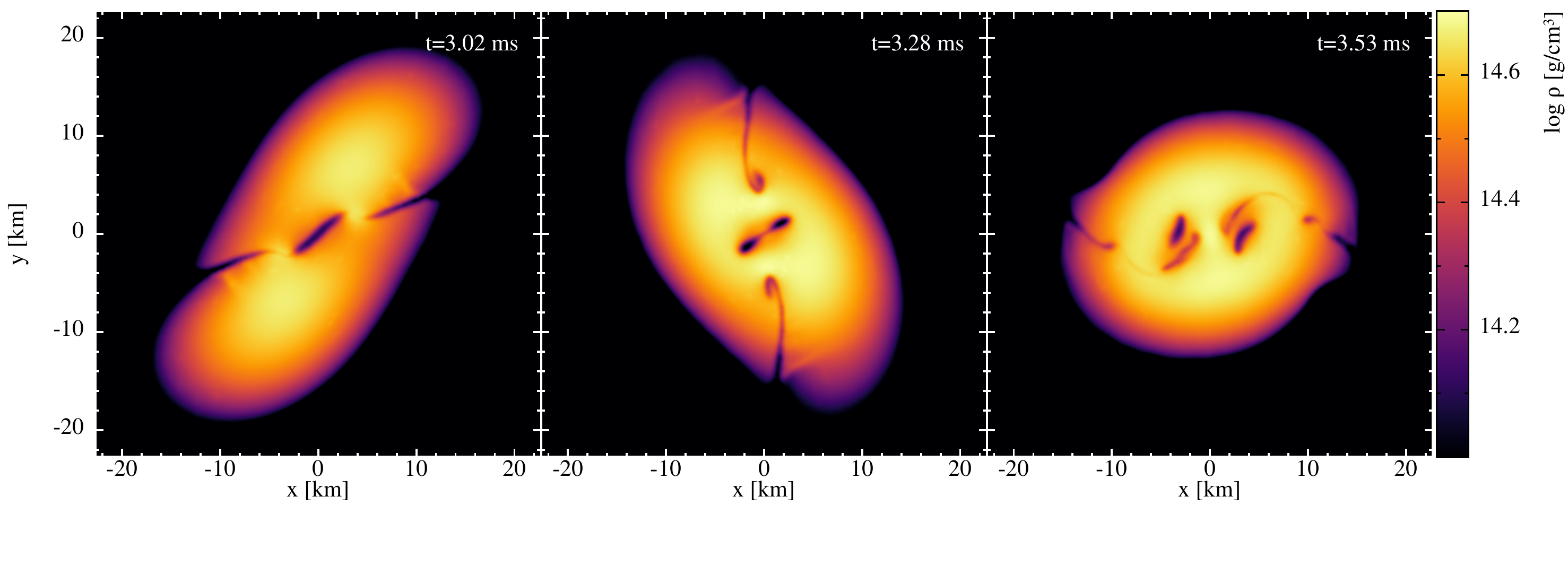}
  
  \vspace*{-0.6cm}
  
\caption{Evolution of the Kelvin--Helmholtz vortices in simulation \texttt{HR\_2x1.4\_G2.75}, see \autoref{tab:run_params}.}
\label{fig:KH}       
\end{figure*}

At the interface where the two neutron stars come into contact, a 
Kelvin--Helmholtz unstable shear layer forms. Although we are presently
not modelling magnetic fields, it is worth pointing out that this
 shear layer is important for
magnetic field amplification. In the resulting
Kelvin--Helmholtz vortices initially present magnetic fields can be 
efficiently amplified beyond magnetar field strength 
\cite{price06,kiuchi14,kiuchi15,kiuchi18}, which obviously has a large
astrophysical relevance for potentially increasing the maximum remnant mass, 
magnetically ejecting matter and for launching  GRBs. Traditional
SPH-approaches have been found to be challenged in resolving weakly
triggered Kelvin--Helmholtz instabilities \cite{agertz07,mcnally12}, 
but these problems are absent in the modern high-accuracy
SPH-approach that is implemented in \spB, 
mostly due to the reconstruction procedure in the dissipative terms 
and to the much more accurate gradients than those used in ``old school SPH." \footnote{See Sec. 3.5 in the paper describing the
\texttt{MAGMA2}-code \cite{rosswog20a} for a detailed discussion of Kelvin--Helmholtz
instabilities within high-accuracy SPH.}
To illustrate the Kelvin--Helmholtz
instabilities that  emerge in our \SpB simulations, we show the density near the shear interface between the two stars for our simulation \texttt{HR\_2x1.4\_G2.75} in Fig.~\ref{fig:KH}. In this simulation four vortices are initially triggered, that subsequently move inwards and finally merge. 
While the initial stages show an almost perfect symmetry, see Fig.~\ref{fig:KH}, tiny asymmetries, seeded
by the numerics, are amplified by the Kelvin--Helmholtz instability and finally lead to a
breaking of exact symmetry.  It goes without saying that such a breaking of perfect symmetry will also occur in nature.

The Kelvin--Helmholtz instability also seeds physical odd-$m$ instabilities in
the merger remnant. In fact, \cite{radice16b} found, in a dedicated study,
that several odd-$m$ modes are seeded, among which the $m=1$ is the most
pronounced one. These modes grow exponentially and saturate on a time scale
of $\sim 10$ ms. The study concluded  that the appearance of the $m=1$
one-armed spiral instability is a generic outcome of a neutron star merger
for both ``soft" and ``stiff" equations of state. They found, however,  the
instability to be very pronounced for their stiff EOS (MS1b
\cite{mueller96}) and hardly noticeable for their soft EOS
(SLy \cite{douchin01}). These findings are consistent with our results
shown in Fig.~\ref{fig:dens_evol}, where the $\Gamma= 2.75$ cases show
noticeable deviations from perfect symmetry, whereas the $\Gamma=2.00$
cases show no obvious deviations.

\begin{figure*}
  \includegraphics[width=0.51\textwidth]{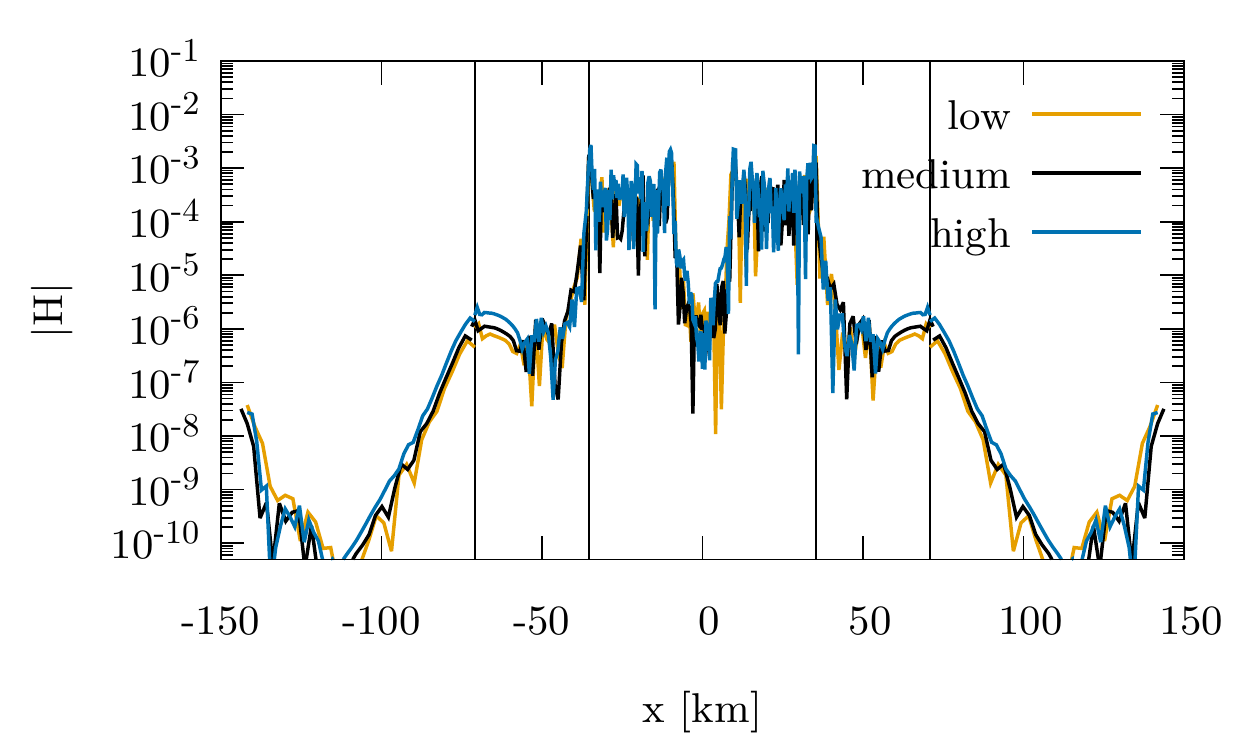}\hspace{-1em}
  \includegraphics[width=0.51\textwidth]{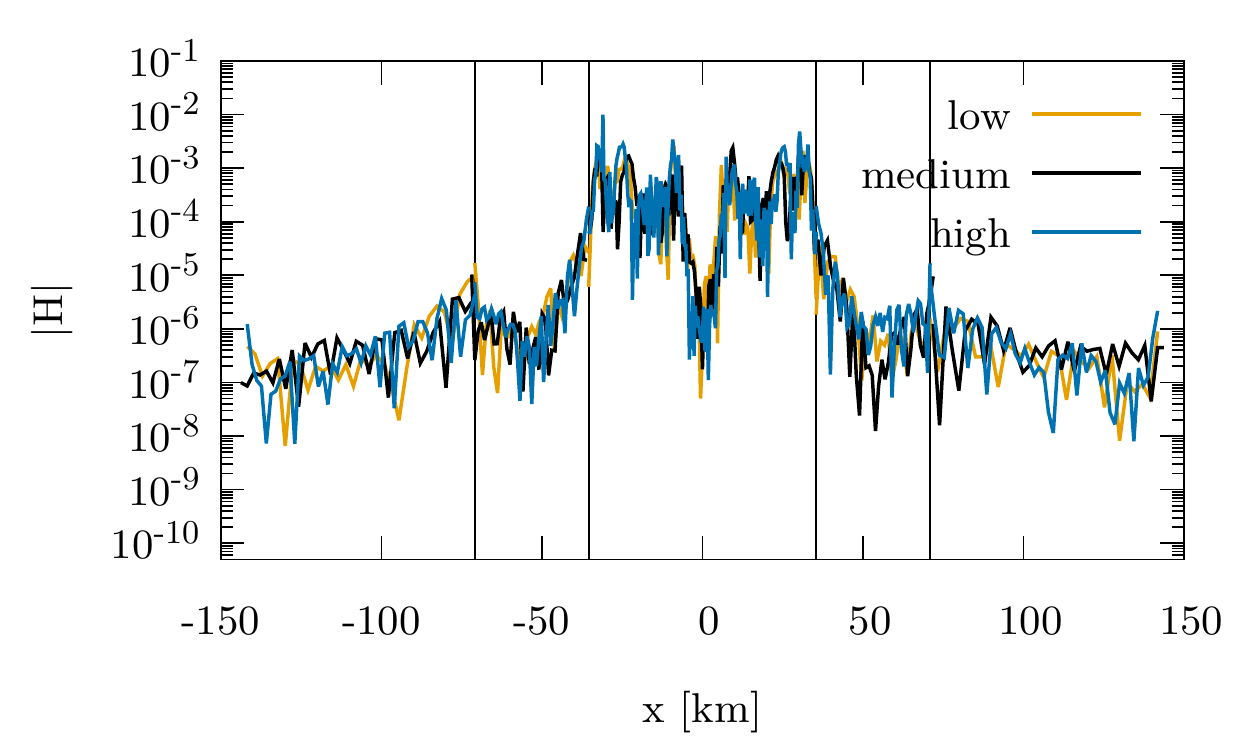}\\
  \includegraphics[width=0.51\textwidth]{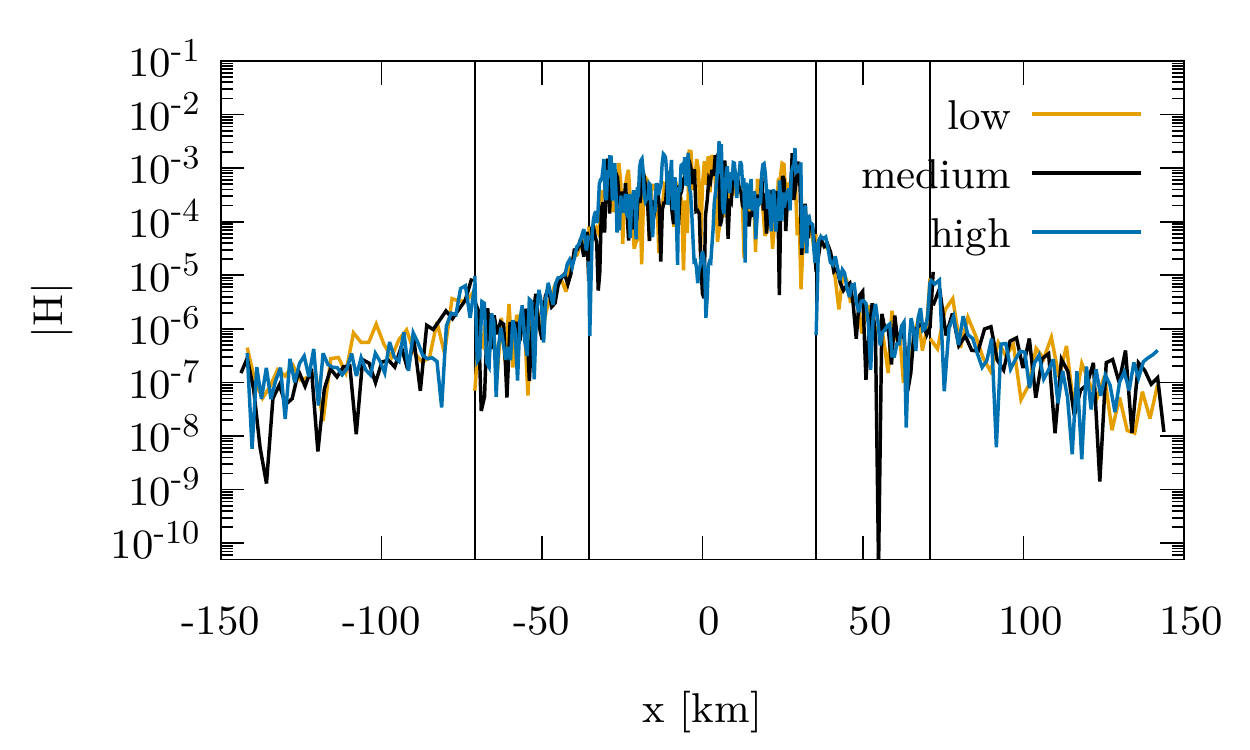}\hspace{-1em}
  \includegraphics[width=0.51\textwidth]{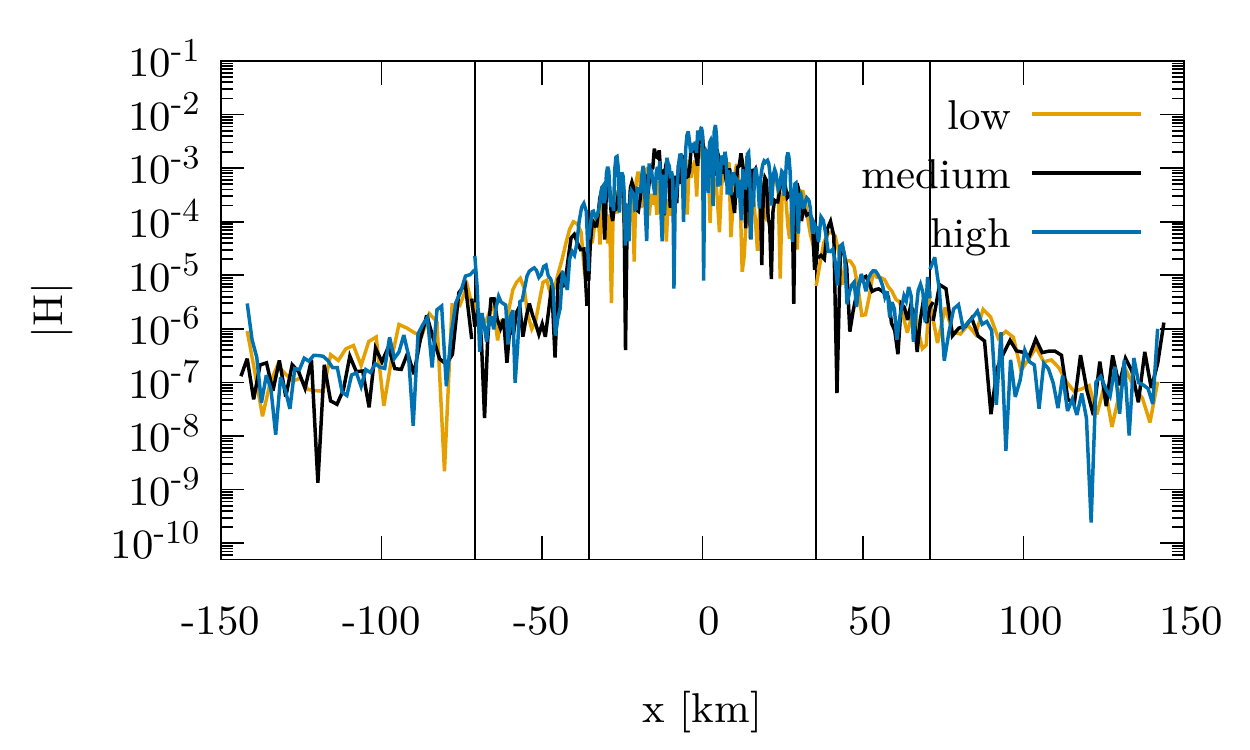}
\caption{The Hamiltonian constraint along the $x$-axis scaled for 
$2^{\rm nd}$ order convergence  for the runs with two 1.3 \Msun stars and the 
soft equation of state ($\Gamma=2.00$) at different times in the 
simulations. We have only plotted the three finest refinement levels 
and have indicated the boundaries between refinement levels with
vertical thin lines. The top left plot is at $t=0.05$ ms, that is, shortly 
after the beginning of the simulation. The top right plot
is at the $t=1.81$ ms (low resolution), $t=1.84$ ms (medium 
resolution) and $t=1.87$ ms (high resolution), when the neutron
stars have returned to  the $x$-axis after half an orbit. The 
bottom left plot is at $t=3.13$ ms (low resolution), $T=3.23$ ms 
(medium resolution) and $t=3.33$ ms (high resolution), when
the stars have again returned to the $x$-axis after having completed
a full orbit. The final bottom right plot is at a $t=13.1$ ms, that is,\
a time significantly after the merger.}
\label{fig:Ham}       
\end{figure*}

Last, but not least, we show in Fig.~\ref{fig:Ham} the 
Hamiltonian constraint along the
$x$-axis for representative simulations (two 1.3 \Msun stars with the
$\Gamma=2.0$ equation of state) at 4 different times: 
shortly after the start of the simulations, after half
an orbit, a full orbit (shortly before the neutron
stars touch) and a significant time after the merger.
In each plot 3 resolutions are shown, scaled for $2^{\rm nd}$ order convergence. In 
the top left plot (early in the simulations) the 
accuracy of the initial data limits the convergence for $|x|>50$ km. In the 
three remaining plots, it is clear that the constraint violations are largest 
where the matter is located, but remain low in the exterior and converge at
about $2^{\rm nd}$ order. See \ref{app:constraints} for the computation of
the constraint violations for the initial data.

\section{Summary and outlook}
\label{sec:summary}

In this paper, we have described the current status of our freshly developed
general-relativistic, Lagrangian hydrodynamics code \spB. Since the focus of our original
paper \cite{rosswog21a} was to demonstrate the ability to accurately
handle fully dynamical, general relativistic single stars, our main
focus here is on those new methodological elements that are needed
to simulate relativistic binary neutron star mergers.
These elements include, in particular,   a structured mesh
(``fixed mesh refinement"; see Sec~\ref{sec:spacetime}) on which we evolve the
spacetime. This, in turn, requires improvements to the particle--mesh coupling, 
as explained in detail in Sec.~\ref{sec:particle_mesh}. Finally, we describe a new method to 
accurately set up SPH particles based on the results from initial 
data solvers such as \Lo, see Sec.~\ref{subsec:ID}. The method is implemented in a new code called \spl.

Given the central role of Numerical Relativity for relativistic astrophysics 
and for gravitational wave astronomy, we consider this (to the best of our knowledge)
{\em first} Lagrangian hydrodynamics approach that solves the full 
set of Einstein equations, as an important step forward. It will 
allow to validate both Eulerian Numerical Relativity approaches 
as well as Lagrangian ones that use only approximate strong-field gravity.

Despite this progress, it is clear that our current code version only contains 
the most basic physics ingredients, that is, relativistic hydrodynamics, dynamical 
evolution of the spacetime and simple equations of state. The first detection 
of a binary neutron star merger \cite{abbott17b,abbott17c}, however, has 
impressively underlined the multi-physics nature of  neutron star merger events.
It has demonstrated in particular that neutron star mergers are, as it had been expected for
some time \cite{lattimer74,eichler89,rosswog99,freiburghaus99b}, prolific sources
of r-process elements \cite{kasen17,kasliwal17,tanvir17,rosswog18a}. The detection of
the early blue kilonova component \cite{evans17} and the identification of strontium
\cite{watson19}, a very light r-process element that is only synthesized for
moderately large electron fractions, have emphasized that
weak interactions play a crucial role in shaping the observable features of a neutron
star merger event. Moreover,  the short GRB following GW170817
 demonstrated that (at
least some) neutron star mergers are able to launch relativistic jets. Taken together, these observations 
suggest that at least nuclear matter properties, neutrino physics and
likely magnetic fields are major actors in a neutron star merger event.

Our next \SpB development steps will
be geared towards both further technical improvements and towards more realistic
micro-physics. On the technical side, for example, we expect that the amount of dissipation
that is applied in the hydrodynamic evolution can be further reduced and we aim at further increasing 
the code's computational performance. In the current stage, both (artificial) dissipation and finite resolution likely still leave a noticeable imprint on the
simulation outcome. On the micro-physical side, we consider the implementation
of realistic nuclear matter equations of state as they are provided 
in the \comp database as an important and natural 
step forward. In addition 
to increasing the physical accuracy of the
matter description, this ingredient is also indispensable to follow the electron
fraction $Y_e$ of the ejecta and to implement (any kind of) neutrino physics. 
\comp equations of state can be used in \Lo but, in its present form,
\SpB does not yet support tabulated EOSs. The implementation of tabulated 
equations of state requires changes in the very core of our hydrodynamics scheme; in 
particular, we have to change the algorithm for the conversion from conserved to 
primitive variables. This has, of course, been successfully done in Eulerian contexts 
\cite{siegel18,kastaun21}, and is certainly also doable for our equation set, but
it will require some dedicated effort. Once this has been achieved, we will tackle
the implementation of a fast, yet reasonably accurate neutrino treatment such as the 
{\em Advanced Spectral Leakage Scheme} (ASL) \cite{perego16,gizzi19a,gizzi21a}. These issues
will be addressed in future work.

\appendix
\section{Gravitational wave extraction}
\label{sec:App1}

At this early stage of our new code, we can extract gravitational
waves in two different ways. We can either use the quadrupole
approximation directly while the simulation is running or we can
post process data afterwards using the Einstein Toolkit. In the
following we describe both methods.
\subsection{The quadrupole approximation}
The ``standard Einstein--Landau--Lifshitz quadrupole formula"  \cite{misner73,blanchet90}
reads 
\be
h_{ij} (t,\vec{x})= \frac{2G}{c^4 r} P_{ijkl}(\vec{n}) \frac{d^2}{dt^2} \rqm_{kl}(t-r/c).
\label{eq:hij}
\ee
Here $P_{ijkl}$ is the {\em transverse traceless projection operator}
\bea
P_{ijkl}(\vec{n})&\equiv& (\delta_{ik} - n_i n_k)(\delta_{jl} - n_j n_l) -\\ 
                             &&\frac{1}{2}(\delta_{ij} - n_i n_j) (\delta_{kl} - n_k n_l)\\
&=& P_i^k P_j^l - \frac{1}{2} P_{ij} P^{kl}
\eea
with $P_i^k= \delta_i^k - n_i n^k$, $r=|\vec{x}|= (\delta_{ij} x^i x^j)^{1/2}$, $\vec{n}= \vec{x}/r$ and ${\rqm}_{ij}= I_{jk} - \frac{1}{3} \delta_{ij} \sum_k I_{kk}$ is the reduced quadrupole moment.
The amplitude becomes maximal, $h_{\rm max}$, for an observer located on the
$z$-axis [Cartesian coordinates; see \cite{shibata16}, Eq. (1.47)]:
\be
h_{\rm max}= \frac{G}{c^4} \frac{1}{D_{\rm obs}}\left( \ddot{I}_{xx} -  \ddot{I}_{yy} \right),
\ee
where $D_{\rm obs}$ is the distance to the observer.
The total energy radiated in gravitational waves is given by 
\bea
E_{\rm GW}&=& \frac{1}{5} \frac{G}{c^5} \int_{-\infty}^\infty \dddot{\rqm}_{ij} \dddot{\rqm}_{ij} \; dt.
                   \label{eq:E_GW}
\eea

There is no unique way, in General Relativity, to define the quadrupole moment $I_{ij}$.
One could try to follow post-Newtonian (PN) approaches, but they are sometimes ill-defined
in the strong gravity regime and adding further PN-corrections does
not necessarily improve the result. Shibata and Sekiguchi \cite{shibata03b} find the quadrupole approximation
rather useful even in strong-field gravity, they quote  accuracies of $O(\mathcal{C})$
($\mathcal{C}$ being the compactness) for the amplitudes, that is, in practice  $\sim 20\%$, and much better accuracies
for the phase evolutions.

Following \cite{blanchet90,shibata03b} we calculate the quadrupole moment via the
 {\em conserved rest-mass density},
\be
\rho^\ast \equiv \sqrt{-g}\, U^0 \rho_{\rm rest},
\ee
with the rest-mass density, 
\be
\rho_{\rm rest}= n\, m_0.
\ee
The quantity $\rho^\ast$ is  related to our density variable $N$ (Eq.~\ref{eq:N_def}) by
\be
\rho^\ast= m_0 N.
\ee 
The major advantage of this approach is that the first time derivative can be expressed
analytically as
\be
\frac{dI_{ij}}{dt}= \int d^3x \; \rho^\ast (v^i x^j + v^j x^i).
\ee
For the SPH representation, we transform the integral  into a sum using the particles' volume elements 
$V_b= \nu_b/N_b$,
\bea
\frac{dI_{ij}}{dt} &=& m_0 \sum_b \nu_b (v^i_b x^j_b + v^j_b x^i_b).
\label{eq:dI_dt}
\eea
To obtain the second and third time derivatives of $I_{ij}$, we monitor $dI_{ij}/dt$ at a set of instances
in time, calculate a least square fit to them and take the analytical time derivatives of the approximating
least square polynomial.

\subsection{Extraction from the spacetime via the Einstein Toolkit}
\label{app:etk}
During the evolution we regularly write checkpoint files in order to be
able to restart the simulation. There is one such checkpoint file for the particle
(matter) data and another one for the grid (spacetime/BSSN) data. We have written
a file reader that can read one complete refinement level of 
BSSN data from the checkpoint file into Cactus and thereby run
the gravitational wave extraction tools available in the Einstein 
Toolkit. This consists of the thorns \Extract, \WeylScal4 and 
\Multipole. In \Extract one, or several, extraction surfaces can
be set up and the thorn will then estimate a suitable background
metric and extract the Regge--Wheeler--Zerilli gauge invariant
even $Q^+_{\ell m}$ and odd $Q^{\cross}_{\ell m}$ parity master
functions. Thorn \WeylScal4 calculates the Newman--Penrose Weyl
scalar $\Psi_4$ everywhere on the grid. Thorn \Multipole is then used
to decompose $\Psi_4$ into spin-weighted spherical harmonics modes
$\Psi_4^{\ell m}$ on a set of defined coordinate spheres.

Conveniently, \cite{alcubierre08} has collected all the necessary
formulae for recovering the strain and to calculate the radiated
energy and angular momentum (as well as linear momentum, but that
is not needed here). For completeness we list these formulae here.

With the definition
\be
P^{\cross}_{\ell m}\coloneqq\int_{-\infty}^t Q^{\cross}_{\ell m} dt',
\ee
the strain can be recovered from $Q^+_{\ell m}$ and $P^{\cross}_{\ell m}$ as
\begin{equation}
h^+ +i h^{\cross}=\frac{1}{\sqrt{2}r} \sum_{\ell,m}\left [
Q^+_{\ell m}-i P^{\cross}_{\ell m}\right] 
\prescript{}{-2}Y_{\ell m},
\end{equation}
where $r$ is the distance from the source and 
$\prescript{}{-2}Y_{\ell m}$ is the spin-weight $-2$ spherical
harmonic and the sum is over all possible $\ell$ and $m$. The
radiated energy can be calculated as
\begin{equation}
\frac{dE}{dt}=\lim_{r\rightarrow\infty}\frac{1}{32\pi}\sum_{\ell,m}\left (
\left | \dot{Q}^+_{\ell m}\right |^2+\left | Q^{\cross}_{\ell m}\right |^2\right ),
\end{equation}
where a dot indicates the time derivative. The radiated angular
momentum in the $z$-direction is
\begin{equation}
\frac{dJ_z}{dt}=\lim_{r\rightarrow\infty}\frac{i}{32\pi}\sum_{\ell,m}
m\left (\Bar{Q}^+_{\ell m}\dot{Q}^+_{\ell m}+\Bar{P}^{\cross}_{\ell m}Q^{\cross}_{\ell m}\right ),
\end{equation}
where an overbar means a complex conjugate.

Starting from the $\Psi_4^{\ell m}$ modes, the strain can be
recovered as
\begin{equation}
h^+ +i h^{\cross}=-\int_{-\infty}^t \int_{-\infty}^{t'}\sum_{\ell,m}\Psi_4^{\ell m} \prescript{}{-2}Y_{\ell m}\, dt''\, dt'.
\end{equation}
The radiated energy can also be calculated from the $\Psi_4$ modes as
\begin{equation}
\frac{dE}{dt}=\lim_{r\rightarrow\infty}\frac{r^2}{16\pi}\sum_{\ell,m}
\left |\int_{-\infty}^t \Psi_4^{\ell m} dt' \right |^2
\end{equation}
and the radiated $z$-component of the angular momentum as
\begin{eqnarray}
\frac{dJ_z}{dt} & = & -\lim_{r\rightarrow\infty}\frac{i r^2}{16\pi}\Im 
\left [\sum_{\ell,m} m\int_{-\infty}^t\int_{-\infty}^{t'}
\Psi_4^{\ell m}\, dt'\,dt'' \right. \nonumber \\
 & & \left. \times\int_{-\infty}^t\Bar{\Psi}_4^{\ell m}
\right ].
\end{eqnarray}
The analysis of the extracted modes from the Cactus simulation has been performed using \texttt{kuibit} \cite{Bozzola2021}.

\section{Details on the Artificial Pressure Method}
\label{sec:App2}

\subsection{The initial particle distribution on spherical surfaces}
\label{subsec:app-spherical}

The aim is to find an as-good-as-possible 
representation of the \Lo density distribution
by means of a finite number of 
(close to) equal-mass SPH particles. 
We start with a trial particle
distribution and perform iterative APM improvements on
it until an optimal configuration has been found.
Obviously, the better the initial
particle distribution, the fewer APM iterations will be needed to reach the final goal.
We had  experimented with placing 
particles on cubic lattices
as initial positions, but starting 
from spherical surfaces delivered a) a slightly better density accuracy and b) resulted in a locally isotropic particle
distribution while in the cubic lattice
case preferred directions were still visible.
Our method of choice therefore starts
with particles placed on spherical surfaces and is described in more detail below.

 We describe the method for one star only, since it is applied separately and independently for each star. First, we compute the ``desired particle mass" as
\begin{align}
\label{eq:mdes}
    \massp= \dfrac{M_\mathrm{b}}{n_\mathrm{des}},
\end{align}
where $M_\mathrm{b}$ is the baryon mass of the star given by \Lo, and $n_\mathrm{des}$ ``desired particle number" for the star, specified by the user. Second, we integrate the baryon mass density of the star $\rho_\mathrm{b}(r,\theta,\phi)$ given by \Lo, over $(r, \theta,\phi)$ and extract the radial baryon mass profile $M_\mathrm{b}(r)$. Notice that, since the star is not spherically symmetric, and we place particles on the spherical surfaces uniformly, we lose the density information over $(\theta,\phi)$ in this step. Third, in order to set the number of spherical surfaces inside the stars, we estimate the ``radial particle profile" as
\begin{align}
    \mathrm{par}(r)\coloneqq \int_{0}^r\bigg(\dfrac{\rho_\mathrm{b}(\xi)}{\massp}\bigg)^{1/3}\dd \xi ,
\end{align}
Thus, we define the number of spherical surfaces as
\begin{equation}
\label{eq:ns}
    n_\mathrm{s}\coloneqq \ceil*{\mathrm{par}(R)},
\end{equation}
where $\ceil*{\cdot}$ is the ``ceiling" function, and $R$ the larger radius of the star---the equatorial radius on the $x$ axis towards the companion. 
Fourth, we need to set the radii of the spherical surfaces. We would like to have a larger density of surfaces where the baryon mass density is larger, and a lower density of surfaces where the baryon mass density is lower. Therefore, we place the first spherical surface at a radius given by
\begin{equation}
\label{eq:firstr}
    r_1= \left(\dfrac{\rho_\mathrm{b}^\mathrm{center}}{\massp}\right)^{-1/3},
\end{equation}
and the others at radii\footnote{One might think that defining $n_\mathrm{s}$ in \eqref{eq:ns} is redundant, since we can just place the surfaces according to \eqref{eq:firstr},\eqref{eq:rs}. However, it is necessary to decide when to stop placing surfaces, that is, to compute $n_\mathrm{s}$ in some sensible way. We thought that \eqref{eq:ns} was sensible enough to be tried out; since the resulting particle distributions are satisfying to us, we have kept it.}
\begin{equation}
\label{eq:rs}
    r_i= r_{i-1} + \left(\dfrac{\rho_\mathrm{b}(r_{i-1})}{\massp}\right)^{-1/3}, \quad i \in \lbrace 2,...,n_\mathrm{s} \rbrace.
\end{equation}
The density in \eqref{eq:rs} is evaluated along the larger equatorial radius (i.e., along the $x$ axis in the direction of the companion). If, for some $i < n_\mathrm{s}$, $r_i$ falls outside of the star, we rescale all the radii by the same factor $i/n_\mathrm{s} < 1$. It can still happen that the last surface falls outside of the star, hence we let the location of the last surface, $r_\mathrm{last}$, to be specified by the user. Note that this choice does affect how many particles are placed close to the surface and where. We found that placing the last surface at $r_\mathrm{last}\in [0.95R, 0.99R]$ allows us to place enough particles close to the surface without allowing for very low-mass particles. Hence, when we know all the radii, we rescale them as
\begin{equation}
    r_i \rightarrow \dfrac{r_\mathrm{last}}{r_{n_\mathrm{s}}}\, r_i ,
\end{equation}
so that $r_{n_\mathrm{s}}\rightarrow r_\mathrm{last}$.
At this point, we assign to each spherical surface a baryon mass given by
\begin{equation}
\label{eq:surfacemass}
    M_i= 
    M_\mathrm{b}(r_i)-M_\mathrm{b}(r_{i-1}),
    \quad i \in \lbrace 1,...,n_\mathrm{s} - 1 \rbrace.
\end{equation}
where $i$ is the surface index and $r_0= 0$ (center of the star). For the last surface, the baryon mass is assigned as
\begin{equation}
\label{eq:lastmass}
    M_\mathrm{last}= 
    M_\mathrm{b}(R)-M_\mathrm{b}(r_{n_\mathrm{s}-1}),
\end{equation}
In the last step, we place $n_i$ particles on each surface, where
\begin{equation}
\label{eq:surfaceparts}
    n_i= \nint{M_i/\massp},
\end{equation}
where $\nint{x}$ returns the nearest integer to $x$.

The particles are placed uniformly, avoiding their clustering around the poles, according to the algorithm described in \cite{weisstein}. This consists of placing particles uniformly in the azimuth $\phi$ and $\nu$, with $\nu\in (0,1)$ and colatitude $\theta=\arccos (2\nu-1)$. We place the same number of particles on each meridian and each parallel, such that each quarter of meridian (parallel) contains the same number of particles. In other words, the number of particles $n_\mathrm{circle}$ on each meridian and parallel is a multiple of 4. Consider the part of a meridian spanned from $\theta=0$ to $\theta=\pi/2$. The number of particles on this curve is $n_\mathrm{circle}/4$ by construction. Then, the number of particles over the entire northern hemisphere is $n_\mathrm{circle}^2/4$. The total number of particles on the spherical surface is then
\begin{align}
\label{eq:npartsurface}
    n_\mathrm{s}= \dfrac{n_\mathrm{circle}^2}{2}.
\end{align}
In practice, $n_s$ is set by \eqref{eq:surfaceparts}, and $n_\mathrm{circle}$ is computed by inverting \eqref{eq:npartsurface},
\begin{align}
\label{eq:npartcircle}
    n_\mathrm{circle}= \nint{\sqrt{2n_\mathrm{s}}}.
\end{align}
Note that $n_\mathrm{circle}$ has to be a multiple of 4, so we correct it before using it. After correcting $n_\mathrm{circle}$, we consistently recompute $n_\mathrm{s}$ and place the particles at the desired positions, as described above. At each of these positions, we check that the \Lo mass density is positive, and we place a particle only if it is.

We highlight that this algorithm keeps the same mass for particles on the same surface, but due to the described rounding to integers, the particle mass changes a little between surfaces. In addition, for a spherical shell close to the surface of the star, some positions have a zero density since the star is not spherically symmetric. Such positions are not promoted to be particles, hence the final number of particles on the spherical surface changes (i.e., it is not $n_\mathrm{s}$ anymore). Thus, the particle mass on the spherical surface changes, since the mass of the surface in \eqref{eq:surfacemass}, \eqref{eq:lastmass} is kept the same. Since we would like to have (almost) equal mass particles, the change in particle mass between the surfaces should be small enough. To address this issue, we set up an iteration. 
The iteration replaces the particles on each surface, with the exception of the first surface considered, so that the particle mass on a surface is not too much different than the particle mass on the previous surface. The tolerance is specified by the user, and we found that a $2.5\%$ difference is reasonable.\footnote{It can happen that, since the particle number is an integer and the mass of a surface is a real, it's not possible to achieve a $2.5\%$ difference. When this happens, we allow for a larger difference if convergence is not achieved after a certain number of iterations, usually 100. However, we chose the $2.5\%$ difference because it is possible to achieve it for most of the surfaces.}

Note that since we impose $n_\mathrm{des}^\mathrm{star\, 1}/n_\mathrm{des}^\mathrm{star\, 2}=M^1_\mathrm{b}/M^2_\mathrm{b}$, and since the particle number obtained at the end of the iteration is close to the desired particle number, the particle numbers on the stars will have a ratio very close to the binary mass ratio. Thus, the particle mass ratio \emph{across the stars} will be similar to the particle mass ratio within each star, the latter being small thanks to the iteration on the spherical surfaces and, more importantly, to the APM iteration described in Sec.~\ref{subsec:apm}.

Finally, since the particles on spherical surfaces have very similar masses, it is well-justified to assume them to be equal-mass (assumption that the APM uses) and use them as initial condition for the APM iteration. 

\subsection{Placing ghost particles for BNS}
\label{subsec:app-ghost}

We place the ghost particles on a cubic lattice around each star if two conditions are satisfied: the baryon mass density given by \Lo at that point is 0---so we are outside the star---and the position is between two ellipsoidal surfaces. The semi-axes of the innermost ellipsoidal surface are
\begin{subequations}
\label{eq:ellipseradii}
\begin{align}
  \xi_x &= R + \delta\\
  \xi_y &= R_y + \delta\\
  \xi_z &= R_z + \delta
\end{align}
\end{subequations}, 
with $R$ being the equatorial radius towards the companion, $R_y$, $R_z$ the radii in the $y$ and $z$ direction (given by \Lo), and  series $\delta = 0.3$km 
being a constant whose purpose is described next. The semi-axes of the innermost ellipsoidal surface are placed a little outside of the surface of the star---this is achieved by adding $\delta$ in \eqref{eq:ellipseradii}---to allow for the real particles to move towards the surface of the star if the artificial pressures of their neighbours push them there. If the ghost particles were placed exactly on the surface, then the real particles would not move towards it due to the high pressure gradient they would feel. The  actual value of $\delta$ was determined empirically.

\section{Constraint violations and convergence for the BNS ID}
\label{app:constraints}

After setting up the SPH and BSSN ID, we check that they satisfy the constraint equations using a subroutine from the Einstein Toolkit, which is adapted to our framework in \spB.

We compute the Hamiltonian and momentum constraints in two ways. First, using the full (spacetime and hydro) \Lo ID on the refined mesh, we compute the stress--energy tensor and finally the constraints. This method does not involve the SPH ID in any way. Second, we use the SPH ID on the particles and the BSSN ID on the refined mesh. We compute the stress--energy tensor at the particle positions, which needs the metric to be mapped to the particles (M2P step), and we map the stress--energy tensor to the refined mesh (P2M step). The interpolated stress--energy tensor on the mesh is then used together with the BSSN ID to compute the constraints. The two methods provide us also with another way to test the accuracy of the mapping routines.

We perform local convergence tests \cite[Sec.~9.11]{alcubierre08} with the constraints computed in both ways, to test the robustness of our codes.\footnote{Even if this is all at the level of the ID, \spl uses routines from \spB, so the convergence tests also test the robustness of some parts of \spB.} Note that we do not recompute the entire ID with a different resolution when performing the convergence tests, since this is not possible in our framework. Even if we change the grid resolution, the \Lo \emph{spectral} ID will stay the same; namely, it does not improve its accuracy. Hence, we expect to see convergence only if the finite  error when computing spatial derivatives is larger than the error in the spectral solution from \Lo.

We found that keeping low finite difference order (FDO), and to a minor extent, keeping a low mesh resolution, is crucial to keep the finite differencing error larger than the spectral error. With FDO$=2$, we see convergence for some sequences of resolutions $\lbrace \Delta $, $\Delta/r$, $\Delta/r^2 \rbrace$, with $r\in \lbrace 4/3,3/2,2 \rbrace$ (same resolution on each Cartesian axis). With FDO$=4,6$, the \Lo error dominates and it is harder to see convergence since the constraint violations change little, or not at all, when changing the resolution.

\begin{figure*}[!ht]
    \centering
    \subfloat{\hspace{-2.25mm}\includegraphics[width=0.49875\textwidth,valign=b]{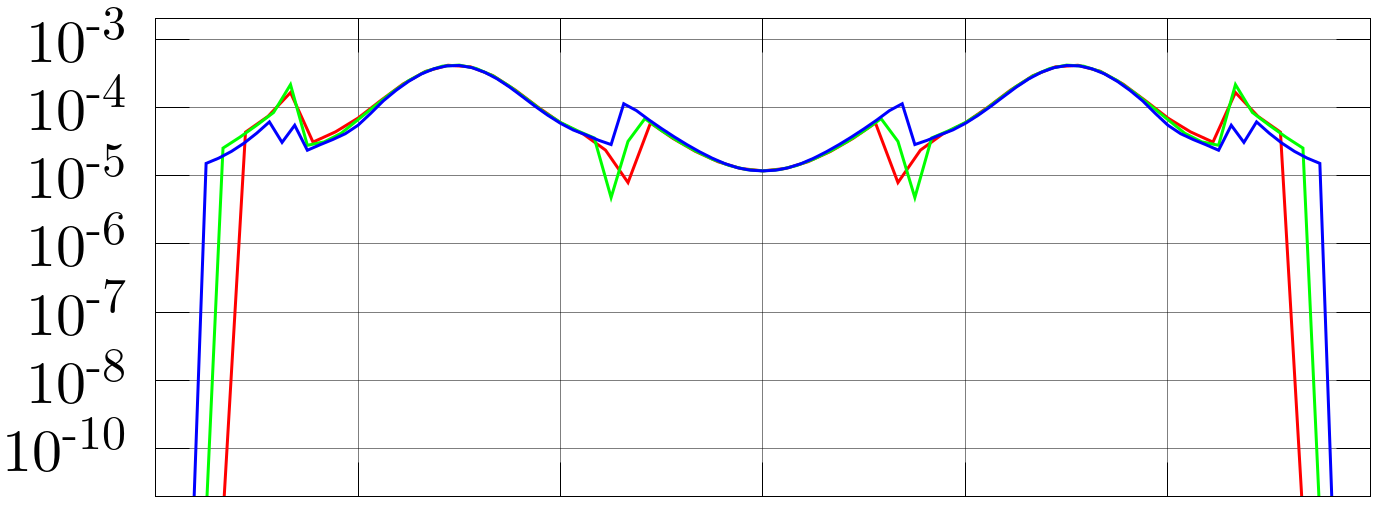}}
    \subfloat{\hspace{4mm}\includegraphics[width=0.4525\textwidth,valign=b]{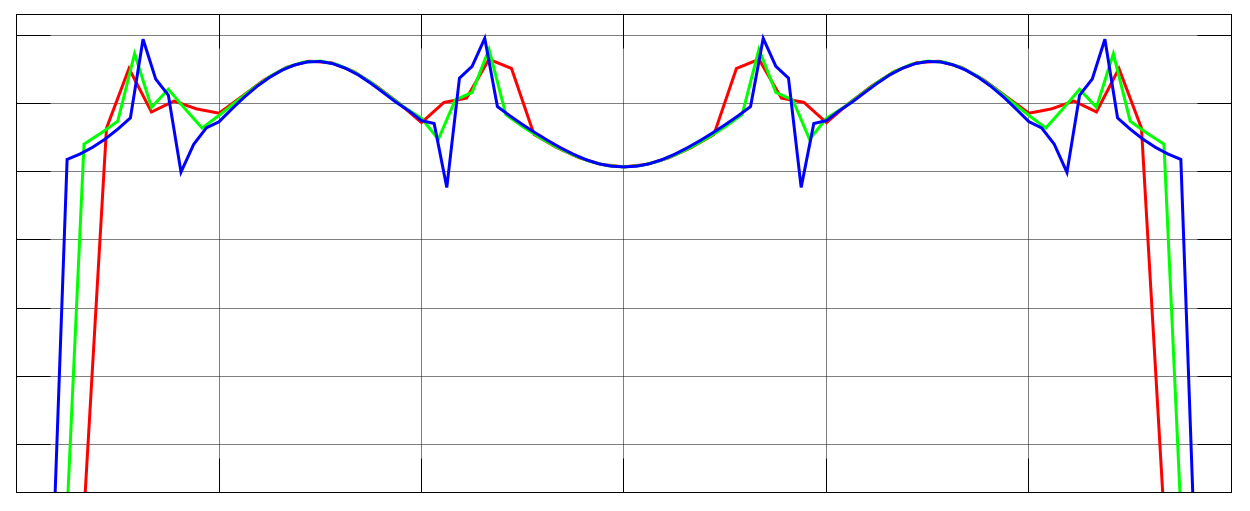}} \\[-3mm]
    \subfloat{\hspace{-2.25mm}\includegraphics[width=0.49875\textwidth,valign=t]{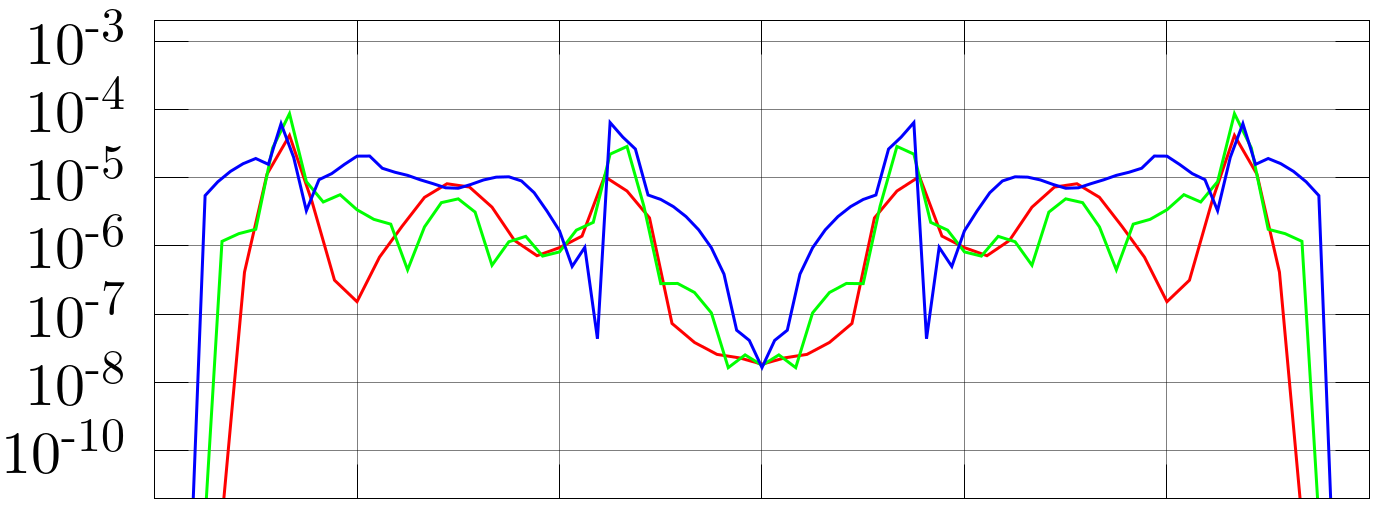}}
    \subfloat{\hspace{4mm}\includegraphics[width=0.455\textwidth,valign=t]{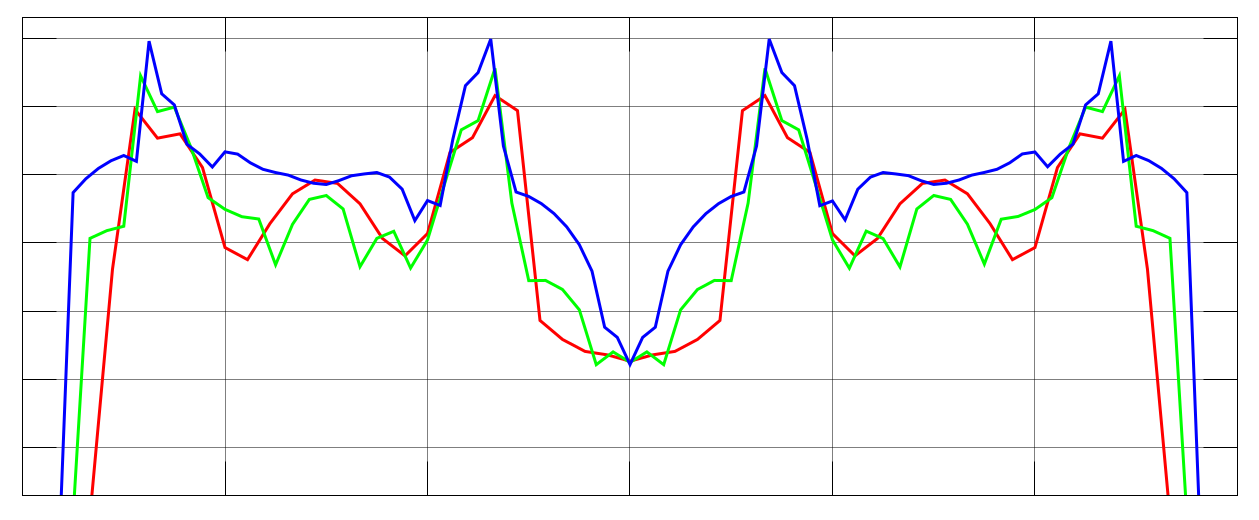}} \\[-3mm]
    \subfloat{\hspace{-0.25mm}\includegraphics[width=0.51125\textwidth,valign=t]{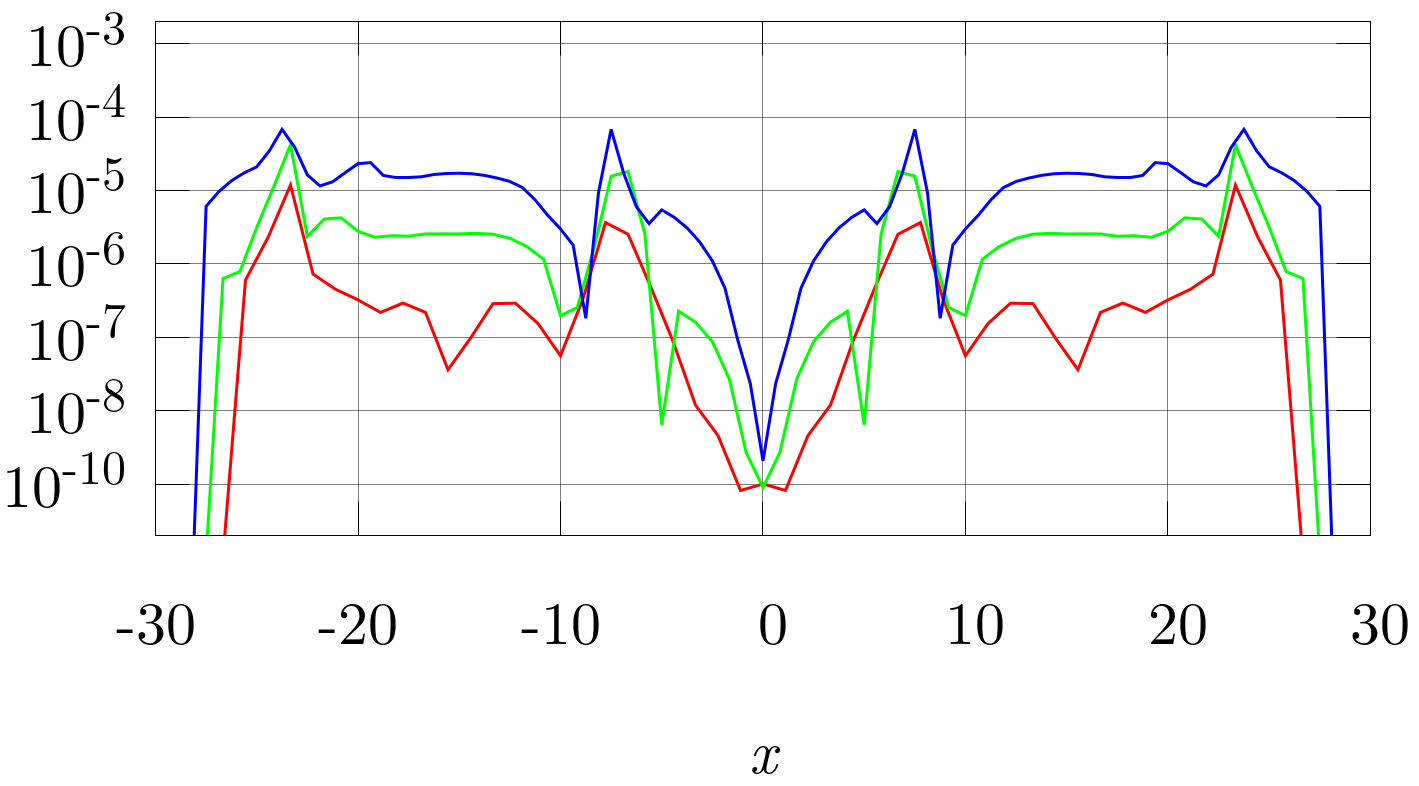}}
    \subfloat{\hspace{0.0mm}\includegraphics[width=0.47925\textwidth, valign=t]{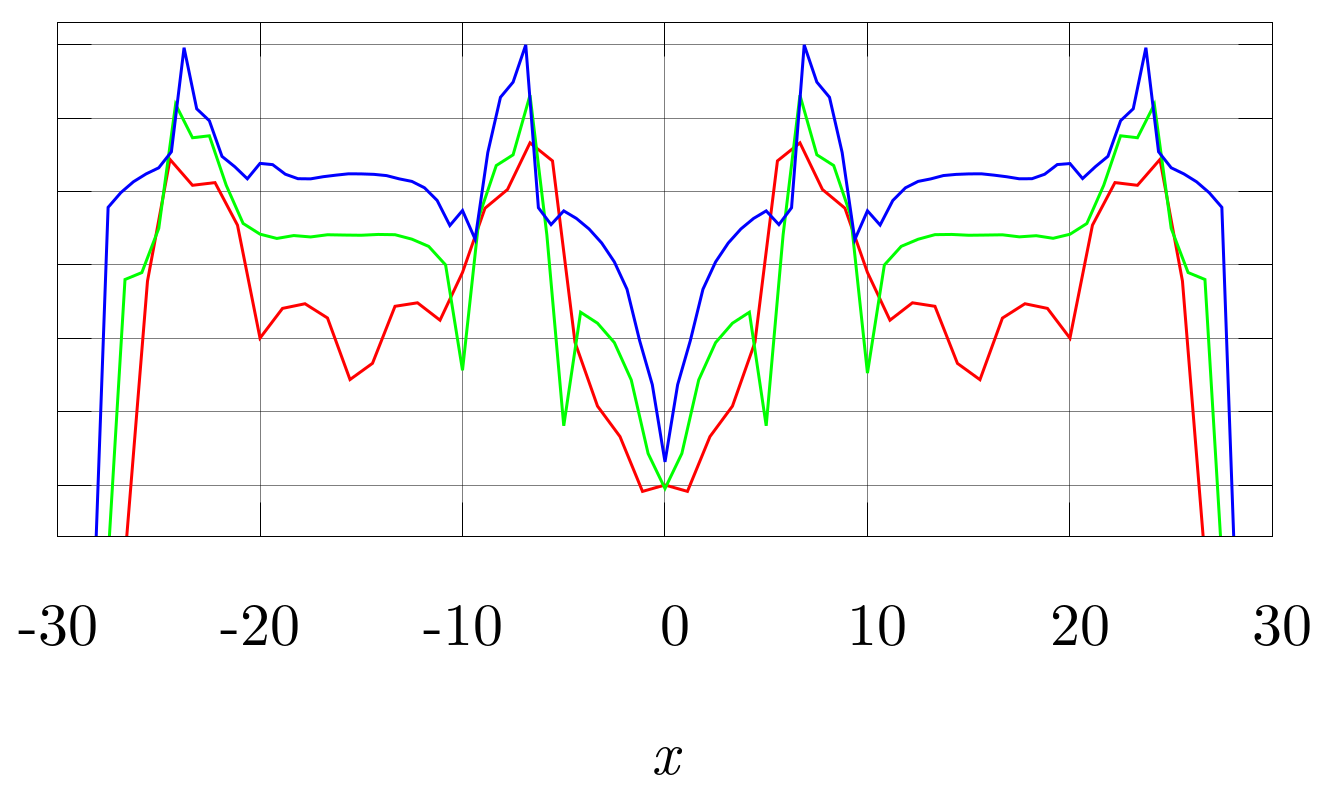}}
    \caption{Rescaled absolute value of the Hamiltonian constraint violations on  $x\in [-44.3,44.3]$km, computed using 3 different mesh resolutions. In the left panels, the full \Lo ID (spacetime and hydro) on the mesh (with one refinement level) is used; in the right panels, the computation uses the spacetime \Lo ID on the mesh (with one refinement level) and the stress--energy tensor computed at the particle positions and mapped to the mesh. The ID is the one for the run \texttt{LR\_2x1.3\_G2.00}. The coarse resolution (red) has $55^3$ grid points and $\Delta \simeq 1.64$km; the medium resolution (green) has $73^3$ grid points and $\Delta \simeq 1.23$km; the fine resolution (blue) has $97^3$ grid points and $\Delta \simeq 923$m. The ratio between the grid spacings is $4/3$, hence the Hamiltonian constraint  violations on the medium resolution are divided by $(4/3)^{\rm FDO}$, and the one on the coarse resolution are divided by $(4/3)^{2\rm FDO}$. In the panels, from top to bottom, the FDO is $2$, $4$, $6$. The 3 rescaled Hamiltonian constraint violations in each panel should overlap to show convergence; this is seen only when FDO=2 since, for higher FDO, the error produced by \Lo is larger than the error produced by the finite difference approximation. Note that increasing the mesh resolution would decrease the finite differencing error too, the \Lo error would dominate, and there would be no convergence. Compared to the constraints violations computed using the ID on the mesh (left panels), those computed using the mapping (right panels) are larger at the surfaces of the stars, but comparable inside the stars. See text for more details. The lengths on the $x$ axis are given in units of the solar mass in geometrized units.}
    \label{fig:cauchy-conv-local}
\end{figure*}

\autoref{fig:cauchy-conv-local} shows the rescaled absolute value of the Hamiltonian constraint  violations on the $x$ axis for a sequence of resolutions with $r=4/3$, computed with the two methods described above, for the \texttt{LR\_2x1.3\_G2.75} ID. In the examples shown in the figures, we see convergence only when FDO=2. Comparing the left and right panels of the figure, we note that the constraint violations are larger at the surfaces of the stars when mapping the metric from the mesh to the particles, and then the stress--energy tensor from the particles to the mesh, rather than just importing the \Lo ID on the mesh. This means that the error introduced by the mapping is a lot larger at the surfaces of the stars, compared to inside the stars. Specifically, in \autoref{fig:cauchy-conv-local}, the mapping induces constraint violations of order between $10^{-4}$ and $10^{-3}$ at the surface of the star, irrespective of the FDO order and the mesh resolution. Lastly, even though we see convergence with FDO=2 only, it is evident from the figures that using a larger FDO leads to lower constraint violations.

\begin{acknowledgements}
We thank E. Gourgoulhon, R. Haas and J. Novak for useful clarifications concerning \Lo; M. Oertel for rewarding discussions on how to use the \comp database and software, and \Lo;  C. Lundman for interesting discussions and S.V. Chaurasia for sharing his
insights into the BAM code.  It is also a pleasure to thank Koutarou Kyutoku for his valuable feedback on the manuscript.
SR has been 
supported by the Swedish Research Council (VR) under grant number 2016- 03657\_3, by 
the Swedish National Space Board under grant number  Dnr. 107/16,  by the 
research environment grant ``Gravitational Radiation and Electromagnetic Astrophysical
Transients (GREAT)'' funded by the Swedish Research 
council (VR) under Dnr 2016-06012, by which also FT is supported, and by the Knut and Alice Wallenberg Foundation
under grant Dnr. KAW 2019.0112. 
We gratefully 
acknowledge inspiring interactions via the COST Action CA16104 
``Gravitational waves, black holes and fundamental physics'' (GWverse) and  COST Action CA16214
``The multi-messenger physics and astrophysics of neutron stars'' (PHAROS).
PD would like to thank the Astronomy Department at SU and the Oscar Klein Centre for their hospitality during
numerous visits in the course of the development of \spB.
The simulations for this paper were performed on the facilities of the North-German Supercomputing Alliance (HLRN),
 on the resources provided by the Swedish National Infrastructure for Computing (SNIC) 
in Link\"oping  partially funded by the Swedish Research Council through grant agreement no. 2016-07213
and on the {\em SUNRISE} HPC facility supported by the Technical Division at the Department of Physics, 
Stockholm University.
Portions of this research were also conducted with high
performance computational resources provided by the Louisiana Optical Network
Infrastructure (http://www.loni.org).
\end{acknowledgements}

\bibliography{biblio}
\bibliographystyle{spphys}       

\end{document}

%% file: own_latex_commands_i.tex
\def\p{\partial}
\def\msun{M$_{\odot}$}
\def\Msun{M$_{\odot}$ }
\def\be{\begin{equation}}
\def\ee{\end{equation}}
\def\bi{\begin{itemize}}

\def\ei{\end{itemize}}
\def\ben{\begin{enumerate}}
\def\een{\end{enumerate}}
\def\bea{\begin{eqnarray}}
\def\eea{\end{eqnarray}}
\def\bt{\begin{tabbing}}
\def\et{\end{tabbing}}

\def\edo{\end{document}}

\def\M4{M$_4$}
\def\M6{M$_6$}
\def\lcm6{LCM$_6$}
\def\qcm6{QCM$_6$}

\def\wh3{W$_{\rm H,3}$}
\def\wh4{W$_{\rm H,4}$}
\def\wh5{W$_{\rm H,5}$}
\def\wh6{W$_{\rm H,6}$}
\def\wh7{W$_{\rm H,7}$}

\newcommand{\rqm}{{%
      \declareslashed{}{\text{-}}{0.04}{0}{I}\slashed{I}}}